\documentclass[a4paper,12pt,english]{article}

\usepackage{amsfonts,bm,amssymb,euscript,array,babel,cite}
\usepackage{amsmath}
\usepackage{mathtools}
\usepackage{mathrsfs}
\usepackage{stmaryrd}
\usepackage{tensor,accents}
\usepackage{hhline}
\usepackage[amsmath]{empheq}
\usepackage{hyperref}
\usepackage{cancel}
\usepackage{multirow}
\usepackage{pdfsync}
\usepackage{tikz}
\usepackage{tikz-cd}

\DeclareRobustCommand{\mathbullet}{\accentset{\sbullet}}

\numberwithin{equation}{section}

\makeatletter

\newcommand{\dd}{\mathrm{d}}
\newcommand{\DD}{\mathrm{D}}

\def\nn{\nonumber}
\def\mc{\mathcal}
\def\E{\textit{\tiny{E}}} 
\newcommand{\sbullet}{%
  \hbox{\fontfamily{lmr}\fontsize{.8\dimexpr(\f@size pt)}{0}\selectfont\textbullet}}
\DeclareRobustCommand{\mathbullet}{\accentset{\sbullet}}
\newcommand{\cQ}{\mathcal{Q}}
\newcommand{\Contravariant}{{\nabla^{\scriptscriptstyle{T^{\ast}\!M}}}}
\newcommand{\Functions}{\mathscr{C}^\infty}
\newcommand{\Koszul}[1]{[#1]_{\text{KS}}}
\newcommand{\snabla}{\scriptscriptstyle{\nabla}}

\makeatother

\def\a{\alpha} \def\b{\beta}  \def\G{\Gamma}

\def\R{{\mathbb R}}    
 
 \def\Z{{\mathbb Z}}

\begin{document}

\makeatother
\parindent=0cm
\renewcommand{\title}[1]{%
    \vspace{10mm}
    \noindent{\Large{\bf #1}}
    \vspace{8mm}
}
\newcommand{\authors}[1]{%
    \noindent{\large #1}
    \vspace{5mm}
} 
\newcommand{\address}[1]{%
    {\itshape #1\vspace{2mm}}
}

\begin{titlepage}
	
\begin{center}
	
    \title{%
        {\Large Supersymmetric Poisson and \\[4pt]
        Poisson-supersymmetric sigma models}%
    }
    
    \vskip 3mm
    
    {\large%
        Thomas Basile$^{\,\a}$,
        Athanasios Chatzistavrakidis$^{\,\beta,\gamma}$,
        Sylvain Lavau$^{\,\delta}$
    }

    \vskip 18pt

    {\em 
        $^{\a}$Service de Physique de l’Univers, Champs et Gravitation,
        Universit\'e de Mons \\
        20 place du Parc, 7000 Mons, Belgium
        \vskip 5pt
        $^\beta$Division of Theoretical Physics,
        Rudjer Bo\v skovi\'c Institute \\
        Bijeni\v cka 54, 10000 Zagreb, Croatia.
        \vskip 5pt
        $^\gamma$Institute for Theoretical Physics,
        University of Wroclaw \\
        pl. Maxa Borna 9, 50-204 Wroclaw, Poland. 
        \vskip 5pt
        $^\delta$Department of Mathematics,
        Galatasaray \"Universitesi \\
        \c{C}\i ra\u{g}an Caddesi No:36, 34349 \.{I}stanbul, Turkey
    }
	
	\vskip 2cm

    \begin{abstract}
        We revisit and construct new examples of supersymmetric
        2D topological sigma models whose target space
        is a Poisson supermanifold. Inspired by the AKSZ construction
        of topological field theories, we follow a graded-geometric 
        approach and identify two commuting homological vector fields 
        compatible with the graded symplectic structure,
        which control the gauge symmetries and the supersymmetries
        of the sigma models. Exemplifying the general structure,
        we show that two distinguished cases exist,
        one being the differential Poisson sigma model constructed
        before by Arias, Boulanger, Sundell and Torres-Gomez
        and the other a contravariant differential Poisson sigma model.
        The new model features nonlinear supersymmetry transformations
        that are generated by the Poisson structure
        on the body of the target supermanifold,
        giving rise to a Poisson supersymmetry. Further examples are characterised by supersymmetry transformations controlled by the anchor map of a Lie algebroid, when this map is invertible, in which case we determine the geometric conditions for invariance under supersymmetry and closure of the supersymmetry algebra.
        Moreover, we show that the common thread through this type
        of models is that their supersymmetry-generating vector field
        is the coadjoint representation up to homotopy
        of a Lie algebroid.   
    \end{abstract}
    
\end{center}

\end{titlepage}

\setcounter{footnote}{0}
\tableofcontents

\newpage 

\section{Introduction} 
\label{sec1}
The Poisson sigma model is directly related to two important
results of 90s mathematical physics: the solution
to the problem of deformation quantization
for Poisson manifolds by Kontsevich \cite{Kontsevich:1997vb} 
and the geometrization of the classical master equation
for topological field theory by Alexandrov, Kontsevich,
Schwarz and Zaboronsky (AKSZ) \cite{Alexandrov:1995kv}.
In the former case, the diagrammatic expression
for the star product of functions
given in \cite{Kontsevich:1997vb} was found to correspond
to a correlation function between two observables
of the Poisson sigma model on a disc \cite{Cattaneo:1999fm},
in other words the diagrams in the star product expansion
can actually be interpreted as Feynman diagrams.
In the second case, the Poisson sigma model was embedded
within the general construction of \cite{Alexandrov:1995kv} 
(inspired by previous work of A. Schwarz \cite{Schwarz:1992nx})
in \cite{Cattaneo:2001ys}.
The underlying mathematical structure
behind $(n+1)$-dimensional AKSZ sigma models,
namely graded symplectic supermanifolds
with symplectic structure of degree $n$
(sometimes called QP$_{n}$ manifolds),%
{\footnote{Referred as $\Sigma_{n}$-manifolds
in \cite{Severa:2001tze}, aka symplectic Lie $n$-algebroids.}} 
corresponds exactly to Poisson geometry
for $n=1$ \cite{Roytenberg:2002nu}, making the Poisson
sigma model \emph{the} 2D AKSZ sigma model. 

The above statements refer to the bosonic Poisson sigma model. 
When sigma models with fermions and supersymmetry
are considered, one may ask whether the above statements carry over in one way or another. 
However, on the one hand, the literature on the supersymmetric 
version of the AKSZ construction is limited, see for instance 
\cite{Salnikov:2016bny} and \cite{Hulik:2022hpb}. On the other hand, deformation quantization 
on superspace in a spirit similar to \cite{Cattaneo:1999fm}
was studied in \cite{Chepelev:2003ga}, see also \cite{Borthwick:1993, Tyutin:2001iz, Ferrara:2003xy} for alternative approaches.
More recently, a different supersymmetric Poisson sigma model 
was constructed in \cite{Arias:2015wha, Arias:2016agc},
within the context of differential Poisson algebras.
The quantization of such differential Poisson sigma models, 
which is expected to yield a star product on the algebra
of differential forms, has not been performed yet,
let alone compared to previous proposals. 

These suggest that much less is known about supersymmetric 
extensions of Poisson sigma models and that the topic
deserves a fresh look from a modern and systematic perspective. 
Nonetheless, the general form of a supersymmetric Poisson
sigma model was already described as a nonlinear super-gauge 
theory in the original publication of Ikeda \cite{Ikeda:1993fh}.%
{\footnote{The bosonic model was also described
around the same time in \cite{Schaller:1994es}.} 
Indeed, the model of \cite{Arias:2015wha} is an interesting 
special case of Ikeda's model, essentially an example of it, 
different from 2D supergravity, which was the example
studied in \cite{Ikeda:1993fh} (and further analyzed in, e.g. \cite{Ikeda:1993dr,Strobl:1999zz, Ertl:2000si, Kummer:2001up, Bergamin:2004us}). 
More specifically, it is an example of supersymmetric
Poisson sigma model wherein the target space
is the shifted cotangent bundle of the parity-reversed
tangent bundle $\Pi TM$ of a smooth manifold $M$,
endowed with a structure of \emph{Poisson supermanifold}.
In particular,  this means that $M$ is itself
an ordinary Poisson manifold, and that its Poisson bracket
extends to the algebra of differential forms
(isomorphic to the algebra of functions on the parity-reversed
tangent bundle).

A closer look to this model reveals an interesting interplay
between two different homological vector fields:
the one associated with the structure of Poisson supermanifold
on $\Pi TM$ controls the gauge symmetry of the model,
as in the bosonic Poisson sigma model, whereas the one
associated with the de Rham differential on $M$
generates its supersymmetry. We may ask whether
this is an isolated model or there exist other examples
of such differential Poisson sigma models featuring other types
of supersymmetry. We answer this question affirmatively in this paper, 
showing that there exists a contravariant model with target space 
the shifted cotangent bundle of the parity-reversed \emph{cotangent} 
bundle of a smooth manifold and that the two models are selected
based on general consistency conditions. This also reveals
their structural kinship through the relation of their construction
to the concept of representations up to homotopy for Lie algebroids,
a notion that extends usual representations from vector bundles
to graded vector bundles (chain complexes) \cite{Abad:2009},
in particular to the coadjoint representation.

In more detail, we set up our analysis by first studying
non-negatively graded (NQ) supermanifolds in Section \ref{sec:Q-smfd}.
A supermanifold carries by construction a $\Z_2$-grading,
and it can be identified with a parity-reversed vector bundle $\Pi E$ 
over a smooth manifold $M$, which is the body of the supermanifold. 
Endowing it with an additional $\Z$-grading, we can consider
super-vector bundles over the supermanifold.
In this very general setting, where we have a $(\Z\times\Z_2)$-bidegree, 
two different degree $1$ vector fields can be considered,
namely one with $\Z$-degree $1$ (denoted $\cQ$)
and one with $\Z_2$-degree $1$ (denoted $\cQ_{S}$).
Asking that their graded commutator vanishes{\footnote{The graded commutator of vector fields
in degree $1$ is the anticommutator, yet we denote it
with ordinary brackets to avoid confusion with the Poisson bracket of functions.}},  
$$[\cQ,\cQ]=0 \quad \text{and} \quad [\cQ_{S},\cQ_{S}]=0\,,$$
imposes consistency conditions that correspond to the notions
of a Lie superalgebroid \cite{Mehta:2006,Mehta:2007}
and of a differential super-vector bundle 
\cite{Kotov:2007nr,Mehta:2006}. The latter is closely related
to the concept of representation of a Lie algebroid,
notably to the richer notion of representation up to homotopy,
where the module is a chain complex (a graded vector bundle).
This notion allows for a generalization of the adjoint
and coadjoint representations for Lie algebras to the Lie algebroid 
realm \cite{Abad:2009}. A central element in these constructions,
which we describe in Section \ref{sec: ruths} following \cite{Abad:2009} 
and adjusting accordingly for supermanifolds, is the so-called
basic connection together with the accompanying notion
of basic curvature. We then discuss how these elements
are elegantly encoded in the homological vector field $\cQ_{S}$.  

In Section \ref{sec:Psmfld}, we specialize to the case of Poisson supermanifolds, where the graded super-vector bundle is a shifted cotangent bundle over the Poisson supermanifold. The presence of an additional graded symplectic structure $\omega$ on the cotangent bundle prompts us to impose the compatibility conditions 
$${\cal L}_{\cQ}\,\omega=0 \quad \text{and}\quad  {\cal L}_{\cQ_{S}}\,\omega=0\,.$$ 
These conditions allow us to define Hamiltonian functions $\mc H$ and $\mc H_{S}$ that can be used to construct a general form of supersymmetric Poisson sigma models. In particular, the first condition together with the fact that $\cQ$ is homological result in the Jacobi identity for the super-Poisson bracket on the algebra of functions on the supermanifold. Working toward finding explicit solutions, in Section \ref{sec: expansion} we expand the generic super-Poisson bracket up to second order in the fermionic coordinate of the Poisson supermanifold and report the conditions stemming from the Jacobi identity up to order $5$ in the fermionic coordinate in manifestly covariant form. Notably, this requires that the body of the supermanifold is an ordinary Poisson manifold $M$, that there exists a $T^{\ast}M$-connection on the dual bundle~$E^{\ast}$, and an inner metric on $E$, not necessarily nondegenerate, which is covariantly constant with respect to this connection. 

In Section \ref{sec:sPSM} we take the final step toward constructing explicit examples of supersymmetric Poisson sigma models in this fashion, which requires to impose that the graded commutator of the two homological vector fields vanishes too, 
$$[\cQ,\cQ_{S}]=0\,,$$
alternatively that the corresponding Hamiltonian functions have vanishing Poisson bracket, 
$$\{\mc H,\mc H_{S}\}=0\,.$$
Then the general form of the action functional for a supersymmetric Poisson sigma model becomes 
$$S[\Phi]=\int\, \omega_{AB}\,\Phi^{A}\wedge\dd \Phi^{B}+\Phi^{\ast}(\mc H)\,,$$
where $\Phi=(\Phi^{A})$ is a map from the parity-reversed tangent bundle of a 2D world sheet to the parity-shifted cotangent bundle of the Poisson supermanifold, and $\Phi^{\ast}$ the pull-back by this map. By virtue of the consistency conditions, the model has gauge symmetries generated by $\cQ$ and supersymmetries generated by $\cQ_{S}$, it is a Cartan integrable Hamiltonian system, and it corresponds to the nonlinear super-gauge theory of \cite{Ikeda:1993fh}. 

The two distinguished examples of the general construction follow directly for the choices $E=TM$ and $E=T^{\ast}M$ for a Poisson manifold $M$, and they are presented in Sections \ref{sec: ABSTG} and \ref{sec:contravariant PSM}. The first is identical to the one constructed in Ref. \cite{Arias:2015wha} using differential Poisson algebras and the second is a nontrivial ``dual'' model. In both cases, the Hamiltonian features a quadratic part controlled by the Poisson structure on the body of the supermanifold and a quartic part whose coefficient is associated with the basic curvature of a connection on the Poisson supermanifold. This results in a supersymmetry-generating homological vector field that can be identified with the coadjoint representation of the tangent and cotangent Lie algebroids, providing a common ground for the two models. This is justified in Section~\ref{sec:sPSM_algd}, where we analyze further the general conditions for invariance under supersymmetry transformations and closure of their algebra for arbitrary Lie algebroids $E$. We show that the only other option includes models based on Lie algebroids whose anchor is invertible, for instance arising from a Lie algebra action or from endomorphisms of the tangent bundle with vanishing Nijenhuis tensor.

\paragraph{A note on notation.}
We will routinely work with geometrical objects in a basis independent
fashion and also present local coordinate expressions in a chosen basis. 
To keep the notation as transparent and light as possible, 
we use the following rules. For ``$E$-on-$V$'' connections,
which are $\R$-linear maps $\Gamma(E) \times \Gamma(V) \to \Gamma(V)$,
we use the symbol $\nabla^{\E}$. For their torsion and curvature 
tensors, we use $T^{\nabla^{\E}}$ and $R^{\nabla^{\E}}$.
When $E=TM$ (ordinary connections), we simplify to the symbol $\nabla$, 
and for dual ordinary connections ($TM$-on-$V^{\ast}$)
we use $\nabla^{\ast}$. When entries or indices are included,
we strip the symbols from additional superscripts
and let it be understood from the entries/indices which objects
they are. For example, for a $T^{\ast}M$-on-$E$ connection
we write any of the following: $\Contravariant$, $\nabla_{\eta}$,
$\nabla^{\mu}$, where $\eta=\eta_{\mu}\dd x^{\mu}$ is an $1$-form.
For its curvature, we write $R^{\Contravariant}$
or $R^{\mu\nu}{}^{a}{}_{b}$ in a basis $\mathfrak{e}_{a}$ of $E$, etc.

\section{NQ-supermanifolds}
\label{sec:Q-smfd} 
\subsection{Homological vector fields}
Any $\Z_2$-graded manifold, aka supermanifold,
$\mc M$ is (non-canonically) diffeomorphic
to a parity-shifted vector bundle $\Pi E$
where $E \twoheadrightarrow M$  is an ordinary
(meaning non-graded, smooth) vector bundle
over a smooth manifold $M$ which is the body
of the supermanifold~$\mc M$
\cite{Batchelor:1979, Batchelor:1980}.
We are interested in supermanifolds equipped with
an additional $\Z$-grading, i.e. we will consider
$(\Z \times \Z_2)$-graded manifolds in this work.
We will denote the bidegree of various objects
on such a $\Z$-graded supermanifold, and the corresponding 
shift functor, as $(\cdot,\cdot)$ and $[\cdot,\cdot]$
respectively, where the first placeholder corresponds
to the $\Z$ degree and the second one to the $\Z_2$ degree.
In this context, the aforementioned diffeomorphism
$\mc M \cong \Pi E$ should be rewritten as
$\mc M \cong E[0,1]$, thereby highlighting the fact
that the coordinates of $E$ along its fibres
are parity-shifted and given $\Z_2$-degree $1$, 
and the whole supermanifold is thought of
as being concentrated in $\Z$-degree $0$.
Put differently, local coordinates 
$(x^{\a})=(x^{\mu},\theta^a)$ on $E[0,1]$
are of bidegrees $(0,0)$ and $(0,1)$ respectively. The indices run through $\mu=1,\dots,\text{dim}\,M$ and $a=1,\dots,\text{rk}\,E$.
We shall use the sign convention of
summing the $\Z$ and $\Z_2$ degrees to a total degree
$|\cdot|$ and using 
\begin{equation}
    \varphi^{\a}\varphi^{\b}
    =(-1)^{|\varphi^{\a}|\cdot|\varphi^{\b}|}\,
    \varphi^{\b}\varphi^{\a}\,,
\end{equation}
for any elements $\varphi$ (coordinates, fields, etc.)

Applying the shift functor $[1,0]$ to supermanifolds,
we therefore obtain $\Z$-graded supermanifolds,
concentrated in $\Z$-degree $0$ and $1$.
Consider for instance a super-vector bundle
$\pi:\mc V \twoheadrightarrow \mc M$ over a supermanifold
$\mc M$. Its suspension $\mc V[1,0]$ in the $\Z$ degree
makes it a $(\Z \times \Z_2)$-graded manifold,
with local coordinates $x^{\alpha}=(x^{\mu},\theta^{a})$ on $\mc M$,
of the same bidegrees as before, and coordinates
$a^{A}=(a^{m},\chi^{I})$ along the fibres of $\mc V[1,0]$,
of bidegrees $(1,0)$ and~$(1,1)$ respectively. Indices $m$ and $I$ run from $1$ to each of the ranks of $\mc V$.\,{\footnote{This is a double vector bundle structure \cite{Mackenzie:2005}. For instance, if $\mc V$ is the tangent bundle over $E$, then $TE$ is a vector bundle over $E$ and also a vector bundle over $TM$, each having its own rank.}}
In this coordinate system, the most general vector field
of degree $(1,0)$ takes the form
\begin{equation}
    \cQ = a^{A}\,\rho_{A}{}^\alpha(x)\,
    \tfrac{\partial}{\partial x^\alpha}
    - \tfrac12\,a^{B} a^{C}\,f_{BC}{}^{A}(x)\,
    \tfrac{\partial}{\partial a^{A}}\,,
\end{equation}
where both $\rho_{A}{}^{\alpha}(x)$ and $f_{AB}{}^C(x)$
are functions of $x^\alpha$, and hence of $\Z$-degree $0$,
and of the same parity as
$a^{A}\,\tfrac{\partial}{\partial x^{\alpha}}$
and $a^{B}\,a^{C}\,\tfrac{\partial}{\partial a^{C}}$ respectively. 
More explicitly, the components of such a vector field
take the form 
\begin{subequations}
\begin{align}
    \cQ^{\mu} & = a^{m}\,\rho_{m}{}^\mu(x,\theta^2)
    + \chi^{I} \theta^{a}\,\rho_{I\,a}{}^\mu(x,\theta^2)\,, \\[4pt]
    \cQ^{a} & = a^{m} \theta^{b}\,\rho_{m\,b}{}^a(x,\theta^2)
    + \chi^{I}\,\rho_{I}{}^{a}(x,\theta^2)\, \\[4pt]
    \cQ^{p} & = -\tfrac12\,a^{m}\,a^{n}\,f_{mn}{}^{p}(x,\theta^2)
    - a^{m} \chi^{I} \theta^{a}\,f_{m\,I\,a}{}^{p}(x,\theta^2)
    -\tfrac12\,\chi^{I} \chi^{J}\,f_{IJ}{}^{p}(x,\theta^2)\,, \\[4pt]
    \cQ^{I} & = -\tfrac12\,a^{m} a^{n} \theta^{a}\,
    f_{mn\,a}{}^I(x,\theta^2)
    - a^{m} \chi^{J}\,f_{m\,J}{}^{I}(x,\theta^2)
    -\tfrac12\,\chi^{J} \chi^{K} \theta^{a}\,f_{JK\,a}{}^{I}(x,\theta^2)\,,
\end{align}
\end{subequations}
where now \emph{all} structure functions that appear
depend smoothly on the degree $(0,0)$ coordinate $x^\mu$
and are \emph{even} polynomials in the degree $(0,1)$
coordinates $\theta^a$ --- as suggested by the notation
$(x,\theta^2)$. The equations resulting from the requirement
that $\cQ$ is homological, 
\begin{equation}
    \tfrac 12 [\cQ,\cQ]= \cQ^2=0\,,
\end{equation} 
i.e. that it `squares'
to zero, simply read
\begin{subequations}
\label{eq:gauge_Q}
\begin{align}
    0 & = \,(-1)^{|B|}\,\rho_{A}{}^{\beta}\,
    \partial_\beta\rho_{B}{}^{\alpha}
    + (-1)^{|A||B|+|A|}\,\rho_{B}{}^{\beta}\,
    \partial_\beta\rho_{A}{}^{\alpha}
    - (-1)^{|A||B|}f_{AB}{}^{C}\,\rho_{C}{}^{\alpha}\,, \\[4pt]
    0 & = (-1)^{|B|(|C|+1)}\,
    \big(\rho_{B}{}^\alpha\,\partial_\alpha f_{CD}{}^{A}
    -(-1)^{|D|}\,f_{DB}{}^E\,f_{CE}{}^{A}\big) \nonumber \\[4pt]
    & \quad + (-1)^{|C|(|D|+1)}\,
    \big(\rho_{C}{}^{\alpha}\,\partial_{\alpha} f_{DB}{}^{A}
    - (-1)^{|B|}\,f_{BC}{}^{E}\,f_{DE}{}^{A}\big) \\[4pt]
    & \quad + (-1)^{|D|(|B|+1)}\,
    \big(\rho_{D}{}^{\alpha}\,\partial_{\alpha} f_{BC}{}^{A}
    - (-1)^{|C|}\,f_{CD}{}^{E}\,f_{BE}{}^{A}\big)\,, \nonumber
\end{align}
\end{subequations}
where $|A|$ denotes the \emph{total degree}
of the fibre coordinate $a^A$. These are the defining 
conditions of a Lie superalgebroid, that is to say,
a Lie algebroid structure on a super-vector bundle.
In other words, the standard result of Vaintrob
establishing a bijective correspondence
between NQ-manifolds (graded manifolds equipped
with a homological vector field and concentrated
in non-negative degrees) of degree $1$ and Lie algebroids 
\cite{Vaintrob:1997} carries over to the category
of supermanifolds, as shown by Mehta in
\cite[Sec. 2.4]{Mehta:2006} and \cite[Sec. 4]{Mehta:2007}.
An expanded form of these conditions is presented
in Appendix \ref{app:Qgauge}.

Apart from a $\Z$-degree $1$ homological vector field,
it is also possible to consider one with $\Z_{2}$-degree $1$. 
The most general vector field of degree $(0,1)$ reads
\begin{equation}
    \cQ_{S} = V^{\alpha}(x)\,
    \tfrac{\partial}{\partial x^{\alpha}}
    + a^{A}\,U_{A}{}^{B}(x)\,\tfrac{\partial}{\partial a^{B}}\,,
\end{equation}
where the components $V^{\alpha}(x)$ and $U_{A}{}^{B}(x)$
are functions of the $\Z$-degree $0$ coordinates~$x^{\alpha}$,
and have parity \emph{opposite} to
$\tfrac{\partial}{\partial x^{\alpha}}$
and $a^{A}\,\tfrac{\partial}{\partial a^{B}}$
respectively. The subscript $S$ refers to ``supersymmetry'', 
since this vector field will generate supersymmetry 
transformations for the sigma models we will consider
in later sections. More explicitly, its components
can be parametrized as
\begin{subequations}\label{Q susy general}
\begin{align}
    \cQ_{S}^{\mu} & = \theta^{a}\,t_{a}{}^{\mu}(x,\theta^2)\,, \\[4pt]
    \cQ_{S}^{a} & = V^{a}(x,\theta^2)\,, \\[4pt]
    \cQ_{S}^{m} & = a^{n} \theta^{a}\,U_{n\,a}{}^{m}(x,\theta^2)
    + \chi^{I}\,W_{I}{}^{m}(x,\theta^2)\,, \\[4pt]
    \cQ_{S}^{I} & = a^{m}\,Y_{m}{}^{I}(x,\theta^2)
    + \chi^{J} \theta^{a}\,Z_{J\,a}{}^{I}(x,\theta^2)\,,
\end{align}
\end{subequations}
where, as previously, all structure functions 
depend in a smooth manner on the degree $(0,0)$ coordinates
and are even polynomially in the degree $(0,1)$ ones.
Since this vector field is also of total degree $1$
on $\mc V$, it also makes sense to require it to be homological,
which yields the conditions
\begin{subequations}
\begin{align}
    0 & = V^{\beta}\,
    \tfrac{\partial}{\partial x^{\beta}} V^{\alpha}\,,\\[4pt]
    0 & = U_{B}{}^{C}\,U_{C}{}^{A} + (-1)^{|B|}\,V^{\alpha}\,
    \tfrac{\partial}{\partial x^{\alpha}} U_{B}{}^{A}\,.
\end{align}
\end{subequations}
These are analyzed in more detail in Appendix \ref{app:Q_susy}.
A first thing one can notice is that if $\cQ_S$
is homological on $\mc V$, it induces a homological
vector field on $\mc M$ via $\cQ_{\mc M} := \pi_{\ast}\cQ_S$
where $\pi$ is the projection of $\mc V$ onto $\mc M$.
In the above coordinate system, it is simply given by
$V^{\alpha}\,\tfrac{\partial}{\partial x^{\alpha}}$.
In other words, $\big(\mc V, \cQ_{S}\big)$ is
a \emph{differential} super-vector bundle, meaning
a super-vector bundle whose total space and base
are both equipped with a homological vector field,
and such that the projection preserves the latter.
One can think of this as the counterpart of
a \emph{differential graded} (dg for short) vector bundle 
(also known as $\cQ$-bundle
\cite{Kotov:2007nr, Mehta:2006}, see e.g. 
\cite{Grigoriev:2019ojp, Grigoriev:2022zlq, Grigoriev:2024ncm, Dneprov:2024cvt} for recent developments
in the context of gauge theories).

\subsection{Representations up to homotopy}
\label{sec: ruths}
$\cQ$-bundles are closely related to the notion
of representation up to homotopy \cite{Abad:2009}.
In the simplest case, the data $\big(\mc V, \cQ_{S}\big)$
can be considered as a \emph{representation}
of the $\cQ$-supermanifold $\big(\mc M, \cQ_{\mc M}\big)$,
in the sense of \cite[Sec. 2]{Bonechi:2011kz}, and based on
Vaintrob's insight. Indeed, recall that the paradigmatic
example of a $\cQ$-manifold is the suspension $E[1]$
of a Lie algebroid $E \twoheadrightarrow M$,
with its Chevalley--Eilenberg differential
as homological vector field. We denote by $\rho$ and $[\cdot,\cdot]$ the anchor and the Lie bracket of the Lie algebroid, and by $\rho_{a}{}^{\mu}$ and $f_{ab}{}^{c}$ their components in a local basis. 
A representation
of this Lie algebroid is another vector bundle
$V \twoheadrightarrow M$ equipped with a \emph{flat}
$E$-on-$V$ connection{\footnote{We recall from the introduction that the statement
``$E$-on-$V$ connection'' refers to a connection
$\nabla^{\E}:\Gamma(E)\times \Gamma(V)\to \Gamma(V)$
with the usual properties of linearity, homogeneity
and Leibniz rule for functions. We will simply use
the symbol $\nabla$ in case $E=TM$ and refer to this
as an ``ordinary'' connection on $V$. Similarly,
we will use $\nabla^{\ast}$ in case $E=T^{\ast}M$
and refer to this as a ``contravariant'' connection
(not to be confused with the dual connection,
which instead refers to connections on $V^{\ast}$.)}}
$\nabla^{\E}$. Correspondingly,
the $\cQ$-manifold $E[1] \oplus V$
with the Chevalley--Eilenberg differential
associated with the representation $V$ as $\cQ$-vector
defines a $\cQ$-bundle over $E[1]$. In this case,
we consider the homological vector field with components
\begin{equation}
    \cQ_{S}^{\mu} = \theta^{a}\,\rho_{a}{}^{\mu}(x)\,,
    \qquad 
    \cQ_{S}^{a} = -\tfrac12\,\theta^{b} \theta^{c}\,f_{bc}{}^{a}(x)\,,
    \qquad 
    \cQ_{S}^{m} = 0\,,
    \qquad 
    \cQ_{S}^{I} = -\chi^{J} \theta^{a}\,\Gamma_{a\,J}^{I}(x)\,,
\end{equation}
where $\Gamma_{a\,J}^{I}$ are the components of the flat $E$-on-$V$
connection. A direct computation yields 
\begin{equation}
    (\cQ_{S})^2 = 0
    \qquad\Longleftrightarrow\qquad 
    \left\{
    \begin{aligned}
        0 & = \rho_{[a}{}^{\nu}\,\partial_{\nu} \rho_{b]}{}^{\mu}
        - \tfrac12\,f_{ab}{}^{c}\,\rho_{c}{}^{\mu}\,, \\[4pt]
        0 & = \rho_{[a}{}^{\mu}\,\partial_{\mu} f_{bc]}{}^{d}
        + f_{[ab}{}^{e} f_{c]e}{}^{d}\,, \\[4pt]
        0 & = \rho_{[a}{}^{\mu}\,\partial_{\mu} \Gamma_{b]J}^{I}
        - \Gamma_{[a\,J}^{K}\,\Gamma_{b]K}^{I}
        - \tfrac12\,f_{ab}{}^{c}\,\Gamma_{c\,J}^{I}
        =: \tfrac 12 R_{ab}{}^{I}{}_{J}\,,
    \end{aligned}
    \right.
\end{equation}
the first two equations being the condition
that the anchor defines a morphism of Lie algebras
and the Jacobi identity for $E$ (which define a Lie algebroid), 
while the third is the vanishing of the curvature $R^{\nabla^{\E}}$
of the connection $\nabla^{\E}$ with components $R_{ab}{}^{I}{}_{J}$ (which defines a representation).

More generally, one can encode representations
\emph{up to homotopy} in $\cQ_{S}$. We first recall
the original definition from \cite{Abad:2009} in our notation
(for more details, see e.g. \cite[Sec. 2.1]{Batakidis:2024}). 
Given the Lie algebroid $E$,
a representation up to homotopy is a graded vector bundle
$\mc V$ together with 
\begin{enumerate}
    \item[(i)] An operator $\partial:\mc V\to \mc V$
    of degree $1$ that turns $(\mc V,\partial)$ into a complex.
    \item[(ii)] An $E$-connection
    ${\nabla}^{\scriptscriptstyle{E}}$
    on the complex $(\mc V,\partial)$.
    \item[(iii)] A total degree $1$,
    endomorphism-valued $2$-form of $E$,
    $\omega_2 \in \Omega^{2}(E;\text{End}^{\,-1}(\mc V))$
    such that the curvature
    of ${\nabla}^{\scriptscriptstyle{E}}$
    is counterbalanced by $[\partial,\omega_2]$ as in 
    \begin{equation}
        [\partial, \omega_2]
        + R^{{\nabla}^{\scriptscriptstyle{E}}}=0\,,
        \label{S vs R}
    \end{equation} 
    where the brackets $[\cdot,\cdot]$ above denote
    the graded-commutator.
    \item[(iv)] A collection of total degree $1$,
    endomorphism-valued $n$-forms on $E$ with $n>2$, denoted
    $\omega_n \in \Omega^{n}(E;\text{End}^{\,1-n}(\mc V))$ 
    which satisfy for every $n>2$ the recursive relations 
    \begin{equation}
        [\partial,\omega_n]
        + \dd_{{\nabla}^{\scriptscriptstyle{E}}}\,\omega_{n-1}
        + \omega_2 \circ \omega_{n-2} + \dots
        + \omega_{n-2} \circ \omega_2 = 0\,,
        \label{omega n}
    \end{equation}
    where $\dd_{{\nabla}^{\scriptscriptstyle{E}}}$
    is the differential on the space
    $\Omega(E,\mc V)=\G(\wedge^{\bullet}E^{\ast}\otimes\mc V)$
    of $\mc V$-valued differential forms on $E$ defined
    via the Koszul formula using the $E$-connection
    ${\nabla}^{\scriptscriptstyle{E}}$:
    \begin{eqnarray}
        (\dd_{{\nabla}^{\scriptscriptstyle{E}}}\,\eta)
        (e^{(1)},\dots,e^{(n+1)}) 
        & = & \sum_{i<j}(-1)^{i+j}
        \eta\big([e^{(i)},e^{(j)}],e^{(1)},\dots,e^{(n+1)}\big) 
        \nn\\[4pt] && +\, \sum_{i=1}^{n+1}(-1)^{i+1}
        {\nabla}^{\scriptscriptstyle{E}}_{e^{(i)}}
        \big(\eta(e^{(1)},\dots,e^{(n+1)})\big)\,.
        \label{d nabla}
    \end{eqnarray}
\end{enumerate}
It was shown in \cite{Abad:2009} that this definition
is  equivalent to the pair $(\mc V,\DD)$ with the degree $1$ 
structure operator $\DD:\Omega(E,\mc V)\to \Omega(E,\mc V)$ 
being a differential, namely $\DD^2=0$,
and a graded derivation, namely for $\omega\in\Omega^{n}(E)$ 
and $\eta\in\Omega(E;\mc V)$ it satisfies
\begin{equation}
    \DD(\omega\,\eta) = \dd_{\E}\,\omega\,\eta
    +(-1)^{n} \omega\,\DD\,\eta\,,
\end{equation}
where $\dd_{\E}$ is the differential on the Lie algebroid
given through the Koszul formula 
\begin{eqnarray}
    (\dd_{\E}\,\omega)(e^{(1)},\dots,e^{(n+1)})
    & = & \sum_{i<j}(-1)^{i+j}
    \omega\big([e^{(i)},e^{(j)}],e^{(1)},\dots,e^{(n+1)}\big) 
    \nn\\[4pt] && +\, \sum_{i=1}^{n+1}(-1)^{i+1}
    {\cal L}_{\rho(e^{(i)})}
    \big(\omega(e^{(1)},\dots,e^{(n+1)})\big)\,.
\end{eqnarray}
The relation of the two definitions is given through 
\begin{equation}
    \DD\,\eta = \partial\eta
    + \dd_{{\nabla}^{\scriptscriptstyle{E}}}\,\eta
    + \sum_{n \ge 2} \omega_n \wedge \eta\,.
\end{equation}
The usual notion of representation of a Lie algebroid
that we recalled earlier is included in this definition
in degree $0$. 

The adjoint and coadjoint representations of a Lie algebroid $E$
are the paradigmatic examples of this notion.
Although it will turn out that the coadjoint one
will be related to the field theories we will study below,
we begin with the adjoint representation,
which is defined as follows: given any $TM$-on-$E$ connection
$\nabla$, one can construct
an $E$-on-$(E[0] \oplus TM[-1])$ connection
$\overline{\nabla}^{\scriptscriptstyle{E}}$, via
\begin{equation}\label{eq:basic}
    \overline{\nabla}^{\scriptscriptstyle{E}}_e \begin{pmatrix}
        e' \\ X
    \end{pmatrix} 
    = \begin{pmatrix}
        \nabla_{\rho(e')} e + [e, e'] \\[4pt]
        \rho(\nabla_X e) + [\rho(e), X]
    \end{pmatrix}\,,
\end{equation}
for any vector field $X \in \Gamma(TM)$
and any pair of sections $e,e'\in\Gamma(E)$,
which is called the \emph{basic connection}.
Note that, as suggested by the notation,
the direct sum $E[0] \oplus TM[-1]$ is considered
as a \emph{complex} of vector bundles over $M$, with $E$ in degree $0$ and $TM$ in degree $1$ and
with degree $1$ differential given by the anchor
$\rho:E \longrightarrow TM$. Denoting by
\begin{equation}
    \Gamma_{\mu a}^b\,\mathfrak{e}_b
    = \nabla_{\partial_\mu} \mathfrak{e}_a\,,
\end{equation}
the components of $\nabla$ in a basis
$\{\mathfrak{e}_a\}$ of the fibres of $E$,
the components of the basic connections read
\begin{equation}
    \overline{\Gamma}_{a b}^c
    = \rho_b{}^\mu\,\Gamma_{\mu a}^c + f_{ab}{}^c\,,
    \qquad\qquad
    \overline{\Gamma}_{a \mu}^\nu
    = \Gamma_{\mu a}^b\,\rho_b{}^\nu
    - \partial_\mu \rho_a{}^\nu\,. 
    \label{basic connection coefficients}
\end{equation}
These suggest that we should choose $E[0]\oplus TM[-1]$
as the graded vector bundle $\mc V$ in the definition
of a representation up to homotopy,
together with $\rho=\partial$ and the basic connection
$\overline\nabla^{\E}$ as the $E$-connection on the complex. 
That this choice is compatible with the complex is guaranteed 
by the fact that the basic connection commutes with the anchor:
\begin{equation}\label{ruth anchor}
    [\overline{\nabla}^{\E},\rho]
    := \overline{\nabla}^{\E} \circ \rho
    - \rho \circ \overline{\nabla}^{\E} = 0\,.
\end{equation} 
This is easily proven using the definition
of the basic connection and the fact that $\rho$
is a (Lie algebra) homomorphism.

The basic connection is not flat in general,
so that it does not allow one to define a representation
of $E$ on $E[0] \oplus TM[-1]$ in the usual sense, 
however it satisfies
\begin{equation}
    R^{\overline{\nabla}^{\scriptscriptstyle{E}}}
    + [\rho, S^\nabla] = 0\,,
    \label{R and S}
\end{equation}
where $R^{\overline{\nabla}^{\scriptscriptstyle{E}}}$
is the curvature 
of $\overline{\nabla}^{\scriptscriptstyle{E}}$ and
\begin{equation}
    S^\nabla: \Gamma(E \wedge E) \otimes \Gamma(TM)
    \longrightarrow \Gamma(E)\,,
\end{equation}
is called the basic curvature, and it is defined as
\begin{equation}
    S^\nabla(e,e') X
    = \nabla_{\overline{\nabla}^{\scriptscriptstyle{E}}_e X} e'
    - \nabla_{\overline{\nabla}^{\scriptscriptstyle{E}}_{e'} X} e
    + \nabla_X [e,e'] - [\nabla_X e, e']
    - [e, \nabla_X e']\,,
\end{equation}
for any $e,e' \in \Gamma(E)$ and $X \in \Gamma(TM)$.
The concept was implicitly introduced in Ref. \cite{Blaom:2004}
in the study of Cartan connections for Lie algebroids
(see also \cite{Blaom:2005}) and later defined
in Ref. \cite{Abad:2009} in the present context. 
Eq. \eqref{R and S} is a short version
of the following two equations: 
\begin{subequations}
\begin{align}
   & R^{\overline{\nabla}^{\scriptscriptstyle{E}}}(e,e')X
   =-\rho(S^{\scriptscriptstyle{\nabla}}(e,e')X)\,, \\[4pt] 
   & R^{\overline{\nabla}^{\scriptscriptstyle{E}}}(e,e')e''
   =-S^{\scriptscriptstyle{\nabla}}(e,e')\rho(e'')\,,
\end{align}
\end{subequations}
for $e,e',e''\in \Gamma(E)$ and $X\in\mathfrak{X}(M)$.
In the above basis, the components of the basic curvature
\begin{equation}
    S_{ab\,\mu}{}^c\,\mathfrak{e}_c
    := S^\nabla(\mathfrak{e}_a,\mathfrak{e}_b)\partial_\mu\,,
\end{equation}
are given by
\begin{equation}
    S_{ab\,\mu}{}^c = \partial_\mu f_{ab}{}^c
    + f_{ab}{}^d\,\Gamma_{\mu d}^c
    + 2\,\Gamma_{\mu [a}^d\,f_{b]d}{}^c
    - 2\,\rho_{[a}{}^\nu\partial_\nu\Gamma_{\mu b]}^c
    - 2\,\big(\partial_\mu\rho_{[a}{}^\nu
    - \rho_d{}^\nu\Gamma_{\mu [a}^d\big)\,\Gamma_{\nu b]}^c\,.
\end{equation}
These suggest that we should make the choice
that the $2$-form $\omega_2$ is equal to the basic curvature
of the ordinary connection $\nabla$ on $E$,
$\omega_2=S^{\nabla}$, since this satisfies
the requirement~\eqref{S vs R} in item (iii) of the definition. 
Finally, the basic curvature
is $\dd_{\overline\nabla^{\E}}$-closed \cite{Abad:2009},
\begin{equation}\label{closed S}
    \dd_{\overline\nabla^{\E}}S^{\nabla}=0\,,
\end{equation}
and the requirement \eqref{omega n} is satisfied
with all $\omega_{n>2}=0$. This representation up to homotopy, 
alternatively presented via the structure operator as 
\begin{equation}\label{ruth E}
    \DD=\rho+\overline\nabla^{\E}+S^{\nabla}\,,
\end{equation}
is called the adjoint representation of the Lie algebroid $E$. 

Returning to the homological vector field $\cQ_{S}$,
we can essentially identify it with the structure operator,
a \emph{differential}
on the space of `$E$-forms' valued in $E \oplus TM[-1]$,
that is the space of sections
$\Gamma\big(\wedge E^{\ast} \otimes (E \oplus TM[-1])\big)$.
The latter can be identified with the space of sections
of the super-vector bundle
$E[0,1] \times (E[0,0] \oplus T[0,1]M)
\twoheadrightarrow E[0,1]$, where as before the second entry
of the suspension functor is reserved for the $\Z_2$-grading.
Upon shifting the fibres of this super-vector bundle
by one in the $\Z$-degree, as outlined before, 
we end up with 
\begin{equation}
    \pi: E[0,1] \times (E[1,0] \oplus T[1,1]M)
    \twoheadrightarrow E[0,1]\,, \label{adjoint map}
\end{equation}
which becomes a differential super-vector bundle,
with homological vector field given by
\begin{equation}\label{Qs ruth}
    \begin{aligned}
        \cQ_S & = \theta^a\,\rho_a{}^\mu(x)\,
        \tfrac{\partial}{\partial x^\mu}
        -\tfrac12\,\theta^b \theta^c\,f_{bc}{}^a(x)\,
        \tfrac{\partial}{\partial \theta^a}\\[4pt]
        & \qquad - \big(\theta^b a^c\,\overline{\Gamma}_{bc}^a(x)\,
        + \tfrac12\,\theta^b \theta^c \chi^\mu S_{bc\,\mu}{}^a\big)\,
        \tfrac{\partial}{\partial a^a}
        + \big(a^a\,\rho_a{}^\mu - \theta^a \chi^\nu\,
        \overline{\Gamma}_{a\,\nu}^\mu\big)\,
        \tfrac{\partial}{\partial \chi^\mu}\,,
    \end{aligned}
\end{equation}
and where $\theta^a$ are degree $(0,1)$ coordinates
on $E[0,1]$, while $a^a$ are degree $(1,0)$ coordinates 
on $E[1,0]$ and $\chi^\mu$ are degree $(1,1)$ 
coordinates along the fibres of $T[1,1]M$.
The projection of this vector field on the base,
which is $E[0,1]$ here, concretely amounts to setting
the fibre coordinates $a^a$ and $\chi^\mu$
of the super-vector bundle to zero, which leaves only
the first line of the above equation. Explicitly,
\begin{equation}
    \pi_\ast\cQ_S = \cQ_E \in \Gamma(TE[0,1])
    \qquad\text{with}\qquad
    \cQ_E = \theta^a\,\rho_a{}^\mu(x)\,
    \tfrac{\partial}{\partial x^\mu}
    -\tfrac12\,\theta^b \theta^c\,f_{bc}{}^a(x)\,
    \tfrac{\partial}{\partial \theta^a}\,.
\end{equation}
Finally, to confirm that $\cQ_{S}$, as presented
in local coordinates in \eqref{Qs ruth},
is the structure operator for the representation
up to homotopy, we should compute its square.
It is useful to restructure it as follows 
\begin{equation}
    \cQ_{S}=\cQ_{S}^{(0)}+\cQ_{S}^{(1)}+\cQ_{S}^{(2)}\,,
\end{equation}
where we decomposed the vector field in three pieces
according to the number of graded generators, $0$, $1$
and $2$ respectively,
(called the \emph{arity} \cite{Bonavolonta:2012fh}): 
\begin{subequations}
\begin{align}
    & \cQ_{S}^{(0)} = a^{a}\rho_{a}{}^{\mu}
    \tfrac{\partial}{\partial \chi^\mu}\,, \\[4pt] 
    & \cQ_{S}^{(1)} = \theta^a\,\rho_a{}^\mu(x)\,
        \tfrac{\partial}{\partial x^\mu}
        -\tfrac12\,\theta^b \theta^c\,f_{bc}{}^a(x)\,
        \tfrac{\partial}{\partial \theta^a}
        - \theta^b a^c\,\overline{\Gamma}_{bc}^a(x)
        \tfrac{\partial}{\partial a^a}
         - \theta^a \chi^\nu\,
        \overline{\Gamma}_{a\,\nu}^\mu
        \tfrac{\partial}{\partial \chi^\mu}\,,\\[4pt] 
   & \cQ_{S}^{(2)} = - \tfrac12\,\theta^b \theta^c \chi^\mu 
   S_{bc\,\mu}{}^a \tfrac{\partial}{\partial a^a}\,.
\end{align}
\end{subequations}
Note immediately that $(\cQ_{S}^{(0)})^{2}=0
=(\cQ_{S}^{(2)})^{2}$. Further calculation shows that:
\begin{equation}\label{rho nabla coordinates}
    [\cQ_{S}^{(0)}, \cQ_{S}^{(1)}] = 0
    \qquad\iff\qquad
    \overline{\Gamma}^{\mu}_{a \nu} \rho_{b}{}^{\nu}
    - \overline{\Gamma}^{c}_{ab}\rho_{c}{}^{\mu}
    +\rho_{a}{}^{\nu}\partial_{\nu}\rho_{b}{}^{\mu} = 0\,.
\end{equation}
This condition is the local coordinate expression
of Eq. \eqref{ruth anchor}, i.e. the basic connection
commuting with the anchor. Crucially, we also find that
\begin{equation}
    [\cQ_{S}^{(1)},\cQ_{S}^{(1)}]
    + [\cQ_{S}^{(0)}, \cQ_{S}^{(2)}] = 0
    \iff R_{ab}{}^{\nu}{}_{\mu}
    +\rho_{c}{}^{\nu} S_{ab\,\mu}{}^{c}=0\,,
    \quad 
    R_{ab}{}^{d}{}_{c}
    + \rho_{c}{}^{\mu}S_{ab\,\mu}{}^{d} = 0\,,
\end{equation}
which are the coordinate expressions that correspond
to Eq. \eqref{R and S} with the curvature components
being those of the basic connection. 
Finally, the closure condition \eqref{closed S} is found in:
\begin{equation}\label{q s coordinates}
    [\cQ_{S}^{(1)}, \cQ_{S}^{(2)}] = 0
    \iff \rho_{[a}{}^{\nu}\partial_{|\nu|}S_{bc]\mu}{}^{d}
    -f_{[ab}{}^{e} S_{|e|c]\,\mu}{}^{d}
    -\,\overline{\Gamma}^{\nu}_{[a|\mu|} S_{ b c]\nu}{}^{d}
    +S_{[ab|\mu|}{}^{e}\overline{\Gamma}^{d}_{c]e} = 0\,,
\end{equation}
with the last equation being its component form. 

Similarly to Lie algebras, there exists a dual representation 
to the adjoint, the coadjoint representation. In this case,
we consider the dual complex 
\begin{equation}
    \begin{tikzcd}
        T^{\ast}M \ar[r, "\rho^{\ast}"]
        & E^{\ast}\,,
    \end{tikzcd}
\end{equation}
with the cotangent bundle $T^{\ast}M$ in degree $-1$
and the dual bundle $E^{\ast}$ in degree $0$.
The degree $1$ operator is now the dual anchor,
with components $\rho^{\mu}{}_{a}$, which are equal
to the components $\rho_{a}{}^{\mu}$ of the anchor.
The connection in the present case is the dual one
to the basic connection, which is
an $E$-on-$(T^{\ast}M[1]\oplus E^{\ast}[0])$ connection.
It is defined using the canonical duality of the bundles, 
collectively denoted as $\langle\cdot,\cdot\rangle$
for any vector bundle, through 
\begin{equation}
    \langle\overline\nabla^{\E^{\ast}}_{e'}e^{\ast},e\rangle
    +\langle e^{\ast},\overline\nabla^{\E}_{e'}e\rangle
    =\langle \dd_{\E}\langle e,e^{\ast}\rangle,e'\rangle\,,    
\end{equation}
for any $e, e' \in \Gamma(E)$
and $e^{\ast} \in \Gamma(E^{\ast})$. The components
of this dual basic connection are opposite to the ones
of the straight basic connection,
namely $-\overline\Gamma_{ab}^{c}$
and $-\overline\Gamma_{a\mu}^{\nu}$ as they appear
in~\eqref{basic connection coefficients}. Finally,
the dual of the basic curvature is the tensor operator 
\begin{equation}
    (S^\nabla)^{\ast}: \Gamma(E \wedge E) \otimes \Gamma(E^{\ast})
    \longrightarrow \Gamma(T^{\ast}M)\,,
\end{equation}
defined via
\begin{equation}
    (S^{\nabla})^{\ast}(e,e')e^{\ast}
    =-\,e^{\ast}\circ S^{\nabla}(e,e')\,.
\end{equation}
In components, this tells us that $S_{ab}{}^{c}{}_{\mu}=-S_{ab\mu}{}^{c}$, where the left hand side refers to the dual basic curvature and the right hand side to the straight one. The statement that these data collect into a representation up to homotopy is equivalent to the dual properties to the ones of the adjoint representation, namely 
\begin{subequations}
\begin{align}
    [\overline\nabla^{\E^{\ast}},\rho^{\ast}]&=0\,, \\[4pt] 
    R^{\overline\nabla^{\E^{\ast}}}+[\rho^{\ast},(S^{\nabla})^{\ast}]&=0\,, \\[4pt] 
    \dd_{\overline\nabla^{\E^{\ast}}}(S^{\nabla})^{\ast}&=0\,.
\end{align} 
\end{subequations}
We recall that the curvature of the dual connection is given as 
\begin{equation}
    \langle R^{\overline\nabla^{\E^{\ast}}}(e,e')e^{\ast},e''\rangle=-\,\langle e^{\ast},R^{\overline\nabla^{\E}}(e,e')e''\rangle\,,
\end{equation}
and similarly for the other leg of the connection.
To encode the coadjoint representation in a homological
vector field, we follow the same logic as before, replacing
the differential super-vector bundle \eqref{adjoint map} with 
\begin{equation}
    \pi: E[0,1] \times (T^{\ast}[1,0]M \oplus E^{\ast}[1,1])
    \twoheadrightarrow E[0,1]\,. \label{coadjoint map}
\end{equation}
The corresponding coordinates are $x^{\mu}, \theta^{a}$
of bidegrees $(0,0)$ and $(0,1)$ and $a_{\mu}, \chi_{a}$
of bidegrees~$(1,0)$ and $(1,1)$ respectively and we can write
the homological vector field 
\begin{equation}\label{Qs coruth}
    \begin{aligned}
        \cQ_S & = \theta^a\,\rho_a{}^\mu(x)\,
        \tfrac{\partial}{\partial x^\mu}
        -\tfrac12\,\theta^b \theta^c\,f_{bc}{}^a(x)\,
        \tfrac{\partial}{\partial \theta^a}\\[4pt]
        & \qquad + \big(\theta^b a_{\nu}\,
        \overline{\Gamma}_{b\mu}^{\nu}(x)\,
        + \tfrac12\,\theta^b \theta^c \chi_a S_{bc}{}^{a}{}_{\mu}\big)\,
        \tfrac{\partial}{\partial a_{\mu}}
        - \big(a_{\mu}\,\rho^{\mu}{}_a - \theta^c \chi_b\,
        \overline{\Gamma}_{ca}^{b}\big)\,
        \tfrac{\partial}{\partial \chi_{a}}\,,
    \end{aligned}
\end{equation}
where we recognize the components of the dual anchor,
the dual basic curvature and the dual basic connection
(the latter written in terms of the straight components,
but with the opposite sign, as they should be).
That this vector field is homological is proven
using the same local coordinate formulas as before
and the relation between straight and dual objects.

\section{Poisson supermanifolds}
\label{sec:Psmfld}

\subsection{General structure}
For every ordinary Poisson manifold
$(M,\Pi)$, $\Pi\in\G(\wedge^2TM)$ denoting the Poisson bivector, 
its cotangent bundle $T^{\ast}M$ has a canonical
Lie algebroid structure. The anchor is given by
the induced map $\Pi^{\sharp}: T^{\ast}M\to TM$
with $\Pi^{\sharp}(\eta)=\Pi(\eta,\cdot)$
for $\eta\in\G(T^{\ast}M)$, and the Lie bracket
is given by the Koszul--Schouten bracket on $1$-forms 
\begin{equation}\label{eq:KS}
    \Koszul{\eta,\eta'} = {\cal L}_{\Pi^{\sharp}(\eta)}\eta'
    - {\cal L}_{\Pi^{\sharp}(\eta')}\eta
    - \dd\,\Pi(\eta,\eta')\,.
\end{equation}
In this case we may describe the Lie algebroid in terms
of the graded manifold $T^{\ast}[1]M$
with a suitable homological vector field.
This is a special case since there is also
an additional symplectic structure on the graded manifold,
since it is a cotangent bundle. Requiring that
the graded symplectic form is invariant under the flow
of the homological vector field fixes the latter in terms
of the Poisson structure on $M$. In fact, any $\Z$-graded
manifold concentrated in degrees $0$ and $1$, and equipped
with a symplectic $2$-form of degree $1$ invariant under
a homological vector field $\cQ$ arises in this manner,
i.e. is obtained as the shifted cotangent bundle
of an ordinary Poisson manifold \cite{Roytenberg:2002nu}.
We do not provide additional details yet,
since we will now address all these in a broader context.

It turns out that these statements naturally
carry over in the category of supermanifolds. 
Adding a bidegree $(1,0)$ symplectic structure $\omega$
on $\mc V[1,0]$ makes it diffeomorphic
to the shifted cotangent bundle  of $\mc M= E[0,1]$
(e.g. \cite{Schwarz:1992nx}),
i.e. $\mc V[1,0] \cong T^{\ast}[1,0]\mc M$.
Denoting by~$p_\alpha$ the fibre coordinate
of $\Z$-degree $1$, and by $\alpha \equiv |x^\alpha|$
the \emph{total} degree of the coordinate~$x^\alpha$
(so that $|p_\alpha| = 1+\alpha$),
the symplectic structure $\omega$
and homological field $\cQ$ locally take the form
\begin{equation}
    \omega = \dd p_\alpha\,\dd x^\alpha\,,
    \qquad 
    \cQ = p_\alpha\,\rho^{\alpha\beta}\,
    \tfrac{\partial}{\partial x^\beta}
    -\tfrac12\,p_\alpha p_\beta\,f_\gamma{}^{\alpha\beta}\,
    \tfrac{\partial}{\partial p_\gamma}\,,
\end{equation}
in Darboux coordinates. The compatibility condition of $\cQ$
with $\omega$ yields\footnote{We suppress a third
condition which is trivially satisfied as a consequence
of the two equations in \eqref{eq:Q_omega}. The expanded form of these conditions appears in Appendix \ref{app:Q_omega}.}
\begin{equation}\label{eq:Q_omega}
    {\cal L}_\cQ \,\omega = 0
    \qquad\Longleftrightarrow\qquad 
    \left\{
    \begin{aligned}
        0 & = \rho^{\alpha\beta}
        -(-1)^{(\alpha+1)(\beta+1)}\,\rho^{\beta\alpha}\,,\\[4pt]
        \, f_{\gamma}{}^{\alpha\beta}
        & = -(-1)^{(\alpha+\beta)\gamma+(\alpha+1)(\beta+1)}
        \partial_{\gamma}\rho^{\alpha\beta}\,, 
    \end{aligned}
    \right.
\end{equation}
which tell us that the homological vector field $\cQ$
is completely determined by the graded-symmetric part
of the `anchor', and squares to zero if and only if
\begin{equation}
    (-1)^{\gamma(\alpha+1)}\,\rho^{\alpha\delta}\,
    \partial_{\delta} \rho^{\beta\gamma} 
    + (-1)^{\alpha(\beta+1)}\,\rho^{\beta\delta}\,
    \partial_{\delta} \rho^{\gamma\alpha}
    + (-1)^{\beta(\gamma+1)}\,\rho^{\gamma\delta}\,
    \partial_{\delta} \rho^{\alpha\beta} = 0\,,
\end{equation}
as can be verified upon plugging the components 
of this $\omega$-compatible $\cQ$-vector in the conditions
\eqref{eq:gauge_Q}. The above identity is nothing
but the graded version of the Jacobi identity
for a super-Poisson bivector on $\mc M$. Indeed, defining
\begin{equation}
    \mc P^{\alpha\beta} := (-1)^{\alpha}\rho^{\alpha\beta}
    \qquad\implies\qquad 
    \mc P^{\beta\alpha}
        = -(-1)^{\alpha\beta}\,\mc P^{\alpha\beta}\,,
\end{equation}
the homological $\cQ$ reads
\begin{equation}\label{eq:Q}
    \cQ = (-1)^{\alpha}\big(p_\alpha\,\mc P^{\alpha\beta}\,
    \tfrac{\partial}{\partial x^\beta}
    - \tfrac12\,(-1)^{\alpha\beta}\,
    \partial_{\gamma}\mc P^{\alpha\beta}\,p_\alpha p_\beta\,
    \tfrac{\partial}{\partial p_\gamma}\big)\,,
\end{equation}
and the above identity becomes
\begin{equation}\label{eq:Jacobi_super-Poisson}
    \cQ^2 = 0 
    \quad\iff\quad
    (-1)^{\gamma\alpha}\,\mc P^{\alpha\delta}\,
    \partial_{\delta} \mc P^{\beta\gamma} 
    + (-1)^{\alpha\beta}\,\mc P^{\beta\delta}\,
    \partial_{\delta} \mc P^{\gamma\alpha}
    + (-1)^{\beta\gamma}\,\mc P^{\gamma\delta}\,
    \partial_{\delta} \mc P^{\alpha\beta} = 0\,,
\end{equation}
which is precisely the Jacobi identity satisfied
by the components of a super-Poisson bivector on $\mc M$.
In other words, requiring that $\cQ$ is compatible
with the canonical symplectic structure $\omega$
on the shifted cotangent bundle $T^{\ast}[1,0]\mc M$
implies that $\mc M$ is a Poisson supermanifold,
i.e. a supermanifold whose algebra of functions
is endowed with a structure of Poisson algebra,
with Poisson bracket $\{\cdot,\cdot\}$
determined by $\{x^{\alpha}, x^{\beta}\} = \mc P^{\alpha\beta}$.
This type of supermanifolds were discussed 
in e.g. \cite{Cantrijn:1991},
and were also identified in \cite{Ikeda:1993fh}
as the target space of the supersymmetric version of the Poisson sigma model.{\footnote{It is worth mentioning that another recent appearance of graded Poisson structures in physics and of their relation to generalised geometry is found in Ref. \cite{Boffo:2019zus}.}

We can repeat the exercise of requiring the compatibility
between the symplectic structure on $T^{\ast}[1,0]\mc M$
and the parity odd vector field $\cQ_S$, to find%
\footnote{Here again, we suppressed another condition
that comes out of this computation, as it is redundant%
---trivially satisfied as a consequence of \eqref{eq:QS_omega}.}
\begin{equation}\label{eq:QS_omega}
    \mc L_{\cQ_S} \omega = 0
    \qquad\iff\qquad 
    U^{\alpha}{}_{\beta} = (-1)^{\beta(\alpha+1)}\,
    \tfrac{\partial}{\partial x^{\beta}} V^{\alpha}\,.
\end{equation}
These conditions, given in more detail
in Appendix \ref{app:QS_omega}, lead to the following
local expression,
\begin{equation}\label{eq:comp-Q_S}
    \cQ_S = V^{\alpha}\,\tfrac{\partial}{\partial x^{\alpha}}
    + (-1)^{\alpha(\beta+1)}\,p_{\beta}\,
    \tfrac{\partial}{\partial x^{\alpha}} V^{\beta}\,
    \tfrac{\partial}{\partial p_{\alpha}}\,,
\end{equation}
which we can recognize as that of the Lie derivative
along $\cQ_{\mc M}$ on \emph{polyvector fields on~$\mc M$},
seen as a homological vector field on $T^{\ast}[1,0]\mc M$.
This result is consistent with the idea that a $\Z$-graded 
supermanifold concentrated in degrees $0$ and $1$,
and equipped with a homological vector field
of bidegree $(0,1)$, is a $\cQ$ super-vector bundle.

In the rest of this section, we will focus on Poisson
supermanifolds. Before presenting examples,
let us rewrite the components of the super-Poisson bivector,
its associated homological vector field and the corresponding
Jacobi identity, in terms of the even and odd coordinates
of the supermanifold $\mc M$, respectively $x^{\mu}$ and $\theta^{a}$,
as well as their momenta, respectively $a_{\mu}$ and $\chi_{a}$.
Accordingly, one distinguishes between three types of components
of the Poisson bivector, namely
\begin{equation}
    \mc P^{\mu\nu} = -\mc P^{\nu\mu}\,,
    \qquad 
    \mc P^{\mu a} = -\mc P^{a \mu}\,,
    \qquad 
    \mc P^{ab} = \mc P^{ba}\,,
\end{equation}
whose dependency on the odd coordinates $\theta^{a}$ reads
\begin{equation}
    \mc P^{\mu\nu}(x,\theta) = \mc P^{\mu\nu}(x,\theta^2)\,,
    \qquad 
    \mc P^{\mu a}(x,\theta)
    = \theta^b\,\mc P_b^{\mu a}(x, \theta^2)\,,
    \qquad 
    \mc P^{ab}(x,\theta) = \mc P^{ab}(x, \theta^2)\,,
\end{equation}
i.e. $\mc P^{\mu\nu}$ and $\mc P^{ab}$ are \emph{even} polynomials
in $\theta$ whereas $\mc P^{\mu a}$ are \emph{odd} ones.
The associated homological vector field on $T^{\ast}[1,0]\mc M$ is,
following \eqref{eq:Q}, given by
\begin{align}\label{eq:Q_details}
    \cQ  = &\big(a_{\mu}\,\mc P^{\mu\nu}
    + \chi_{a}\,\mc P^{\nu a}\big)\,
    \tfrac{\partial}{\partial x^{\nu}}
    + \big(a_{\mu}\,\mc P^{\mu a} - \chi_{b}\,\mc P^{ba}\big)\,
    \tfrac{\partial}{\partial \theta^{a}} \nonumber\\[4pt]
    & -\tfrac12\,\big(a_{\mu} a_{\nu}\,
    \partial_{\lambda}\mc P^{\mu\nu} 
    - 2\,a_{\mu} \chi_{a}\,\partial_{\lambda}\mc P^{\mu a}
    + \chi_{a} \chi_{b}\,\partial_{\lambda} \mc P^{ab}\big)\,
    \tfrac{\partial}{\partial a_\lambda} \nonumber  \\[4pt]
    & -\tfrac12\,\big(a_{\mu} a_{\nu}\,\partial_{a} \mc P^{\mu\nu}
    - 2\,a_{\mu} \chi_{b}\,\partial_{a} \mc P^{\mu b}
    + \chi_{b} \chi_{c}\,\partial_{a} \mc P^{bc}\big)\,
    \tfrac{\partial}{\partial \chi_{a}}\,. 
\end{align}
The Jacobi identity \eqref{eq:Jacobi_super-Poisson}
splits into four identities,
\begin{subequations}
\begin{align}
    0 & = \mc P^{[\mu|\kappa}\,\partial_{\kappa}\mc P^{\nu\lambda]}
    + \mc P^{[\mu|a}\,\partial_{a}\mc P^{\nu\lambda]}\,,
    \label{eq:Jac1} \\
    0 & = -\mc P^{\kappa a}\,\partial_{\kappa}\mc P^{\mu\nu}
    + \mc P^{ab}\,\partial_{b}\mc P^{\mu\nu}
    + 2\,\mc P^{[\mu|\kappa}\,\partial_{\kappa}\mc P^{\nu]a}
    + 2\,\mc P^{[\mu|b}\,\partial_{b}\mc P^{\nu]a}\,,
    \label{eq:Jac2} \\
    0 & = \mc P^{\mu\nu}\,\partial_{\nu}\mc P^{ab}
    + \mc P^{\mu c}\,\partial_{c}\mc P^{ab}
    + 2\,\mc P^{\nu(a}\,\partial_{\nu}\mc P^{\mu|b)}
    - 2\,\mc P^{c(a}\,\partial_{c}\mc P^{\mu|b)}\,,
    \label{eq:Jac3} \\
    0 & = -\mc P^{\mu(a}\,\partial_\mu\mc P^{bc)}
    + \mc P^{d(a}\,\partial_d\mc P^{bc)}\,,
    \label{eq:Jac4} 
\end{align}
\end{subequations}
relating the three different types of components
of $\mc P^{\alpha\beta}$.
Note that each one of these four relations can in fact
generate more than one identities since they have to be evaluated
order by order in the odd coordinates $\theta^a$,
as we will see in examples spelled out below. A quick counting
shows that Eqs. \eqref{eq:Jac1} and \eqref{eq:Jac3} solely
contain even powers of $\theta$, while Eqs.~\eqref{eq:Jac2}
and \eqref{eq:Jac4} only odd powers.

The homological vector fields $\cQ$ and $\cQ_{S}$
being compatible with the symplectic form
on $T^{\ast}[1,0]\mc M$, they admit each a Hamiltonian function,
that is a degree $2$ and parity-even function
$\mc H \in \mc C_2(T^{\ast}[1,0]\mc M)$
and $\mc H_{S} \in \mc C_2(T^{\ast}[1,0]\mc M)$ such that
\begin{equation}
    \cQ = \bm{\{} \mc H, - \bm{\}}\,,
    \qquad\qquad
    \cQ_{S} = \bm{\{} \mc H_{S}, - \bm{\}}
\end{equation}
where $\bm{\{}-,-\bm{\}}$ denotes the degree $-1$
Poisson bracket on $T^{\ast}[1,0]\mc M$ induced by the canonical
symplectic form. More explicitly, this bracket reads
\begin{equation}
    \bm{\{} f, g \bm{\}} = (-1)^{\alpha\,|f|+1}\,
    \big(\tfrac{\partial f}{\partial x^{\alpha}}\,
    \tfrac{\partial g}{\partial p_{\alpha}}
    +(-1)^{|f|}\,\tfrac{\partial f}{\partial p_{\alpha}}\,
    \tfrac{\partial g}{\partial x^{\alpha}}\big)\,,
    \qquad f, g \in \mc C(T^{\ast}[1,0]\mc M)\,, 
\end{equation}
which leads to 
\begin{subequations}
\begin{align}\label{eq: Hamiltonian}
    \mc H & = \tfrac12\,(-1)^{\alpha(\beta+1)}\,
    \mc P^{\alpha\beta}\,p_{\alpha}\,p_{\beta}
     = \tfrac12\,\mc P^{\mu\nu}\,a_{\mu} a_{\nu}
     + \mc P^{\mu a}\,a_{\mu} \chi_{a}
     + \tfrac12\,\mc P^{ab}\,\chi_{a} \chi_{b}\,, \\[4pt]
    \mc H_{S} & = (-1)^{\alpha}\,V^{\alpha}\,p_{\alpha}
    = V^\mu\,a_{\mu} - V^{a}\,\chi_{a}\,,
\end{align}
\end{subequations}
for the Hamiltonian functions associated with $\cQ$
and $\cQ_{S}$ respectively.

\subsection{Expansion up to second order
in the fermionic coordinate}
\label{sec: expansion}
Let us unpack some of the structures encoded in a generic
super-Poisson bracket, focusing on the case where all of its components are at most quadratic
in the odd coordinates $\theta^{a}$,
that we write in suggestive notation as
\begin{subequations}
    \label{super Poisson expanded}
    \begin{align}
        \mc P^{\mu\nu}(x,\theta) & = \Pi^{\mu\nu}(x)
        + \tfrac12\,\theta^{a}\theta^{b}\,
        \mc P_{ab}{}^{\mu\nu}(x)\,, \\[4pt]
        \mc P^{\mu a}(x,\theta)
        & = \theta^{b}\,\Gamma_b^{\mu a}(x)\,, \\[4pt]
        \mc P^{ab}(x,\theta) & = g^{ab}(x)
        + \tfrac12\,\theta^{c} \theta^{d}\,
        \mc P_{cd}{}^{ab}(x)\,.
    \end{align}
\end{subequations}
Without loss of generality, we will consider the supermanifold
to be a shifted vector bundle, i.e. $\mc M \cong E[0,1]$,
with $E \twoheadrightarrow M$ and $M$ being an ordinary
(non-graded, smooth) manifold, so that $\theta^a$
can be thought of as a basis of the fibers of $E^{\ast}$
and the algebra of functions on~$\mc M$ is isomorphic
to the exterior algebra of sections
of $E^{\ast} \twoheadrightarrow M$. Next,
we will study the Jacobi identity for $\mc P^{\alpha\beta}$,
order by order in $\theta^{a}$.

\paragraph{Order 0.}
We expect two conditions at this order,
one from \eqref{eq:Jac1} and one from \eqref{eq:Jac3}.
The order $0$ component in $\theta$ of \eqref{eq:Jac1} reads
\begin{equation}
    \Pi^{\kappa[\mu}\,\partial_\kappa\Pi^{\nu\lambda]} = 0\,,
\end{equation}
and therefore simply tells us that $\Pi$ is a Poisson bivector
on the base manifold $M$. 

To understand the second condition, we need to clarify 
the geometrical meaning of the mixed component $\mc P^{\mu a}$.  
As the notation suggests, the coefficients $\Gamma_{b}^{\mu a}$ 
are those of a \emph{contravariant} connection. In other words,
the Poisson bracket between a function $f \in \Functions(M)$
and a section $e^{\ast} \in \Gamma(E^{\ast})$
defines a $T^{\ast}M$-on-$E^{\ast}$ connection, i.e. a map
\begin{equation}\label{eq:connection}
    \Contravariant: \Gamma(T^{\ast}M) \times \Gamma(E^{\ast})
    \longrightarrow \Gamma(E^{\ast})\,,
\end{equation}
satisfying all standard properties of homogeneity,
linearity and Leibniz rule for a vector bundle connection, 
via{\footnote{In accordance with our general practice
in this paper, we simplify the notation for the connection
when it is clear from its entries. Therefore $\nabla_{\dd f}$ 
refers to $\Contravariant$, as it is clear from its 1-form
entry $\dd f$.}}
\begin{equation}
    \nabla_{\dd f} = \{f,-\}\,,
    \qquad \forall f \in \Functions(M)\,.
\end{equation}

This identification follows from the fact that \emph{any} derivation
$\Phi: \Functions(M) \longrightarrow \Gamma(E^{\ast})$ 
\emph{factors through} the $\Functions(M)$-bimodule of $1$-forms
$\Omega^1(M)$ via the de Rham differential $\dd$.
To be more precise, given the derivation $\Phi$,
there exists a (unique, up to an isomorphism)
$\Functions(M)$-linear map
$\phi: \Omega^1(M) \longrightarrow \Gamma(E^{\ast})$
such that the diagram
\begin{equation}
    \begin{tikzcd}[sep=large]
        \Functions(M) \ar[r, "\dd"] \ar[dr, swap, "\Phi"]
        & \Omega^1(M) \ar[d, "\phi"] \\ & \Gamma(E^{\ast})
    \end{tikzcd}
\end{equation}
commutes (for more details about this property,
in more general contexts, see e.g. \cite{Dubuc:1984, Carchedi:2012} 
and references therein).

Now let us come back to the Poisson bracket
between a function and a section $e^{\ast} \in \Gamma(E^{\ast})$,
and consider the map defined by
$\Phi_{e^{\ast}} := \{-,e^{\ast}\}: \Functions(M) \to \Gamma(E^{\ast})$.
Since the Poisson bracket obeys the Leibniz rule
in both arguments, this map is a derivation
of the $\Functions(M)$-bimodule $\Gamma(E^{\ast})$,
and hence by the previous universal property,
there exists a $\Functions(M)$-linear endomorphism
$\phi_{e^{\ast}}: \Omega^1(M) \to \Gamma(E^{\ast})$ such that
$\Phi_{e^{\ast}}(f) = \{f,e^{\ast}\} = \phi_{e^{\ast}}(\dd f)$,
for any $f \in \Functions(M)$ and $e^{\ast} \in \Gamma(E^{\ast})$.
The Leibniz rule of the Poisson bracket also yields
$\Phi_{f\,e^{\ast}}(g) = f\,\Phi_{e^{\ast}}(g) + \{g,f\}\,e^{\ast}$
when multiplying $e^{\ast}$ by a function $f$. The second term
on the right hand side can be written as
$\{g,f\} = \Pi^\sharp(\dd g)(f)$, i.e. recognized 
as the action on $f$ of the vector field given by the anchor
of the cotangent Lie algebroid applied to $\dd g$.
In summary, $\phi_{e^{\ast}}(\dd f)$ is $\Functions(M)$-linear
in $\dd f$ and obeys the Leibniz rule in $e^{\ast}$
with respect to the cotangent Lie algebroid anchor, 
which are the defining properties
of a $T^{\ast}M$-on-$E^{\ast}$ connection,
hence the identification \eqref{eq:connection}.%
\footnote{Note that this type of connection
is known to arise naturally as the semiclassical data
of the formal deformation (meaning first order) 
of a bimodule over an algebra, which is itself deformed
(and hence naturally gives rise to a Poisson structure),
see \cite{Reshetikhin:1996, Bursztyn:2001, Hawkins:2002rf, Beggs:2003ne,Beggs:2017yzq}.}

Given a basis $\{\mathfrak{e}^a\}$ of the fiber of $E^{\ast}$,
the components of $\Contravariant$ are defined as
\begin{equation}
    \nabla_{\dd x^{\mu}} \mathfrak{e}^{a}
    = \Gamma^{\mu a}_{b}\,\mathfrak{e}^{b}\,,
\end{equation}
and its action on a section $e^{\ast}=e^{\ast}_a\,\mathfrak{e}^{a} \in \Gamma(E^{\ast})$ for $\eta \in \Omega^1(M)$ reads
\begin{equation}
    \nabla_\eta e^{\ast} = \eta_\mu\,
    \big(\Pi^{\mu\nu}\partial_\nu e^{\ast}_{a}
    + \Gamma^{\mu b}_{a}\,e^{\ast}_{b}\big)\,\mathfrak{e}^{a}
    \equiv \eta_{\mu}\,\nabla^{\mu}e^{\ast}_{a}\,\mathfrak{e}^{a}\,,
\end{equation}
which is exactly what one finds when computing $\{f,e^{\ast}\}$
upon identifying the section $e^{\ast}$ with the odd function 
$e^{\ast}_a(x)\,\theta^{a}$. The curvature of such a connection
is defined as
\begin{equation}
    R^{\Contravariant}(\eta,\eta') e^{\ast}
    := [\nabla_{\eta}, \nabla_{\eta'}]\,e^{\ast}
    - \nabla_{[\eta,\eta']_{\text{KS}}} e^{\ast}\,,
    \qquad 
    \eta,\eta' \in \Omega^1(M)\,,
    \quad e^{\ast} \in \Gamma(E^{\ast})\,,
\end{equation}
whose components are given by
\begin{equation}
    R^{\mu\nu\,a}{}_{b}
    := \langle R^{\Contravariant}(\dd x^{\mu}, \dd x^{\nu})
    \mathfrak{e}^{a}, \mathfrak{e}_{b} \rangle
    = 2\,\big(\Pi^{[\mu|\lambda}\,
    \partial_{\lambda} \Gamma_{b}^{\nu] a}
    + \Gamma_{b}^{[\mu|c}\,\Gamma_{c}^{\nu]|a}
    - \tfrac12\,\partial_{\lambda}\Pi^{\mu\nu}\,
    \Gamma_{b}^{\lambda a}\big)\,,
\end{equation}
with $\mathfrak{e}_{a}$ a basis of the fibers of $E$, 
and $\langle\cdot,\cdot\rangle$ the canonical pairing
between $E$ and $E^{\ast}$. Note that any ordinary connection
$\nabla$ on $E$, meaning a $TM$-on-$E$ linear connection,
gives rise to a $T^{\ast}M$-on-$E^{\ast}$ connection by taking
\begin{equation}\label{eq:induced_connection}
    \mathbullet{\nabla}^{\scriptscriptstyle{T^{\ast}\!M}}_{\eta}
    = \nabla^{\ast}_{\Pi^\sharp(\eta)}\,, 
    \qquad \eta \in \Omega^1(M)\,,
\end{equation}
with $\nabla^{\ast}$ the dual connection of $\nabla$. We denoted this connection with a bullet to highlight the fact that it is canonically induced by an ordinary connection, since not all such connections arise this way. Concretely, this means taking
\begin{equation}
    \Gamma_b^{\mu a} =\mathbullet{\Gamma}^{\mu a}_{b}:= -\Pi^{\mu\nu}\,\Gamma_{\nu b}^a\,,
\end{equation}
in which case, 
\begin{equation}
    R^{\mathbullet{\nabla}} = R^{\snabla}(\Pi^{\sharp},\Pi^{\sharp})
    \qquad\iff\qquad 
    R^{\mu\nu\,a}{}_{b} \equiv \Pi^{\mu\kappa}\Pi^{\nu\lambda}\,
    R_{\kappa\lambda}{}^{a}{}_{b}\,,
\end{equation}
with $R^{\snabla}$ being the curvature of $\nabla$,
as noted in e.g. \cite{Arias:2015wha, Chatzistavrakidis:2023otk}.

We are now ready to turn our attention to the degree $0$ component
of Eq. \eqref{eq:Jac3}, which we can write as
\begin{equation}
    \nabla^\mu g^{ab} \equiv \Pi^{\mu\nu}\,
    \partial_\nu g^{ab} - 2\,\Gamma_c^{\mu (a}\,g^{b)c} = 0\,,
\end{equation}
i.e. we recognize it as the condition that the symmetric tensor
$g \in \Gamma(S^2E)$ be covariantly constant with respect
to the dual of the $T^{\ast}M$-on-$E^{\ast}$ connection identified 
before. Note that we have not made any assumption on the nondegeneracy of this symmetric tensor. Quite on the contrary, we will see later that in most of the examples we will assume that it vanishes completely. For the induced connection $\Contravariant=\mathbullet{\nabla}$,
this condition boils down to
\begin{equation}\label{eq: Jacobi deg0b}
    \mathbullet{\nabla}^\mu g^{ab}
   : = \Pi^{\mu\nu}\nabla_{\nu} g^{ab}=0\,,
\end{equation}
i.e. $g$ need not be preserved `on the nose' by $\nabla$,
but its covariant derivative should sit in the kernel
of $\Pi^{\sharp}$.

\paragraph{Splitting the supermanifold.}
At this point, a couple of remarks are in order.
First, note that any contravariant connection on $E$
can be written as
\begin{equation}
    \Contravariant = \mathbullet{\nabla}^{\scriptscriptstyle{T^{\ast}\!M}} + \phi\,,
    \qquad\text{with}\qquad
    \phi: \Omega^1(M)
        \longrightarrow \Gamma\big({\rm End}(E^{*})\big)\,,
\end{equation}
since the difference between any two linear connections
is a tensor. In other words, we can always choose
the induced connection $\mathbullet\nabla$ as reference point
(on the affine space of contravariant connections),
and parameterize any other contravariant connection 
$\Contravariant$ `in reference to the latter',
i.e. by specifying the ${\rm End}(E^{\ast})$-valued vector field
$\phi \in \Gamma(TM \otimes {\rm End}\,E^{\ast})$\,.
This suggests that we may write 
\begin{equation}
    \mc P^{\mu a} = \big(-\Pi^{\mu\nu}\,\Gamma_{\nu b}^{a}
    +\phi^{\mu}{}_{b}{}^{a}\big)\,\theta^{b}+\mc{O}(\theta^{3})\,,
\end{equation}
thereby extracting a tensorial piece, $\phi$,
out of the lowest order $\theta$-component of $\mc P^{\mu a}$,
and express the non-tensorial piece in terms
of the lowest order component of $\mc P^{\mu\nu}$
and a $TM$-on-$E$ connection $\nabla$.

Then it is useful to write the various identities that are extracted from \eqref{eq:Jac1}--\eqref{eq:Jac4} in terms of tensors for $E$ and $TM$ (and their duals).
A choice of $TM$-on-$E$ connection defines
an isomorphism
\begin{equation}
    \Gamma(TE[0,1]) \overset{\snabla}{\cong} \Gamma(\wedge E^{\ast})\,
    \underset{\Functions(M)}{\otimes}\,
    \Big(\Gamma(TM) \oplus \Gamma(E)\Big)\,,
\end{equation}
as modules over $\Gamma(\wedge E^{\ast})$.
Under this isomorphism the components of $\mc P$,
that we shall denote with a subscript $\nabla$,
are related to the initial ones via
\begin{subequations}
    \label{eq: P covariant}
    \begin{align}
        \mc P_{\snabla}^{\mu a} & = \mc P^{\mu a}
        + \mc P^{\mu\nu}\,\Gamma_{\nu b}^{a}\,\theta^b\,,\\[4pt]
        \mc P_{\snabla}^{ab} & = \mc P^{ab}
        - 2\,\mc P^{\mu(a}\,\Gamma_{\mu c}^{b)}\,\theta^{c}
        + \mc P^{\mu\nu}\,\Gamma_{\mu c}^{a}\Gamma_{\nu d}^{b}\,
        \theta^{c} \theta^{d}\,,
    \end{align}
\end{subequations}
and their expansion in powers of $\theta$ consist of tensors
on the body of $\mc M$, which is also the base $M$
of the model vector bundle $E$. In terms of these new components,
the Jacobi identity becomes
\begin{subequations}
\begin{align}
    0 & = \mc P^{[\mu|\kappa}\nabla_{\kappa}\mc P^{\nu\lambda]}
    + \mc P_{\snabla}^{[\mu|a}\partial_{a}\mc P^{\nu\lambda]}\,,\\[4pt]
    0 & = -\mc P_{\snabla}^{\kappa a}\nabla_{\kappa}\mc P^{\mu\nu}
    + \mc P_{\snabla}^{ab}\partial_{b}\mc P^{\mu\nu}
    + 2\,\mc P^{[\mu|\kappa}\nabla_{\kappa}\mc P_{\snabla}^{\nu]a} 
    + 2\,\mc P_{\snabla}^{[\mu|b}\partial_{b}\mc P_{\snabla}^{\nu]a}
    - \mc P^{\mu\kappa} \mc P^{\nu\lambda}\,
    R_{\kappa\lambda}{}^{a}{}_{b}\,\theta^{b}\,,\\[4pt]
    0 & = \mc P^{\mu\nu}\nabla_{\nu}\mc P^{ab}_{\snabla}
    + \mc P^{\mu c}_{\snabla} \partial_{c}\mc P^{ab}_{\snabla}
    + 2\,\mc P^{\nu (a}_{\snabla} \nabla_{\nu} \mc P^{\mu|b)}_{\snabla}
    - 2\,\mc P^{c(a}_{\snabla} \partial_{c}\mc P^{\mu|b)}_{\snabla}
    +2\,\mc P^{\mu\kappa} \mc P^{\lambda(a}_{\snabla} R_{\kappa\lambda}{}^{b)}{}_{c}\,\theta^{c}\,, \\[4pt]
    0 & = -\mc P^{\mu(a}_{\snabla} \nabla_{\mu} \mc P^{bc)}_{\snabla} 
    + \mc P^{d(a}_{\snabla} \partial_{d} \mc P^{bc)}
    - \mc P^{\mu(a|}_{\snabla} \mc P^{\nu|b}_{\snabla}
    R_{\mu\nu}{}^{c)}{}_{d} \theta^{d}\,,
\end{align}
\end{subequations}
where $\nabla$ acts on sections of tensors products
of $E$ and $\wedge E^{\ast}$, the latter part being considered
as polynomials in $\theta^a$, as
\begin{equation}
    \nabla_{\mu} T^{a_1 \cdots a_k}(x,\theta)
    = \big(\partial_{\mu} - \Gamma_{\mu b}^{c}\,
    \theta^{b}\,\tfrac{\partial}{\partial\theta^{c}}\big)\,
    T^{a_1 \cdots a_k}
    + \sum_{l=1}^{k} \Gamma^{a_l}_{\mu b}\,
    T^{\cdots a_{l-1}\,b\,a_{l+1} \cdots}\,.
\end{equation}
Under this isomorphism, the `covariant' components of $\mc P$
read
\begin{equation}\label{eq: P covariant 2}
    \mc P_{\snabla}^{\mu a}
    = \phi^{\mu}{}_{b}{}^{a}\,\theta^{b}\,,
    \qquad 
    \mc P_{\snabla}^{ab} = g^{ab}
    + \tfrac12\,\theta^{c} \theta^{d}\,R_{cd}{}^{ab}\,,
\end{equation}
up to second order in $\theta$, where $\phi^{\mu}{}_{b}{}^{a}$
are the components of the ${\rm End}(E)$-valued vector field
relating an arbitrary contravariant connection to the induced one
$\mathbullet\nabla$ as discussed in the previous paragraphs,
and where $R_{ab}{}^{cd}$ are the components of a section
of $\wedge^2 E^{\ast} \otimes S^{2} E$.

\paragraph{Higher orders.}
Now that we have re-expressed the content
of $\mc P^{\alpha\beta}$ in terms of tensors over $M$,
let us go back to the higher orders in $\theta^{a}$
of its Jacobi identity. We collect the conditions order by order in $\theta$ in Table \ref{Table1}. 

\begin{table}
	\begin{center}	\begin{tabular}{| c | c | }
			\hline 
			\multirow{3}{3em}{\centering{Order}} & \multirow{3}{10.5em}{\centering{Conditions}}\\  &      \\ & \\ \hline
			\multirow{3}{3em}{} &  \multirow{3}{10.5em}{}   \\  & $ \Pi^{\kappa[\mu}\,\partial_\kappa\Pi^{\nu\lambda]} = 0$    \\ $0$ & \\ & $ \nabla^\mu g^{ab}=0$\\ & \\\hline 
			\multirow{3}{3em}{} & \multirow{3}{10.5em}{}\\ & $\quad \Pi^{\mu\kappa}\,\Pi^{\nu\lambda}\,
    R_{\kappa\lambda}{}^{a}{}_{b}
    + \phi^{\kappa}{}_{b}{}^{a}\,\nabla_{\kappa} \Pi^{\mu\nu}
    + 2\,\Pi^{\kappa[\mu}\,\nabla_{\kappa}\phi^{\nu]}{}_{b}{}^{a}
    -2\,\phi^{[\mu}{}_{b}{}^{c}\,\phi^{\nu]}{}_{c}{}^{a}
    + g^{ac}\,\mc P_{bc}{}^{\mu\nu}=0\quad $\\  $\theta$ & \\ & $\phi^{\mu}{}_{d}{}^{(a}\,\nabla_{\mu} g^{bc)}
    + g^{e(a}\,R_{de}{}^{bc)}=0$   \\ &
			\\\hline 
   \multirow{3}{3em}{} & \multirow{3}{10.5em}{} \\ & $\Pi^{\kappa[\mu}\,\nabla_{\kappa}\mc P_{ab}{}^{\nu\lambda]}
    + \mc P_{ab}{}^{\kappa[\mu}\,\nabla_{\kappa}\Pi^{\nu\lambda]}
    + 2\,\phi^{[\mu}{}_{[a}{}^{c}\,\mc P_{b]c}{}^{\nu\lambda]}=0$ \\ $\theta^2$ & \\ & $\tfrac12\,\Pi^{\mu\nu}\,\nabla_{\nu} R_{cd}{}^{ab}
    + \tfrac12\,\mc P_{cd}{}^{\mu\nu}\,\nabla_{\nu} g^{ab}
    - \phi^{\mu}{}_{[c}{}^{e}\,R_{d]e}{}^{ab}
    \,+$ \\[4pt] &  $+\,2\,\phi^{\nu}{}_{[c}{}^{(a}\,
        \nabla_{\nu} \phi^{\mu}{}_{d]}{}^{b)}- R_{cd}{}^{e(a}\,\phi^{\mu}{}_{e}{}^{b)}
    + 2\,\Pi^{\mu\kappa}\,\phi^{\lambda}{}_{[c}{}^{(a}\,
    R_{\kappa\lambda}{}^{b)}{}_{d]}=0$ \\ & \\ \hline
   \multirow{3}{3em}{} & \multirow{3}{10.5em}{} \\ & $\phi^{\kappa}_{[b}{}^{a}\,
    \nabla_{\kappa} \mc P_{cd]}{}^{\mu\nu}
    - 2\,\mc P_{[bc}{}^{[\mu|\kappa}\,
    \nabla_{\kappa} \phi^{|\nu]}{}_{d]}{}^{a}
    + 2\,\Pi^{[\mu|\kappa}\,\mc P_{[bc}{}^{|\nu]\lambda}\,
    R_{\kappa\lambda}{}^{a}{}_{d]}=0$\\ $\theta^3$ & \\ & $\phi^{\mu}_{[d}{}^{(a}\,\nabla_{\mu} R_{ef]}{}^{bc)}
    + R_{[de}{}^{g(a}\,R_{f]g}{}^{bc)}
    + 2\,\phi^{\mu}{}_{[d}{}^{(a} \phi^{\nu}{}_{e}{}^{b}\,
    R_{\mu\nu}{}^{c)}{}_{f]}=0$  \\ & \\ \hline
   \multirow{3}{3em}{} & \multirow{3}{10.5em}{} \\  & $\mc P_{[ab}{}^{[\mu|\kappa}\,
    \nabla_{\kappa} \mc P_{cd]}{}^{|\nu\lambda]}=0$    \\ $\theta^4$ & \\ & $\mc P_{[cd}{}^{\mu\nu}\,\nabla_{\nu} R_{ef]}{}^{ab}
    + 2\,\mc P_{[cd}{}^{\mu\kappa}\,\phi^{\lambda}{}_{e}{}^{(a}\,
    R_{\kappa\lambda}{}^{b)}{}_{f]}=0$\\ & \\ \hline
   \multirow{3}{3em}{} & \multirow{3}{10.5em}{} \\ $\theta^5$ & $\mc P_{[bc}{}^{\mu\kappa}\,\mc P_{de}{}^{\nu\lambda}\,
    R_{\kappa\lambda}{}^{a}{}_{f]}=0$ \\  & \\  \hline\end{tabular}\end{center}\caption{Covariant form of the general conditions obtained from the Jacobi identity of the super-Poisson structure with components at most quadratic in the odd coordinates $\theta^{a}$.}\label{Table1}\end{table}

In the remainder of the paper, we will be focusing on the case
where $\phi=0$, i.e. we will assume that the contravariant
connection induced by the super-Poisson bivector is induced
from an ordinary $TM$-on-$E$ connection, and where
the quadratic component (and in fact,
all higher order components) in $\mc P^{\mu\nu}$ vanishes,
i.e. $\mc P_{ab}{}^{\mu\nu}=0$. In this situation,
the above identities boil down to the flatness
of the induced contravariant connection,
\begin{equation}
    \Pi^{\mu\kappa}\,\Pi^{\nu\lambda}\,
    R_{\kappa\lambda}{}^{a}{}_{b} = 0
    \qquad\iff\qquad
    R^{\snabla}\big(\Pi^{\sharp},\Pi^{\sharp}\big) = 0\,,
\end{equation}
and the conditions
\begin{subequations}
\label{eq:Bianchi-like}
\begin{align}
    0 & = g^{e(a}\,R_{de}{}^{bc)}\,, 
        \label{eq: Jacobi deg1b} \\[4pt]
    0 & = \Pi^{\mu\nu}\,\nabla_{\nu} R_{cd}{}^{ab}\,,
        \label{eq: Jacobi deg2b}\\[4pt]
    0 & = R_{[de}{}^{g(a}\,R_{f]g}{}^{bc)}\,,
        \label{eq: Jacobi deg3b}
\end{align}
\end{subequations}
on the tensor $R_{cd}{}^{ab}$,
which are reminiscent of the algebraic and differential
Bianchi identities. We will see in Section \ref{sec:sPSM_algd}
that these constraints allow one to define a supersymmetric
Poisson sigma model, with supersymmetry transformations
being encoded in a Lie algebroid structure on $E$.

\section{Supersymmetric Poisson sigma models}
\label{sec:sPSM}
\subsection{The general Ikeda supermodel}
\label{ikeda}
A nonlinear supergauge theory was defined constructively
in Ref. \cite{Ikeda:1993fh}. Here we revisit this theory
within the context presented in the previous sections.
We are interested in 2D supersymmetric topological sigma models,
where the target space is an NQ-supermanifold with a degree $1$ 
compatible symplectic structure. 
In general, the non-supersymmetric version of such models
requires the existence of a QP structure on the target space
$\mc M$, which induces a corresponding QP structure
on the mapping space ${\sf Maps(\mc X,\mc M)}$,
where $\mc X$ is the source dg manifold; this is the backbone
of the AKSZ construction \cite{Alexandrov:1995kv}.
We would like to reconstruct the classical action 
of the model from the geometrical data,
therefore we focus on degree-preserving maps in this work.
In the supersymmetric case, the geometrical data
are the source data $(\mc X,\dd)$ and the target data
$(\mc M, \omega, \mc Q, \mc Q_{S})$. There are two homological 
vector fields, $\mc Q$ of degree $(1,0)$ that controls
the gauge symmetries of the model and $\mc Q_{S}$ 
of degree $(0,1)$ that controls the supersymmetries.
To construct an action that is both gauge invariant
and supersymmetric, we require that both homological vector fields 
are compatible with the graded symplectic structure $\omega$
and that they are mutually commuting, 
\begin{equation}
    [\mc Q,\mc Q]=0=[\mc Q_{S},\mc Q_{S}]\,,
    \qquad
    [\mc Q,\mc Q_{S}]=0\,,
    \qquad 
    {\mc L}_{\mc Q}\omega = 0 = {\mc L}_{\mc Q_{S}}\omega\,.
\end{equation}
The expanded form of all these conditions appears
in Appendix \ref{appa}. 

The general supersymmetric sigma model is constructed
with target space $T^{\ast}[1,0]E[0,1]$. We recall that
we have assigned to it the $\Z$-degree $0$ coordinates
$x^{\alpha}=(x^{\mu},\theta^{a})$ and the $\Z$-degree $1$ 
coordinates $p_{\alpha}=(a_{\mu},\chi_{a})$. We will denote
the corresponding fields in the mapping space as
$X^{\alpha}=(X^{\mu},\theta^{a})$ and
$P_{\alpha}=(A_{\mu},\chi_{a})$ respectively,
and when we want to refer collectively to all of them
as $\Phi=(X^{\alpha},P_{\alpha})$ (note that there is
some abuse of notation for the $\Z_2$-odd fields
in that we have refrained from capitalizing them).
The action functional of the supersymmetric sigma model
in our conventions is 
\begin{align}
    S[X^{\alpha},P_{\alpha}]
    & =  \int \bigg(P_{\alpha}\wedge\dd X^{\alpha}
    + \tfrac12 (-1)^{\alpha(\beta+1)}\,\mc P^{\alpha\beta}(X) 
    P_{\alpha} \wedge P_{\beta}\bigg) \\[4pt] \nonumber
    & =  \int \bigg(A_{\mu} \wedge \dd X^{\mu}
    +\chi_{a} \wedge \dd \theta^{a}
    + \tfrac12 \mc P^{\mu\nu} A_{\mu} \wedge A_{\nu}
    + \mc P^{\mu a} A_{\mu} \wedge \chi_{a}
    + \tfrac12\mc P^{ab}\,\chi_{a} \wedge \chi_{b}\bigg)\,,
\end{align}
where the interaction term corresponds to the Hamiltonian
for $\cQ$ given in \eqref{eq: Hamiltonian} above.
This action appeared already in Ref. \cite[Sec. 5]{Ikeda:1993fh}. 
The gauge symmetries of this general model can be found
in a straightforward manner from the homological
vector field $\mc Q$. They are (see e.g. 
\cite{Grutzmann:2014hkn,Chatzistavrakidis:2023otk,Grigoriev:2019ojp}
for detailed explanations) 
\begin{equation}\label{gauge symmetry general}
    \delta_{\varepsilon}\Phi=[\dd +\mc Q,\varepsilon]\,,
\end{equation}
where the gauge symmetry parameters are $\varepsilon=(0,\varepsilon_{\alpha})$ with $|\varepsilon|=1$ and $\varepsilon_{\alpha}=(\varepsilon_{\mu},\varepsilon_{a})$ the even and odd scalar parameters for the fields $A_{\mu}$ and $\chi_{a}$ respectively. The pull-back to the space of fields on the right hand side of \eqref{gauge symmetry general} is implicit. We can calculate the graded commutator for the homological vector field given in the chosen basis of Section \ref{sec:Q-smfd} and obtain the gauge symmetries for each component: 
\begin{subequations}
\begin{align}
    \delta_{\varepsilon}X^{\mu}
    &=\varepsilon_{\nu}{\cal{P}}^{\nu\mu}
    -\varepsilon_{a}\theta^{b}\mc{P}^{a\mu}_{b}\,, \\[4pt] 
    \delta_{\varepsilon}A_{\mu}
    &= \dd\varepsilon_{\mu}
    -\varepsilon_{\nu}A_{\rho}\partial_{\mu}\mc{P}^{\nu\rho}
    -\varepsilon_{\nu}\chi_{a}\theta^{b}\partial_{\mu}\mc{P}^{a\nu}_{b}
    -\varepsilon_{a}A_{\nu}\theta^{b}\partial_{\mu}\mc{P}^{a\nu}_{b}{}
    -\varepsilon_{a}\chi_{b}\partial_{\mu}\mc{P}^{ba}\,,\\[4pt]
    \delta_{\varepsilon}\theta^{a}
    &=\varepsilon_{\mu}\theta^{b}\mc{P}^{\mu a}_{b}
    -\varepsilon_{b}\mc{P}^{ba}\,, \\[4pt] 
    \delta_{\varepsilon}\chi_{a}
    &= \dd\varepsilon_{a}
    -\varepsilon_{\mu}A_{\nu}\partial_{a}\mc{P}^{\mu\nu}
    -\varepsilon_{\mu}\chi_{b}\partial_{a}(\theta^{c}\mc{P}^{\mu b}_{c})
    -\varepsilon_{b}A_{\mu}\partial_{a}(\theta^{c}\mc{P}^{\mu b}_{c})
    -\varepsilon_{b}\chi_{c}\partial_{a}\mc{P}^{cb}\,,
\end{align}
\end{subequations}
where we expressed all transformations in terms of the components of $\mc{P}^{\alpha\beta}$ and we denoted $\partial_{a}\equiv \partial/\partial \theta^{a}$. This form of the gauge symmetries agrees with what was described in Ref. \cite{Ikeda:1993fh}. 

The supersymmetry transformations of the fields that leave the action invariant are controlled by the homological vector field $\mc Q_{S}$ and they are given as 
\begin{equation}
\delta_{S}\Phi=[\mc Q_{S},\Phi]\,.
\end{equation}
We can rewrite them in more explicit terms as 
\begin{subequations}\begin{align}
    \delta_{S}X^{\mu}&= \theta^{a} t_{a}{}^{\mu}\,,\\[4pt]
    \delta_{S}A_{\mu}&= A_{\nu}\theta^{a} U^{\nu}{}_{a\mu}+\chi_{b}W^{b}{}_{\mu}\,,\\[4pt]
    \delta_{S}\theta^{a}&= V^{a}\,, \\[4pt]
    \delta_{S}\chi_{a}&= A_{\mu}Y^{\mu}{}_{a}+\chi_{b}\theta^{c}Z^{b}{}_{ca}\,. 
\end{align}
\end{subequations}
This is in general a nonlinear supersymmetry transformation and we also recall that all coefficients may also depend on even powers of $\theta^{a}$, in addition to their $X^{\mu}$ dependence. Moreover, for a given model, there might exist more than one supersymmetry \cite{Arias:2016agc}, giving rise to extended supersymmetric models. The kinetic sector of the general model is invariant under these supersymmetry transformations provided that the symplectic form in Darboux coordinates is invariant under the supersymmetry-generating homological vector field $\cQ_{S}$. The analysis of this condition is simple, see Appendix \ref{app:QS_omega}, and it fixes all coefficients in terms of only $t_{a}{}^{\mu}$ and $V^{a}$. We repeat the result here:
\begin{align}\label{eq: QQS}
    Y^{\mu}{}_{a} = - t_{a}{}^{\mu}
    + \theta^{b}\partial_{a} t_{b}{}^{\mu}\,,
    \qquad
    U^{\mu}{}_{a\nu} = \partial_{\nu} t_{a}{}^{\mu}\,,
    \qquad
    W^{a}{}_{\mu} = \partial_{\mu}V^{a}\,,
    \qquad
    \theta^{c} Z^{a}{}_{cb} = \partial_{b} V^{a}\,.
\end{align}
The invariance of the interaction sector under supersymmetry 
corresponds to the condition $[\cQ,\cQ_{S}]=0$,
which is more complicated in its general form,
see Appendix \ref{app:QQ}. We will describe specific solutions
later in this section.

As for any gauge theory, the field strengths of the various fields are given in terms of the components of the homological vector field $\mc Q$ by the formula 
\begin{equation}
    F^{\alpha}=\dd\Phi^{\alpha}-\mc Q^{\alpha}\,,
\end{equation}
and similarly for the lower index, or in more detail
\begin{subequations}
\begin{align}
    F^{\mu}&=\dd X^{\mu}-A_{\nu}\mc{P}^{\nu\mu}+\chi_{a}\theta^{b}\,\mc{P}^{a\mu}_{b}\,, \\[4pt] 
    F_{\mu}&= \dd A_{\mu}+\tfrac 12 A_{\nu}\wedge A_{\rho} \,\partial_{\mu}\mc{P}^{\nu\rho}-A_{\nu}\wedge \chi_{b}\,\theta^{a}\,\partial_{\mu}\mc{P}^{\nu b}_{a}+\tfrac 12 \chi_{a}\wedge \chi_{b}\, \partial_{\mu}\mc{P}^{ab}\,,\\[4pt] 
    F^{a}&=\dd\theta^{a}-A_{\mu}\theta^{b}\,\mc{P}^{\mu a}_{b}+\chi_{b}\,\mc{P}^{ba}\,,\\[4pt] 
    F_{a}&=\dd\chi_{a}+\tfrac 12 A_{\mu}\wedge A_{\nu}\,\partial_{a}\mc{P}^{\mu\nu}+A_{\mu}\wedge \chi_{b}\,\partial_{a}\mc{P}^{\mu b}+\tfrac 12 \chi_{b}\wedge \chi_{c}\,\partial_{a}\mc{P}^{bc}\,.
\end{align}
\end{subequations}
The equations of motion for the topological sigma model are obtained by setting all these field strengths to zero. 
These field strengths are often said to be `Cartan integrable',
meaning that setting them to zero is consistent
with the fact that the de Rham differential is nilpotent.
More concretely, this means that they verify
\begin{equation}
    \dd F^\alpha + F^{\beta}\,\partial_{\beta} \cQ^{\alpha}
    = \cQ^{\beta}\,\partial_{\beta} \cQ^{\alpha} \equiv 0\,,
\end{equation}
which can be read as a kind of generalized covariant
constancy condition for the curvature defined by $\cQ$
and hence is often referred to as a Bianchi identity
(see, e.g., \cite[Sec. 3.3]{Boulanger:2008up}).%

The formulation presented so far, which is due to Ikeda,
is direct and simple, but it does not exhibit manifest
target space covariance. This is particularly important
in the supersymmetric case, since there is
a $T^{\ast}M$-on-$E^{\ast}$ connection hidden in the components
$\mc{P}^{\mu a}$, as we discussed in Section \ref{sec:Psmfld}, 
which is not the case for the bosonic model. To account for this, 
we make the assumption that this $T^{\ast}M$-on-$E^{\ast}$ connection is induced by a $TM$-on-$E$ connection $\nabla$
with coefficients $\Gamma_{\mu a}^{b}$ and we perform the field redefinition 
\begin{equation}
A^{\scriptscriptstyle{\nabla}}_{\mu}=A_{\mu}+\Gamma_{\mu a}^{b}\theta^{a}\chi_{b}\,.
\label{field redefinition}
\end{equation}
Writing the action functional in terms of the redefined 1-form, we obtain 
\begin{align}
   S[X,A^{\scriptscriptstyle{\nabla}},\theta,\chi]
   & = \int \bigg(A_{\mu}^{\scriptscriptstyle{\nabla}}\wedge \dd X^{\mu}+\chi_{a}\wedge \nabla\theta^{a}+\tfrac 12 \mc P^{\mu\nu}A^{\scriptscriptstyle{\nabla}}_{\mu}\wedge A_{\nu}^{\scriptscriptstyle{\nabla}}+\big(\mc P^{\mu a}+\mc P^{\mu\nu}\Gamma_{\nu b}^{a}\theta^{b}\big)A^{\scriptscriptstyle{\nabla}}_{\mu}\wedge \chi_a \nonumber\\[4pt] 
    & \hspace{50pt} +\, \tfrac 12 \big(\mc P^{ab}-2\mc P^{\mu a}\Gamma_{\mu c}^{b}\theta^{c}+\mc P^{\mu\nu}\Gamma_{\mu c}^{a}\Gamma_{\nu d}^{b}\theta^{c}\theta^{d}\big)\chi_{a}\wedge\chi_{b}\bigg)\,,
\end{align}
where the covariant exterior derivative is given as 
\begin{equation}
    \nabla\theta^{a}=\dd\theta^{a}+\Gamma_{\mu b}^{a}\dd X^{\mu}\theta^{b}\,.
\end{equation}
This aligns with the definitions we presented
in Section \ref{sec:Psmfld}, in particular we may write
the action in the compact form 
\begin{equation}
    S[X^{\alpha},P^{\scriptscriptstyle{\nabla}}_{\alpha}]
    = \int \bigg(P^{\scriptscriptstyle{\nabla}}_{\alpha} 
    \wedge \nabla X^{\alpha} + \tfrac12\,(-1)^{\alpha(\beta+1)}\,
    \mc P_{\scriptscriptstyle{\nabla}}^{\alpha\beta}(X) 
    P^{\scriptscriptstyle{\nabla}}_{\alpha}
    \wedge P^{\scriptscriptstyle{\nabla}}_{\beta}\bigg)\,,
\end{equation}
with $P^{\scriptscriptstyle{\nabla}}_{\alpha}
=(A^{\scriptscriptstyle{\nabla}}_{\mu},\chi_{a})$
and the covariant components
$\mc P_{\scriptscriptstyle{\nabla}}^{\alpha\beta}$
defined in \eqref{eq: P covariant}.  
This equivalent formulation will be useful in understanding
some specific cases below. Each of the two equivalent formulations, related through
the simple field redefinition \eqref{field redefinition}, 
highlights different features of the model. For example,
although the formulation in terms of the redefined $1$-form 
highlights target space covariance, the original one
is much simpler in the specification of the geometrical data. 
Indeed, the degree $(1,0)$ symplectic $2$-form of the model reads
\begin{subequations}
\begin{align}
    \omega & =  \dd A_{\mu} \wedge \dd X^{\mu}
    + \dd\chi_{a} \wedge \dd\theta^{a} \\[4pt] 
    & =  \dd A^{\scriptscriptstyle{\nabla}}_{\mu}
    \wedge \dd X^{\mu} + \dd\chi_{a} \wedge (\dd\theta^{a}
    +\dd X^{\mu}\Gamma_{\mu b}^{a}\theta^{b})
    \nonumber\\[4pt] & \quad
    -\,\chi_{a}\Gamma^{a}_{\mu b} \dd X^{\mu}
    \wedge \dd\theta^{b}
    -\chi_{a} \partial_{\kappa}\Gamma_{\nu b}^{a}
    \theta^{b} \dd X^{\nu} \wedge\dd X^{\kappa}\,,
\end{align}
\end{subequations}
and it is canonical in the original formulation,
instead containing various off diagonal terms in terms
of the redefined field content, where we do not use Darboux coordinates. It will be useful to express the  
supersymmetry transformations in terms of this new field. 
For the fields $X^{\mu}$ and $\theta^{a}$, they do not get modified, whereas for the remaining two fields the new 
supersymmetry transformations are 
\begin{subequations}
\begin{align}
    \delta_{S} A^{\scriptscriptstyle{\nabla}}_{\mu}
    & =  A^{\scriptscriptstyle{\nabla}}_{\nu}\theta^{a}
    \big(U^{\nu}{}_{a\mu}+\Gamma_{\mu a}^{b}Y^{\nu}{}_{b}\big)
    +\chi_b\big(W^{b}{}_{\mu}+\Gamma_{\mu a}^{b}V^{a}
    +(t_{c}{}^{\nu}\partial_{\nu}\Gamma_{\mu d}^{b}
    -\Gamma_{\mu c}^{e}Z^{b}{}_{de})\theta^{c}\theta^{d}\big)\,,
     \\[4pt] 
    \delta_{S}\chi_{a}
    & =  A_{\mu}^{\scriptscriptstyle{\nabla}}Y^{\mu}{}_{a}
    +\chi_{b}\theta^{c}\big(Z^{b}{}_{ca}
    -Y^{\mu}{}_{a}\Gamma_{\mu c}^{b}\big)\,,
\end{align}
\end{subequations}
with coefficients as in \eqref{eq: QQS}. A characteristic example 
of such a supersymmetric Poisson sigma model is $\mc N=1$
dilaton supergravity in $2$ dimensions,
based on the super-Poincar\'e algebra \cite{Ikeda:1993fh}.
In the rest of this section, we will construct two examples based on Lie super-algebroids and show that they are essentially singled out through an analysis of the general case.

\subsection{The ABST-G model and de Rham supersymmetry}
\label{sec: ABSTG}
The \emph{differential Poisson sigma model},
introduced in \cite{Arias:2015wha, Arias:2016agc},
is an example of super-Poisson sigma model, 
wherein the target space is obtained from a dg-Poisson
manifold concentrated in degrees $0$ and $1$.
As discussed previously, such a target space
is equivalent to a shifted vector bundle $E[1]$,
for which the space of sections of the exterior algebra
of its dual, $\Gamma(\wedge E^{\ast})$, is equipped
with a (degree $0$) Poisson bracket. In fact,
the differential Poisson sigma model is obtained
from the general model of Section \ref{ikeda} 
for the choice $E=TM$, in other words, the target space
is $T^{\ast}[1,0]T[0,1]M$.

To be specific, the fields of this differential Poisson sigma model are as in Section \ref{ikeda} for $E=TM$, namely $(X^{\mu},\theta^{\mu},A_{\mu}^{\scriptscriptstyle{\nabla}},\chi_{\mu})$, and its action functional is  
\begin{equation}
    S_{\,TM} = \int\bigg(A_{\mu}^{\scriptscriptstyle{\nabla}} 
    \wedge \dd X^{\mu} + \chi_{\mu} \wedge \nabla\theta^{\mu}
    + \tfrac12 \Pi^{\mu\nu} A^{\scriptscriptstyle{\nabla}}_{\mu}
    \wedge A^{\scriptscriptstyle{\nabla}}_{\nu}
    + \tfrac14 R_{\kappa\lambda}{}^{\mu\nu} \chi_{\mu} 
    \wedge \chi_{\nu}\,\theta^{\kappa} \theta^{\lambda}\bigg)\,,
\end{equation}
with the coefficient $R_{\kappa\lambda}{}^{\mu\nu}$
to be specified. This action is obtained by the following choice
of components for the structure on the Poisson supermanifold: 
\begin{subequations}\label{eq:Poisson_tangent}
\begin{align}
    \mc P^{\mu\nu}&= \Pi^{\mu\nu}\,, 
    \label{poisson 00 abstg} \\[4pt] 
    \mc P^{\mu a} & = -\Pi^{\mu\kappa}
    \Gamma_{\kappa\lambda}^{a}\theta^{\lambda}\,, \\[4pt] 
    \mc P^{ab} &=  \tfrac 12\big( R_{\rho\sigma}{}^{ab}
    + 2\Pi^{\kappa\lambda}\Gamma^{(a}_{\kappa\rho}\Gamma^{b)}_{\lambda\sigma}\big)\theta^{\rho}\theta^{\sigma}\,,
    \label{poisson 11 abstg}
\end{align}
\end{subequations}
where we kept the Latin indices to distinguish
between different components without introducing
an unnecessary layer of notation, even though in this case
all indices are of the same type and in the position
they appear. We observe that with respect to the expansion~\eqref{super Poisson expanded} of the super-Poisson structure, 
in this case the (inverse) metric $g^{\mu\nu}$
is completely degenerate and there is no quadratic piece
in $\mc P^{\mu\nu}$. Moreover, the $T^{\ast}M$-connection
is chosen to be the canonically induced one
from an ordinary connection and thus the endomorphism~$\phi$ 
vanishes. With these identifications, we can now gradually 
unveil the particular graded geometrical structure
of the target space. Before doing so, it is useful to mention 
that in the original formulation in terms
of the field $A_{\mu}$, the action may be equivalently
written as 
\begin{align}
    S_{\,TM} & = \int\bigg(A_{\mu} \wedge \dd X^{\mu}
    + \chi_{\mu} \wedge \dd \theta^{\mu}
    + \tfrac12 \Pi^{\mu\nu} A_{\mu} \wedge A_{\nu}
    \nonumber \\[4pt] 
    & \quad\qquad - \Pi^{\mu\kappa}\Gamma_{\kappa\lambda}^{\nu}
    \theta^{\lambda}A_{\mu}\wedge\chi_{\nu}
    + \tfrac14 \big( R_{\rho\sigma}{}^{\mu\nu}
    +2\Pi^{\kappa\lambda}\Gamma^{\mu}_{\kappa\rho}
    \Gamma^{\nu}_{\lambda\sigma}\big)\chi_{\mu}\wedge\chi_{\nu}
    \theta^{\rho}\theta^{\sigma}\bigg)\,. \qquad 
\end{align}
Turning to the homological
vector fields that dictate the gauge symmetries
and the supersymmetries respectively, once again
it is much simpler to reveal the structure
in the original formulation. Starting from the $(1,0)$
vector field $\mc Q$, it is completely specified by
the components \eqref{poisson 00 abstg}-\eqref{poisson 11 abstg}, 
together with the requirement that it is compatible
with the graded symplectic structure.
The result in this case is 
\begin{align}
    \mc Q  & = \big(\Pi^{\mu\nu}a_{\mu}
    - \Pi^{\nu\kappa}\Gamma_{\kappa\rho}^{\mu}\,
    \chi_{\mu}\theta^{\rho}\big)
    \tfrac{\partial}{\partial x^{\nu}}\nonumber\\[4pt]
    & + \big(\!-\Pi^{\mu\kappa}\Gamma_{\kappa\lambda}^{\nu}\,
    a_{\mu}\theta^{\lambda} + \tfrac12\chi_{\mu}
    \big({R}_{\rho\sigma}{}^{\mu\nu}
    +2\Pi^{\kappa\lambda}\Gamma_{\kappa\rho}^{(\mu}
    \Gamma_{\lambda\sigma}^{\nu)}\big)
    \theta^{\rho}\theta^{\sigma}\big)
    \tfrac{\partial}{\partial\theta^{\nu}} \nonumber\\[4pt]
    & + \big(\!-\tfrac12\,\partial_{\lambda}\Pi^{\mu\nu}\,
    a_{\mu} a_{\nu} - \Pi^{\mu\kappa}\Gamma_{\kappa\lambda}^{\nu}\,
    a_{\mu} \chi_{\nu} \theta^{\lambda} 
    + \tfrac14\,\partial_{\lambda}\big( {R}_{\rho\sigma}{}^{\mu\nu}+2\Pi^{\kappa\lambda}
    \Gamma_{\kappa\rho}^{(\mu}\Gamma_{\lambda\sigma}^{\nu)}
    \big)
    \chi_{\mu} \chi_{\nu} \theta^{\rho} \theta^{\sigma}\big)
    \tfrac{\partial}{\partial a_{\lambda}} \nonumber\\[4pt]
    &+ \big(\!\, \Pi^{\mu\kappa}
    \Gamma_{\kappa\sigma}^{\nu}\,a_{\mu} \chi_{\nu}
    + \tfrac 12\big({R}_{\sigma\rho}{}^{\mu\nu}+2\Pi^{\kappa\lambda}
    \Gamma_{\kappa\sigma}^{(\mu}\Gamma_{\lambda\rho}^{\nu)}
    \big)
    \chi_{\mu} \chi_{\nu} \theta^{\rho}\big)
    \tfrac{\partial}{\partial\chi_{\sigma}}\,.
\end{align}
The supersymmetry generating homological vector field
of degree $(0,1)$ in this case is
\begin{equation}
    \mc Q_{S} = \theta^{\mu}\tfrac{\partial}{\partial x^{\mu}}
     -a_{\mu}\tfrac{\partial}{\partial\chi_{\mu}}\,.
\end{equation}
It corresponds to the simple and linear
supersymmetry transformations 
\begin{equation}
    \delta_{S}X^{\mu} = \theta^{\mu}\,,
    \qquad \delta_{S}\chi_{\mu} = -A_{\mu}\,,
    \qquad \delta_{S}\theta^{\mu} = 0 = \delta_{S}A_{\mu}\,,
\end{equation}
which is easily checked to be a symmetry
of the action $S_{\,TM}$. As noticed in \cite{Arias:2016agc}, 
this supersymmetry is associated to the de Rham differential 
and we call it de Rham supersymmetry. That $\mc Q_{S}$
is homological is obvious, while its compatibility
with the graded symplectic structure follows in a simple way 
from the fact that the conditions
\eqref{QS omega compatibility} hold. Regarding its commutator 
with the vector field $\cQ$, according to the detailed analysis 
in Appendix \ref{app:QQ}, we get three independent 
conditions,{\footnote{The other two conditions are trivial 
because both $g^{\mu\nu}$ and $V^{\mu}$ vanish
in this example.}} 
which we write directly in terms of the covariant component
of $\mc P_{\scriptscriptstyle{\nabla}}^{ab}$, denoted as 
$R_{\kappa\lambda}{}^{\mu\nu}$ in \eqref{eq: P covariant 2}:
\begin{subequations}
    \begin{align}
        & \overline\nabla_{\kappa}\Pi^{\mu\nu}=0\,, \\[4pt] 
        & R_{\kappa\lambda}{}^{\mu\nu}=-\Pi^{(\mu|\rho}
        R^{\scriptscriptstyle{\overline\nabla}}_{\kappa\lambda}{}^{|\nu)}{}_{\rho}\,, \\[4pt] 
        & \overline\nabla_{[\rho}R_{\kappa\lambda]}{}^{\mu\nu}-{T}^{\scriptscriptstyle{\overline{\nabla}}}_{[\rho\kappa}{}^{\sigma}R_{\lambda]\sigma}{}^{\mu\nu}=0\,,
    \end{align}
\end{subequations}
where we introduced the opposite (or conjugated) connection
$\overline\nabla$ defined by
\begin{equation}
    \overline\nabla_X Y := \nabla_Y X + [X,Y]
    \qquad\iff\qquad 
    \overline\Gamma_{\mu\nu}^\lambda
        = \Gamma_{\nu\mu}^\lambda\,.
\end{equation}
The opposite connection is such that the ``average'' connection
$\tfrac 12(\nabla+\overline{\nabla})$ is torsion-free.
Note that the same result is obtained from
\eqref{basic connection coefficients}, meaning that
the opposite connection is the same as the basic connection 
introduced to facilitate the concept of representations
up to homotopy, which explains why we use the same notation.
We observe that in this example we have obtained the components 
of $\mc P_{\scriptscriptstyle{\nabla}}^{ab}$, which are fixed 
in terms of the Poisson structure on the base and the curvature 
of the basic connection. It is also worth mentioning that
the curvature of the basic connection in the case
of the tangent bundle is opposite to the basic curvature
of the ordinary connection.  

In the covariant formulation, we can write the vector field
in terms of $a^{\scriptscriptstyle{\nabla}}_{\mu}
=a_{\mu}+\Gamma_{\mu\nu}^{\rho}\theta^{\nu}\chi_{\rho}$ as 
\begin{equation}
    \mc Q_{S}=\theta^{\mu}\tfrac{\partial}{\partial x^{\mu}}
    - \big(a^{\scriptscriptstyle{\nabla}}_{\nu}\theta^{\rho}
    \overline{\Gamma}^{\nu}_{\rho\mu} + \tfrac12 \chi_{\nu}
    {R}^{\scriptscriptstyle{\overline{\nabla}}}_{\kappa\lambda}{}^{\nu}{}_{\mu}
    \theta^{\kappa}\theta^{\lambda}\big)
    \tfrac{\partial}{\partial a^{\scriptscriptstyle{\nabla}}_{\mu}}
    - \big(a^{\scriptscriptstyle{\nabla}}_{\mu}
    -\overline{\Gamma}_{\mu\rho}^{\nu}
    \theta^{\rho}\chi_{\nu}\big)
    \tfrac{\partial}{\partial\chi_{\mu}}\,,
\end{equation}
which generates the same supersymmetry transformations
for the redefined fields and it is a symmetry
of the classical action. Comparing this
with the structure operator \eqref{Qs coruth}
in the representation up to homotopy perspective,
we observe that the supersymmetry-generating homological vector field corresponds
to the coadjoint representation of the tangent Lie algebroid 
with unit anchor map and the usual Lie bracket of vector fields.

\subsection{The contravariant model and Poisson supersymmetry}
\label{sec:contravariant PSM}
We now introduce a different supersymmetric Poisson sigma model, 
this time with choice of vector bundle $E=T^{\ast}M$.
Its target space is $T^{\ast}[1,0]T^{\ast}[0,1]M$ 
and the fields of the sigma model are
$(X^{\mu},\theta_{\mu},A_{\mu}^{\scriptscriptstyle{\nabla}},
\chi^{\mu})$.{\footnote{Note that models with a double cotangent bundle target space where constructed in \cite{Chatzistavrakidis:2021nom, Chatzistavrakidis:2022hlu, Chatzistavrakidis:2024utp, Ikeda:2021rir}, dubbed (twisted) R-Poisson sigma models due to their particular geometric structure. These are different models, since they are purely bosonic and they were mainly considered in higher than $2$ dimensions.}}
The main difference is that the fermionic fields take values
in the dual bundle compared to the previous case.
For this reason, we will refer to this case
as the contravariant supersymmetric Poisson sigma model.
Its action functional is 
\begin{align}
    S_{\,T^{\ast}M}
    = \int\bigg(A_{\mu}^{\scriptscriptstyle{\nabla}}
    \wedge \dd X^{\mu} + \chi^{\mu} \wedge \nabla\theta_{\mu}
    + \tfrac12\Pi^{\mu\nu} A^{\scriptscriptstyle{\nabla}}_{\mu}
    \wedge A^{\scriptscriptstyle{\nabla}}_{\nu} 
    + \tfrac14 R^{\kappa\lambda}{}_{\mu\nu}
    \chi^{\mu} \wedge \chi^{\nu}
    \theta_{\kappa}\theta_{\lambda}\bigg)\,,
\end{align}
where the coefficient of the quartic term
$R^{\kappa\lambda}{}_{\mu\nu}$ must be determined anew
(and it is different than in the previous model).
To understand the structure of this dual model,
we should specify the geometric meaning of the coefficients
that appear in it and the new type of supersymmetry it features.
In accordance with our analysis in Section \ref{sec:Psmfld},
the bivector $\Pi^{\mu\nu}$ is again a Poisson structure
on the body of the target supermanifold.
From a non-manifestly covariant perspective,
the choice of coefficients in the sigma model is 
\begin{subequations}\label{eq:Poisson_cotangent}
\begin{align}
    \mc P^{\mu\nu}&= \Pi^{\mu\nu}\,, 
    \label{poisson 00 contra} \\[4pt] 
    \mc P^{\mu}{}_{\nu}&= \Pi^{\mu\kappa}\Gamma_{\kappa\nu}^{\lambda}\theta_{\lambda}\,, \\[4pt] 
    \mc P_{\mu\nu} &=\tfrac 12\big( R^{\kappa\lambda}{}_{\mu\nu}-2\Pi^{\rho\sigma}\Gamma^{\kappa}_{\rho(\mu}\Gamma^{\lambda}_{\sigma\nu)}\big)\theta_{\kappa}\theta_{\lambda}\,,
    \label{poisson 11 contra}
\end{align}
\end{subequations}
corresponding to a degree $0$ Poisson structure.
Note the sign difference in the mixed component ${\mc P}^{\mu}{}_{\nu}$, since this refers to the dual connection coefficients.
This fixes completely the homological vector field $\cQ$
and determines the full set of gauge symmetries of the model.
We do not write their explicit form here, since they are 
similar to the ones of the previous model. 

With regard to supersymmetry, the contravariant model exhibits
a new feature; the supersymmetry transformations are controlled
by the Poisson structure and its derivatives, and they are nonlinear.
To track them correctly, let us first separate the action
into kinetic and interaction terms.
The kinetic sector is the one that contains derivatives,
namely the first two terms in the action $S_{\,T^{\ast}M}$.
The invariance of the kinetic sector under supersymmetry transformations 
is equivalent to the compatibility of the homological vector field
$\cQ_{S}$ with the symplectic structure. Referring to the data
of $\cQ_{S}$ as they appear in Eq. \eqref{Q susy general}
and to the compatibility conditions
Eq. \eqref{QS omega compatibility}, we observe that
in the present case it suffices to determine the coefficients 
$t^{\mu\nu}$ and $V_{\mu}$. We fix these coefficients to be 
\begin{subequations}
\begin{align}
    t^{\mu\nu}&= \Pi^{\mu\nu}\,, \\[4pt] 
    V_{\mu} & =  
    -\tfrac 12\,\partial_{\mu}\Pi^{\kappa\lambda}
    \theta_{\kappa}\theta_{\lambda}\,. 
\end{align}
\end{subequations}
Thus $t^{\mu\nu}$ is identified with the Poisson structure
on the body of the supermanifold, which now plays a dual role, 
inducing both the gauge symmetries and the supersymmetry
of the model. The kinetic sector
of the sigma model respects the supersymmetry transformations 
generated by this $\cQ_{S}$ provided that the rest
of the coefficients are 
\begin{equation}
    Y^{\mu\nu}=\Pi^{\mu\nu}\,,
    \quad
    U^{\nu\rho}{}_{\mu} = \partial_{\mu}\Pi^{\rho\nu}\,,
    \quad
    Z_{\mu}{}^{\nu\rho} = \partial_{\mu}\Pi^{\nu\rho}\,,
    \quad
    W_{\mu\nu} = -\tfrac 12\,\partial_{\mu}\partial_{\nu}\Pi^{\rho\sigma}
    \theta_{\rho}\theta_{\sigma}\,.
\end{equation}
We observe that the supersymmetry is completely determined
by the Poisson structure already at kinetic level. 

In addition, we must address the supersymmetry of the interaction sector. This amounts to the compatibility between the two homological vector fields $\cQ$ and $\cQ_{S}$, which is also studied in full generality in Appendix \ref{appa}. 
There are three conditions that should be met, as in the previous case. 
The first condition reads
\begin{equation}\label{eq: Pi Pi T}
    \Pi^{\mu\kappa}\Pi^{\nu\lambda}(\Gamma^{\rho}_{\kappa\lambda}
    -\Gamma^{\rho}_{\lambda\kappa}) = 0
    \quad \iff \quad
    T^{\nabla}\big(\Pi^{\sharp}(\eta_1),\Pi^{\sharp}(\eta_2)\big)=0\,, 
    \quad \forall \, \eta_1, \eta_2\in \Gamma(T^{\ast}M)\,,
\end{equation}
and it has an interesting geometric interpretation
that ties it together with the previous case.
Considering the basic connection in the present case,
its coefficients are generally given by
Eq. \eqref{basic connection coefficients},
and focusing on its $T^{\ast}M$-on-$T^{\ast}M$ component,
say $\overline{\nabla}^{\ast}$, we obtain 
\begin{equation}\label{basic 1 T*M}
    \overline{\Gamma}^{\mu\nu}{}_{\rho}
    = -\Pi^{\nu\kappa}\Gamma_{\kappa\rho}^{\mu}
    + \partial_{\rho}\Pi^{\mu\nu}\,.
\end{equation}
A straightforward calculation leads to 
\begin{equation}
    T^{\nabla}\big(\Pi^{\sharp}(\eta_1),\Pi^{\sharp}(\eta_2)\big)=0 
    \qquad \iff \qquad
    \overline{\nabla}^{\ast}\Pi(\eta_1,\eta_2) = 0\,.
\end{equation}
We observe that we obtain a condition analogous
to the case of the tangent bundle; the Poisson structure
is covariantly constant with respect to the basic connection
in each case. Note that the other component of the basic connection
is a $T^{\ast}M$-on-$TM$ one with components 
\begin{equation}\label{basic 2 T*M}
    {\overline{\Gamma}}^{\mu}{}_{\rho}{}^{\nu}
    =\Pi^{\nu\kappa}\Gamma_{\rho\kappa}^{\mu}
    -\partial_{\rho}\Pi^{\mu\nu}\,,
\end{equation}
and that \eqref{basic 1 T*M} and \eqref{basic 2 T*M}
are not a pair of dual connections, which is the reason
we carefully placed their indices. The second condition reads 
\begin{equation}
    \Pi^{\nu\sigma}\big(R^{\kappa\lambda}{}_{\sigma\rho}
    + S^{\kappa\lambda}{}_{\sigma\rho}\big) = 0\,,
\end{equation}
where $S$ is the basic curvature in this case,
whose precise local coordinate expression is 
\begin{equation}
    S^{\kappa\lambda}{}_{\mu\nu}
    =\partial_{\mu}\partial_{\nu}\Pi^{\kappa\lambda}
    - 2\Gamma_{\mu\rho}^{[\kappa}\partial_{\nu}\Pi^{\lambda]\rho}
    - \Gamma_{\mu\nu}^{\rho}\partial_{\rho}\Pi^{\kappa\lambda}
    + 2\Gamma_{\sigma\nu}^{[\lambda}\partial_{\mu}\Pi^{\kappa]\sigma}
    - 2\Pi^{\sigma[\kappa}\partial_{\sigma}\Gamma_{\mu\nu}^{\lambda]}
    + 2\Pi^{\rho\sigma}\Gamma_{\mu\rho}^{[\kappa}\Gamma_{\sigma\nu}^{\lambda]}\,.
\end{equation} 
This means that we can take the coefficient
$R^{\kappa\lambda}{}_{\mu\nu}$ to be opposite to the basic curvature
up to terms in the kernel of the map $\Pi^{\sharp}$.
To complete the analysis and fully determine the coefficient,
we determine the third and final condition
from supersymmetric invariance of the interaction sector, which reads
\begin{align}\label{eq:Bianchi_R}
    \Pi^{\rho[\sigma}\partial_{\rho}R^{\kappa\lambda]}{}_{(\mu\nu)}
    + R^{\rho[\sigma}{}_{(\mu\nu)}\partial_{\rho}\Pi^{\kappa\lambda]}
    - 2\overline{\Gamma}^{[\sigma|\rho}{}_{(\mu|}
    R^{|\kappa\lambda]}{}_{\rho|\nu)} = 0\,.
\end{align}
This condition bears a striking resemblance with the Bianchi identity
verified by the basic curvature, which reads
\begin{equation}
    \Pi^{[\sigma|\rho} \partial_{\rho} S^{|\kappa\lambda]}{}_{\mu\nu}
    + S^{[\sigma|\rho}{}_{\mu\nu} \partial_{\rho} \Pi^{|\kappa\lambda]} 
    + \overline{\Gamma}^{[\sigma|\rho}{}_{\nu}\,
        S^{|\kappa\lambda]}{}_{\mu\rho}\,
    - \overline{\Gamma}^{[\sigma|}{}_{\mu}{}^{\rho}\,
        S^{|\kappa\lambda]}{}_{\rho\nu} = 0\,,
\end{equation}
and differs from \eqref{eq:Bianchi_R} only by the last term
involving the component of the $T^{\ast}M$-on-$TM$ part
of the basic connection \eqref{basic 2 T*M}. Comparing the latter
with \eqref{basic 1 T*M}, the component of the $T^{\ast}M$-on-$T^{\ast}M$
part of the basic connection, one finds that they are related by
\begin{equation}\label{eq:rel basic cotangent}
    \overline{\Gamma}^{\mu}{}_{\lambda}{}^{\nu}
    = -\overline{\Gamma}^{\mu\nu}{}_{\lambda}
    - \Pi^{\nu\alpha} T_{\alpha\lambda}{}^{\mu}\,,
\end{equation}
with $T_{\alpha\beta}{}^{\gamma} = 2\,\Gamma_{[\alpha\beta]}^{\gamma}$
the torsion of $\nabla$, and hence the Bianchi identity
for the basic curvature can be re-written as
\begin{equation}
    \Pi^{[\sigma|\rho} \partial_{\rho} S^{|\kappa\lambda]}{}_{\mu\nu}
    + S^{[\sigma|\rho}{}_{\mu\nu} \partial_{\rho} \Pi^{|\kappa\lambda]} 
    + \overline{\Gamma}^{[\sigma|\rho}{}_{\nu}\,
        S^{|\kappa\lambda]}{}_{\mu\rho}\,
    + \overline{\Gamma}^{[\sigma|\rho}{}_{\mu}\,
        S^{|\kappa\lambda]}{}_{\rho\nu}
    + \Pi^{\alpha\beta} T_{\beta\mu}{}^{[\sigma} S^{\kappa\lambda]}{}_{\alpha\nu} = 0\,.
\end{equation}
Knowing that $R$ is opposite to the basic curvature,
up to terms in the kernel, we conclude that we may take the coefficient of the quartic term in the action to be%
\footnote{Let us remark that the symmetric part
of the basic curvature for the cotangent Lie algebroid
(in its lower indices) is given by
\[
     S^{\kappa\lambda}{}_{(\mu\nu)}
     = \mathring{S}^{\kappa\lambda}{}_{\mu\nu}
     - \tfrac12\,\Pi^{\rho\sigma}\,
     T_{\rho\mu}^{[\kappa}\,T_{\sigma\nu}^{\lambda]}\,,
\]
where $\Gamma^{\lambda}_{\mu\nu}
= \mathring{\Gamma}_{\mu\nu}^{\lambda}
+ \tfrac12\,T^{\lambda}_{\mu\nu}$
with $\mathring{\Gamma}_{\mu\nu}^{\lambda}$ the symmetric,
i.e. torsionless, part of the connection $\nabla$,
and $\mathring{S}$ denotes the basic curvature with respect to
this torsion-free connection.}%
\begin{equation}
    R^{\kappa\lambda}{}_{\mu\nu} = -S^{\kappa\lambda}{}_{(\mu\nu)}\,,
\end{equation}
upon also imposing that the torsion of $\nabla$ belongs
to the kernel of $\Pi^{\sharp}$, 
\begin{equation}\label{eq: Pi T}
    \Pi^{\mu\rho} T_{\rho\nu}{}^{\sigma} = 0
    \qquad\iff\qquad 
    T^{\nabla}(\Pi^{\sharp}(\eta),-) = 0\,,
    \quad \forall \eta \in \Gamma(T^{\ast}M)\,,
\end{equation}
thereby implying the previously encountered condition
\eqref{eq: Pi Pi T}. Note that, in light of
Eq.~\eqref{eq:rel basic cotangent}, this condition amounts
to requiring that the basic connection is indeed formed
of a pair of dual connections. Equivalently, this amounts
to the requirement that the basic connection preserves
the non-degenerate bilinear form on the complex $TM \oplus T^{\ast}M$
that is the canonical pairing between the tangent
and cotangent bundle, i.e.
\begin{equation}
    \Pi^{\sharp}(\eta) \langle X, \eta' \rangle
    = \langle \overline{\nabla}_{\eta} X, \eta' \rangle
    + \langle X, \overline{\nabla}_{\eta} \eta' \rangle\,,
    \qquad X \in \Gamma(TM)\,,\ \eta,\eta' \in \Gamma(T^{\ast}M)\,,
\end{equation}
where on the right hand side, the first term
is the $T^{\ast}M$-on-$TM$ part of the basic connection,
and the second term its $T^{\ast}M$-on-$T^{\ast}M$ part.
A direct computation shows that the above identity holds
if and only if \eqref{eq: Pi T} is satisfied.

Summarizing the discussion on supersymmetry and keeping in mind
the identified geometrical conditions, the  homological vector field 
that generates supersymmetry transformations is 
\begin{align}
    \mc Q_{S} & = \Pi^{\mu\nu}\theta_{\mu}
        \tfrac{\partial}{\partial x^{\nu}}
    -\tfrac12 \partial_{\mu}\Pi^{\kappa\lambda}
        \theta_{\kappa}\theta_{\lambda}
        \tfrac{\partial}{\partial\theta_{\mu}}\,+ \nonumber\\[4pt] 
    & \qquad +\,\big(\partial_{\rho}\Pi^{\nu\mu}a_{\mu}\theta_{\nu}
    -\tfrac12 \partial_{\rho}\partial_{\nu}\Pi^{\kappa\lambda}
    \theta_{\kappa}\theta_{\lambda}\chi^{\nu}\big)
    \tfrac{\partial}{\partial a_{\rho}} + \big(\Pi^{\mu\rho}a_{\mu}
    -\partial_{\mu}\Pi^{\rho\nu}\chi^{\mu}\theta_{\nu}\big)
    \tfrac{\partial}{\partial\chi^{\rho}}\,.
\end{align}
Indeed, besides being homological, this $(0,1)$ vector field
is also compatible with the symplectic structure,
since it satisfies the corresponding consistency conditions.
We write the supersymmetry transformations of the various fields 
explicitly:
\begin{subequations}
\begin{align}
    \delta_{S} X^{\mu} & = -\Pi^{\mu\nu}\theta_{\nu}\,, \\[4pt] 
    \delta_{S} \theta_{\mu}
        & = -\tfrac12\, \partial_{\mu}\Pi^{\kappa\lambda}
        \theta_{\kappa}\theta_{\lambda}\,, \\[4pt] 
    \delta_{S} A_{\mu}
        & = \partial_{\mu}\Pi^{\kappa\lambda}A_{\lambda}\theta_{\kappa}
    - \tfrac12\partial_{\mu}\partial_{\nu}\Pi^{\kappa\lambda}
    \theta_{\kappa}\theta_{\lambda}\chi^{\nu}\,, \\[4pt] 
    \delta_{S} \chi^{\mu} & = -\Pi^{\mu\nu}A_{\nu}
    - \partial_{\rho}\Pi^{\mu\nu}\chi^{\rho}\theta_{\nu}\,.
\end{align}
\end{subequations}
When we redefine the bosonic 1-form to $A^{\scriptscriptstyle{\nabla}}$,
the supersymmetry transformations for $A^{\scriptscriptstyle{\nabla}}$ 
and $\chi^{\mu}$ are given in terms of the components
\eqref{basic 1 T*M} and \eqref{basic 2 T*M} of the basic connection
and they become 
\begin{subequations}
\begin{align}
    \delta_{S}A^{\scriptscriptstyle{\nabla}}_{\mu}
    & = -\overline{\Gamma}^{\rho}{}_{\mu}{}^{\nu}
    A^{\scriptscriptstyle{\nabla}}_{\nu}\theta_{\rho}
    - \tfrac12 S^{\kappa\lambda}{}_{\mu\nu}
    \theta_{\kappa}\theta_{\lambda}\chi^{\nu}\,, \\[4pt]
    \delta_{S} \chi^{\mu}
    & = -\Pi^{\mu\nu}A^{\scriptscriptstyle{\nabla}}_{\nu}
    + \overline{\Gamma}^{\rho\mu}{}_{\nu}\chi^{\nu}\theta_{\rho}\,.
\end{align}
\end{subequations}
In this form it becomes clear that the Poisson-supersymmetry 
generating homological vector field corresponds precisely
to the structure operator for the coadjoint representation
(up to homotopy) of the cotangent Lie algebroid.

\subsection{Lie algebroid models and anchor supersymmetry}
\label{sec:sPSM_algd}
We now return to a more general context with the purpose
of determining the common characteristics of the models described
in sections \ref{sec: ABSTG} and \ref{sec:contravariant PSM},
and investigating whether other examples can be constructed
and under which conditions. To this end, we consider
any Lie algebroid $E$, giving rise to a Poisson supermanifold $E[0,1]$ 
and we take $T^{\ast}[1,0]E[0,1]$ as target space of the sigma model. 
The previous two examples correspond to the ``extremal'' cases
of tangent and cotangent Lie algebroids. To keep the discussion
more general, we include an order $0$ component in $\mc P^{ab}$
and $V^{a}$, both of which were vanishing in the examples above. 
Nevertheless, we still assume that the connection
is the induced one and that $\mc P^{\mu\nu}$ has only degree $0$ 
components given by the Poisson structure. Hence we choose: 
\begin{subequations}
\begin{align}
    \mc P^{\mu\nu}&= \Pi^{\mu\nu}\,, 
    \label{poisson 00 E} \\[4pt] 
    \mc P^{\mu a} & = -\Pi^{\mu\nu}\Gamma_{\nu b}^{a}\theta_{b}\,, \\[4pt] 
    \mc P^{ab} & = g^{ab} + \tfrac12 \big(R_{cd}{}^{ab}
    + 2\Pi^{\rho\sigma} \Gamma^{(a}_{\rho c}\Gamma^{b)}_{\sigma d}\big) 
    \theta^{c}\theta^{d}\,,
    \label{poisson 11 E}
\end{align}
\end{subequations}
for some $R_{cd}{}^{ab} = R_{[cd]}{}^{(ab)}$ to be determined by supersymmetry and for a symmetric tensor $g$ with components $g^{ab}$, not necessarily nondegenerate.%
\footnote{Although the position of indices suggests
that we should call this an inverse metric $g^{-1}$,
we refrain from using this notation because we have not assumed 
nondegeneracy and indeed in both previous examples
we had a completely degenerate metric.} Note that for vanishing $g$ the Poisson structure is not just even but actually of degree $0$, a feature shared by the examples in sections \ref{sec: ABSTG} and \ref{sec:contravariant PSM}. On the other hand, when $\Pi^{\mu\nu}$ and $R_{cd}{}^{ab}$ vanish and the only nonvanishing component is $g$, the Poisson structure has degree $-2$. We will mention an example of this type at the end of this section. 
In general, we already know that for gauge invariance of the theory,
the induced connection must satisfy the metricity 
condition~\eqref{eq: Jacobi deg0b} with respect to $g^{ab}$.
Moreover, in the basis that corresponds to the splitting
of the supermanifold, for these choices we have 
\begin{equation}
   \mc P_{\snabla}^{\mu a} = 0
   \qquad \text{and} \qquad
   \mc P_{\snabla}^{ab} = g^{ab}
   + \tfrac12 \theta^{c}\theta^{d} R_{cd}{}^{ab}\,.
\end{equation}
Furthermore, we know from the graded Jacobi identity
that the following algebraic and differential Bianchi identities
hold due to \eqref{eq: Jacobi deg1b} and \eqref{eq: Jacobi deg2b}, 
respectively:
\begin{subequations}
\begin{align}
     g^{d(e} R_{cd}{}^{ab)} & = 0\,, \\[4pt] 
    \Pi^{\mu\nu}\nabla_{\nu} R_{cd}{}^{ab} & = 0\,.
\end{align}
\end{subequations}
These, together with $R^{\snabla}(\Pi^{\sharp},\Pi^{\sharp})=0$,
are all we need for gauge invariance of the model,
based on the detailed analysis of Section \ref{sec:Psmfld}.

We now turn to the supersymmetry of the model.
Regarding the kinetic sector there is nothing new to add
with respect to the general case; its invariance
under supersymmetry fixes most of the coefficients
in $\cQ_{S}$ as in \eqref{eq: QQS}.
The two yet undetermined coefficients, $t_{a}{}^{\mu}$
and $V^{a}$ may be expanded in powers of $\theta^2$,
as explained in Section \ref{sec:Q-smfd}.
We will not deal with the most general case,
and instead we will restrict up to second order and assume that these two coefficients take the form
\begin{subequations}
\label{eq:Q_S-algebroid}
\begin{align}
    t_{a}{}^{\mu}(x,\theta^2) & = t_{a}{}^{\mu}(x)\,, \\[4pt] 
    V^{a}(x,\theta^2) & = \mathring{V}^{a}(x)
    + \tfrac12 C_{bc}{}^{a}(x)\theta^{b}\theta^{c}\,.
\end{align}
\end{subequations}
In plain words, $t_{a}{}^{\mu}$ is only a function of the bosonic
scalar field, and all powers equal to or higher than $\theta^2$ vanish.
Geometrically, these are the components of the map $t: E \to TM$,
the anchor of the Lie algebroid $E \twoheadrightarrow M$.
The coefficient $V^{a}$  contains a constant and a quadratic term
in $\theta$, which respectively correspond to a section
$\mathring{V} \in \Gamma(E)$  with components~$\mathring{V}^{a}$,
and to the structure coefficients of the Lie bracket
on sections of $E$. 
We can now analyze order by order in $\theta$ the conditions
for supersymmetry in the interaction sector of the model.

\paragraph{Order $0$.}
At lowest order we get two equations: 
\begin{subequations}
\begin{align}
     \tfrac 12 t_{a}{}^{\rho}\partial_{\rho}\Pi^{\mu\nu}
    +\Pi^{\rho[\mu}\big(\partial_{\rho}t_{a}{}^{\nu]}
    -\Gamma_{\rho a}^{b}t_{b}{}^{\nu]}\big) & = 0\,,
    \label{eq: susy order 0a} \\[4pt] 
     \Pi^{\mu\nu}\nabla_{\mu}\mathring{V}^{a}
    +g^{ab}t_{b}{}^{\nu} & = 0\,.
    \label{eq: susy order 0b} 
\end{align}
\end{subequations}
To clarify their geometrical meaning, let us first recall
what these conditions amounted to in the two previous examples.
The second condition was trivial in both cases, since both $g$
and $\mathring{V}$ were zero. The first condition had the same 
geometric meaning for both the tangent and the cotangent case:
it said that the Poisson structure $\Pi$ is covariantly constant
with respect to the basic $E$-on-$TM$ connection in each case. 
Since we identify the map $t$ with the anchor of the Lie algebroid,
we can write the first condition in terms of the coefficients
of this connection given in \eqref{basic connection coefficients} as 
\begin{equation}
    \tfrac 12 t_{a}{}^{\rho}\partial_{\rho}\Pi^{\mu\nu}
    -\Pi^{\rho[\mu}\overline{\Gamma}^{\nu]}_{a\rho}=0\,.
\end{equation}
This is precisely the condition of covariant constancy
of the Poisson structure on the base with respect
to the basic $E$-on-$TM$ connection. Turning to the second condition, we observe that the right-hand side
of \eqref{eq: susy order 0b} is the composition of the map
$t: E\to TM$ and the metric $g$, which gives rise to the induced map 
$g^{\sharp}: E^{\ast}\to E$. The left-hand side
is the induced connection acting on the section $\mathring{V}$.
In total, the geometric form of the two lowest order conditions is
\begin{subequations}
    \begin{align}
    \overline\nabla^{\E}\Pi&=0\,, \\[4pt] 
    \mathbullet{\nabla}\mathring{V}&= t\circ g^{\sharp}\,.
\end{align}
\end{subequations}
Note that the covariant constancy of the Poisson bivector $\Pi$
with respect to the basic connection can be re-written as
\begin{equation}
    \overline{\nabla}^{\E}_{t(e)} \Pi
    \equiv \mathcal{L}_{t(e)}\Pi
    - t_{\wedge}(\mathbullet{\nabla} e) = 0\,,
    \qquad e \in \Gamma(E)\,,
\end{equation}
where
\begin{equation}
    \begin{tikzcd}
        t_{\wedge}: \Gamma(TM) \otimes \Gamma(E)
        \ar[r, "{- \otimes t}"]
        & \Gamma(TM) \otimes \Gamma(TM) \ar[r, "\wedge"]
        & \Gamma(\wedge^2 TM)
    \end{tikzcd}
\end{equation}
denotes the application of the anchor map
on the second factor of the tensor product, then antisymmetrization.
In components, it reads
\begin{equation}
    \mathcal{L}_{t_{a}} \Pi^{\mu\nu}
    = 2\,\Pi^{[\mu|\lambda} \Gamma_{\lambda a}^{b} t_{b}^{|\nu]}\,,
\end{equation}
i.e. the Lie derivative of the Poisson bivector $\Pi$
along the distribution defined by the anchor of $E$
does not vanish, but is instead encoded by the covariant derivative
of this distribution by the induced $T^{\ast}M$-on-$E$ connection.

\paragraph{Order $1$.}
There is only one condition at this order. It reads
\begin{equation}\label{eq: susy gauge order 1}
    \Pi^{\mu\nu}\Gamma^{(a}_{\nu c}\partial_{\mu}\mathring{V}^{b)}
    -\tfrac 12 t_{c}{}^{\mu}\partial_{\mu}g^{ab}
    -\tfrac 12 \mathring{V}^{d}\mc{P}_{dc}{}^{ab}
    -g^{d(a} C_{dc}{}^{b)}=0\,.
\end{equation}
It is trivially satisfied in the previous examples
where both $\mathring{V}$ and $g$ were zero.
Using the order $0$ result, we can write this condition
in a more transparent and covariant form, 
\begin{equation}
    \overline{\nabla}_{c} g^{ab}
        + R_{cd}{}^{ab}\,\mathring{V}^{d} = 0
        \qquad \iff \qquad
        \overline{\nabla}^{\E} g = \imath_{\mathring{V}} R\,,
\end{equation}
where $\imath_{\mathring{V}}$ denotes the interior product
with $\mathring{V}$.

\paragraph{Order $2$.}
Under the assumptions we made, there is also
only one condition at order $2$. This is:  
\begin{equation}
    \Pi^{\mu\nu}\,S_{ab\,\nu}{}^{c}
        - R_{ab}{}^{cd}\,t_{d}{^\mu} = 0\,,
\end{equation}
which relates the tensorial part $R_{ab}{}^{cd}$
of the quadratic piece of the Poisson structure
on~$E[0,1]$ to the basic curvature $S_{ab\mu}{}^{c}$.
When evaluated on a pair of sections $e, e' \in \Gamma(E)$,
a $1$-form $\eta \in \Omega^1(M)$ and a section
of the dual vector bundle $e^{\ast} \in \Gamma(E^{\ast})$,
the above identity reads
\begin{equation}\label{eq:relation_basic_R}
    \langle S(e,e') \Pi^{\sharp}(\eta), e^{\ast} \rangle 
        = \langle R(e,e'), t^{\ast}(\eta) \otimes e^{\ast} \rangle\,,
\end{equation}
where $\langle\cdot,\cdot\rangle$ denote the canonical pairing
between sections of $E$ and its dual $E^{\ast}$,
and $t^{\ast}: \Gamma(T^{\ast}M) \to \Gamma(E^{\ast})$
is the dual map of the anchor.
The fact that these two tensors are related
by contraction with the Poisson bivector $\Pi$
of the base manifold $M$ and the anchor~$\rho$
of the Lie algebroid $E$ makes it clear
that the previous two examples, $E=TM$ and $E=T^{\ast}M$, 
are singled out in this framework.
Indeed, for the tangent bundle, the anchor is the identity
so that this condition can be read as a \emph{definition}
of the tensor $R_{ab}{}^{cd}$. In the cotangent bundle case,
the anchor is the base Poisson bivector,
so that one can factor it and read the previous condition
as a definition of the tensor $R_{ab}{}^{cd}$ again,
\emph{up to terms in the kernel of $\Pi^{\sharp}$}.

There is one further possibility that allows for $R_{ab}{}^{cd}$ to be determined. If the anchor $t$ is \emph{invertible}, which means that
$E \cong TM$---the vector bundle $E$ is isomorphic
to the tangent bundle $TM$ as a vector bundle over $M$%
---then the previous identity can also be read as a definition
of the tensor $R$. Note that this implies that $E$ is isomorphic
to $TM$ as a \emph{Lie algebroid}, since the anchor is also
required to be a Lie algebra morphism between sections of $E$
and vector fields on the base.

As an example, one may consider any endomorphism
$J \in \Gamma\big({\rm End}(TM)\big)$
of the tangent bundle whose Nijenhuis tensor, 
\begin{equation}
    N_J : \Gamma(TM) \otimes \Gamma(TM)
        \longrightarrow \Gamma(TM)\,,
\end{equation}
defined on a pair of vector fields $X, Y \in \Gamma(TM)$ as
\begin{equation}
    N_J(X,Y) := -J^2[X,Y] + J\big([J(X),Y] + [X,J(Y)]\big)
    - [J(X),J(Y)]\,, 
\end{equation}
vanishes, and construct a Lie algebroid structure on $TM$,
by taking $J$ as its anchor, and 
\begin{equation}
    [X,Y]_J := [J(X),Y] + [X,J(Y)] - J\big([X,Y]\big)\,,
    \qquad X, Y \in \Gamma(TM)\,.
\end{equation}
for its Lie bracket \cite{Kosmann-Schwarzbach:1990}.
The latter obeys the Jacobi identity as a consequence
of the vanishing of the Nijenhuis tensor for $J$. 
Moreover, as long as $J$ is invertible,
then $\big(TM, J, [\cdot,\cdot]_J\big)$ is a Lie algebroid
with bijective anchor. Examples of such invertible endomorphisms
of particular interest are complex structures on $M$,
namely those endomorphisms $J$ which square to minus the identity, 
$J^2 = -\mathbf{1}_{TM}$, and whose Nijenhuis tensor vanishes.%
\footnote{Note that generalized complex structures,
introduced by Hitchin \cite{Hitchin:2003}, are at the heart
of a number of two-dimensional Poisson sigma models,
see for instance \cite{Lindstrom:2004iw, Zucchini:2004ta, Lindstrom:2004hi, Zucchini:2005rh, Pestun:2006rj, Zucchini:2007ie} and references therein.}

More generally, since the condition \eqref{eq:relation_basic_R}
relates the evaluation of the tensor $R$ on sections of
${\rm Im}(t^{\ast}) \otimes E^{\ast}$ to the composition
of the basic curvature and the sharp-morphism, this identity
becomes sufficient to determine $R$ if the dual anchor $t^{\ast}$
is surjective, or equivalently if the anchor $t$ is injective.
Such Lie algebroids amount to being given an \emph{involutive}
distribution on $M$, corresponding to the image of the anchor.

Action Lie algebroids provide examples where the anchor
can be injective, or even bijective, depending on the setup.
Suppose that a Lie algebra $\mathfrak{g}$ acts on a manifold $M$,
that is, there exists a Lie algebra morphism
$\mathfrak{g} \longrightarrow \Gamma(TM)$ sending elements
of $\mathfrak{g}$ to vector fields on $M$. The trivial bundle
$M \times \mathfrak{g}$ is then endowed with a Lie algebroid structure with anchor
\begin{equation}
    t: \Gamma(M \times \mathfrak{g})
        \cong \Functions(M) \otimes \mathfrak{g}
            \longrightarrow \Gamma(TM)\,,
\end{equation}
being simply the $\Functions(M)$-linear extension
of the Lie algebra action, and bracket
\begin{equation}
    [f \otimes \xi, g \otimes \xi']
        = f\,g \otimes [\xi, \xi']_{\mathfrak{g}}
        + f\,\big(t(\xi)g\big) \otimes \xi'
        - g\,\big(t(\xi')f\big) \otimes \xi\,,
\end{equation}
for $f, g \in \Functions(M)$ and $\xi,\xi' \in \mathfrak{g}$.
The action of $\mathfrak{g}$ on $M$ being (locally) free
is equivalent to the anchor $t$ being injective.
Put differently, at every point of $M$, the representation
of $\mathfrak{g}$ is \emph{faithful} (i.e. no element
is represented trivially). Another, closely related source
of examples are gauge Lie algebroids
(see e.g. \cite{Bekaert:2023jvl} for more details
concerning such Lie algebroids).

\paragraph{Order $3$.}
Finally, there exists a single order $3$ consistency condition
that must be satisfied for the theory to have supersymmetric invariance,
which reads
\begin{equation}
    t_{[a|}{}^{\mu}\,\partial_{\mu} R_{|bc]}{}^{de}
    + C_{[ab}{}^{\times}\,R_{c]\times}{}^{de}
    + 2 R_{[ab}{}^{\times(d}\,\overline{\Gamma}_{c]\times}^{e)}
    = 0\,,
\end{equation}
which is nothing but the component form
of the `Bianchi-like' identity
\begin{equation}
    \dd_{\overline{\nabla}^{\E}} R = 0\,.
\end{equation}
Put differently, the tensorial part of $\mc P^{ab}$
that is $R \in \Gamma(\wedge^2 E^{\ast} \otimes S^2 E)$,
which can be thought of as a cochain in the complex
$\Omega(E,SE)$ of forms on $E$ valued in its symmetric algebra,
is annihilated by the differential $\dd_{\overline{\nabla}^{\E}}$
associated with the basic connection.
The resemblance with the Bianchi identity
for the basic curvature is again noticeable,
although it is important to keep in mind 
that the tensor $R$ does not take values 
in the same bundle as the basic curvature.

\paragraph{Closure of the supersymmetry.}
Apart from the invariance of the action under supersymmetry 
transformations, we would like these transformations to close
into an algebra. For this reason, we impose in addition
that the vector field $\cQ_{S}$ is homological.
Under the assumptions of the present section,%
\footnote{Assumptions that, in particular, include 
that $E$ be a Lie algebroid.}
we obtain an additional set of conditions
on the undetermined coefficients. At order $0$ we obtain 
\begin{equation}
    t_{a}{}^{\mu}\mathring{V}^{a}=0
    \qquad \Longleftrightarrow \qquad
    t(\mathring{V})=0\,,
\end{equation}
i.e. that $\mathring{V}$ is an $E$-section in the kernel
of the anchor map. On top of that, a last condition
appears at order $1$, 
\begin{equation}
    t_{b}{}^{\mu} \partial_{\mu} \mathring{V}^{a}
    + C_{bc}{}^{a} \mathring{V}^{c} = 0\,,
\end{equation}
which, upon using the previous condition, can be re-written as
\begin{equation}
    t_{b}{}^{\mu} \partial_{\mu} \mathring{V}^{a}
    + \big(C_{bc}{}^{a} + t_{c}{}^{\mu} \Gamma_{\mu b}{}^{a}\big) 
    \mathring{V}^{c} \equiv \overline{\nabla}_{b} \mathring{V}^{a} = 0\,,
\end{equation}
i.e. the section $\mathring{V} \in \Gamma(E)$ is covariantly
constant with respect to the basic connection,
\begin{equation}
    \overline{\nabla}^{\E} \mathring{V} = 0\,.
\end{equation}
Note that, under the change of coordinates
$(x^{\mu},\theta^{a},a_{\mu},\chi_{a})
\to (x^{\mu},\theta^{a},a^{\scriptscriptstyle{\nabla}}_{\mu},\chi_{a})$
where, as before
\begin{equation}
    a^{\scriptscriptstyle{\nabla}}_{\mu}
    = a_{\mu} + \Gamma_{\mu a}^{b}\,\theta^{a} \chi_{b}\,,
\end{equation}
the homological vector field $\cQ_{S}$ takes the form
\begin{equation}
    \cQ_{S} = \cQ_{S}^{(-1)} + \cQ_{S}^{(0)}
    + \cQ_{S}^{(+1)} + \cQ_{S}^{(+2)}\,,
\end{equation}
with
\begin{subequations}
\begin{align}
    \cQ_{S}^{(-1)} & = \mathring{V}^{a}\,
        \tfrac{\partial}{\partial \theta^{a}}\,, \\[4pt]
    \cQ_{S}^{(0)} & = \nabla_{\mu}\mathring{V}^{a}\,\chi_{a}\,
    \tfrac{\partial}{\partial a_{\mu}^{\scriptscriptstyle{\nabla}}}
    - a^{\scriptscriptstyle{\nabla}}_{\mu}\,t_{a}{}^{\mu}\,
        \tfrac{\partial}{\partial \chi_{a}}\,, \\[4pt]
    \cQ_{S}^{(1)} & = \theta^{a}\,t_{a}{}^{\mu}\,
        \tfrac{\partial}{\partial x^\mu}
        -\tfrac12\,\theta^{a} \theta^{b}\,C_{ab}{}^{c}\,
        \tfrac{\partial}{\partial \theta^{c}}
        + \theta^{a}\,\overline{\Gamma}_{ab}^{c}\,\chi_{c}\,
        \tfrac{\partial}{\partial \chi_{b}}
        + \theta^{a}\,\overline{\Gamma}_{a\mu}^{\nu}(x)\,
        a^{\scriptscriptstyle{\nabla}}_{\nu}\,
        \tfrac{\partial}{\partial a^{\scriptscriptstyle{\nabla}}_{\mu}} \\[4pt]
        \cQ_{S}^{(2)} & = -\tfrac12\,\theta^{a} \theta^{b}\,
        \chi_{c}\,S_{ab\,\mu}{}^{c}\,
        \tfrac{\partial}{\partial a^{\scriptscriptstyle{\nabla}}_{\mu}}\,.
\end{align}
\end{subequations}
We can recognize, in the above expression, the homological
vector field associated with the coadjoint representation
up to homotopy of the Lie algebroid $E$ spelled out
in \eqref{Qs coruth} previously, together with 
the interior product with $\mathring{V}$,
as the piece $\cQ_{S}^{(-1)}$, and contraction
with the covariant derivative of $\mathring{V}$
as part of the piece $\cQ_{S}^{(0)}$.

To summarize, subject to the above conditions on the geometrical data we have obtained a class of supersymmetric
Poisson sigma models based on a Lie algebroid $E$
with ``anchor supersymmetry'', meaning that the supersymmetry 
transformations are controlled by the anchor of the Lie algebroid
and the structure constants of its Lie bracket, together with the section $\mathring{V}$. This includes the case when the anchor vanishes altogether, as for example in the case of the Lie algebroid that corresponds to a bundle of Lie algebras. A simple yet nontrivial model where this is the case would be to consider a degree $-2$ Poisson structure, in other words by keeping only the metric $g$ in the super-Poisson bivector. This leads to the simple action
\begin{equation}
    S[X,A,\theta,\chi] = \int A_{\mu} \wedge \dd X^{\mu}
    + \chi_{a} \wedge \dd \theta^{a} 
    + \tfrac12 g^{ab}(X)\,\chi_{a} \wedge \chi_{b}\,.
\end{equation}
If we consider in addition that $g$ is nondegenerate, then the compatibility conditions of the supersymmetry with the gauge symmetry result in the supersymmetry transformations
\begin{subequations}
\begin{align}
    \delta_{S} X^{\mu} & = 0\,, \\[4pt]
    \delta_{S} \theta^{a} & = \mathring{V}^{a}(X)
    - \tfrac12 C_{bc}{}^{a}\,\theta^{b} \theta^{c}\,, \\[4pt]
    \delta_{S} A_{\mu} & = \partial_{\mu}\mathring{V}^{a}(X)\,\chi_{a}\,, \\[4pt]
    \delta_{S} \chi_{a} & = - C_{ab}{}^{c}\,\theta^{b} \chi_{c}\,,
\end{align}
\end{subequations}
for arbitrary $\mathring{V}\in\Gamma(X^*E)$, where $C_{ab}{}^{c}$ are the structure constants of the pointwise bracket from the fiber Lie algebras, and $g$ is an ${\rm ad}^{\ast}$-invariant fiberwise metric as a consequence of~\eqref{eq: susy gauge order 1}.

\section{Conclusions and outlook}
\label{sec:conclusion}
In this paper, we revisited the supersymmetric version
of the Poisson sigma model, originally proposed
in \cite{Ikeda:1993fh}, from the point of view of
NQ \emph{supermanifolds}. The latter being equipped
with two gradings, over $\Z$ and over $\Z_2$, two kinds of homological vector fields can be considered, which we identify
as controlling the gauge symmetry and the supersymmetry of the model, respectively. Apart from being homological, gauge and supersymmetry invariance of the kinetic sector requires compatibility with the graded symplectic structure of the target space. Invariance of the interaction sector under supersymmetry transformations imposes the additional condition that the two homological vector fields are graded-commuting. We have analysed this set of conditions in detail throughout the paper and explained the geometric data they entail. Our main conclusions are summarized as follows: 
\begin{itemize}
    \item There exists a distinguished class of supersymmetric Poisson sigma models, where all the above conditions are met, for supermanifolds originating from Lie algebroids. The class comprises (i) the differential Poisson sigma model described in Ref. \cite{Arias:2015wha}, based on the canonical tangent Lie algebroid, (Section \ref{sec: ABSTG}) (ii) a set of models with invertible anchor map (Section \ref{sec:sPSM_algd}), and (iii) a new model, the contravariant supersymmetric Poisson sigma model, based on the cotangent Lie algebroid over the body of the Poisson supermanifold (Section \ref{sec:contravariant PSM}).  
    \item The contravariant model is itself distinguished within this class as the single case where the anchor map is not necessarily invertible, instead being identified with the map induced by the Poisson structure on the body of the supermanifold, which is itself part of the full Poisson structure on the total supermanifold. Since this controls the supersymmetry transformations, this model exhibits a property we call Poisson supersymmetry, which is generically nonlinear.
    \item The common characteristic of all these models is their underlying mathematical structure. The graded-commutativity of the two homological vector fields implies that the Poisson structure on the body of the supermanifold is covariantly constant with respect to the basic $E$-connection on the chain complex $E\overset{\rho}\to TM$. The quartic term in the fermions has a coefficient directly related to the basic curvature tensor of an ordinary connection on the Lie algebroid. This led us to the conclusion that in all cases the supersymmetry-generating vector field is precisely identified with the coadjoint representation up to homotopy of the associated Lie algebroid.    
\end{itemize}

\paragraph{An outlook towards quantization.}
The path integral quantization of the supersymmetric Poisson sigma model considered in \cite{Arias:2015wha}, and of the new example
presented in Section~\ref{sec:contravariant PSM},
would be particularly interesting to understand, 
as one may expect them to produce a deformation quantization 
of the algebra of differential forms or polyvectors, respectively.
This possibility has intriguing implications, starting with
the fact that there seem to be several ways of obtaining
a deformation quantization of such structures. For the sake
of definiteness, let us focus on the case of differential forms.
They appear as functions on the parity-shifted tangent bundle,
which can be made into a Poisson supermanifold,
as explained in \cite{Arias:2015wha} and reviewed
in Section \ref{sec: ABSTG}. The super-Poisson structure
\eqref{eq:Poisson_tangent} is however not the most general
one can consider on $\Pi TM$, as there is in principle 
no reason to discard terms of higher order in the fermionic
coordinates---we made this assumption merely to simplify
our analysis. This begs the question: what is the dependency
of the quantization of $\Omega(M)$ on the a priori different
super-Poisson structures it admits? In the case of ordinary
Poisson manifold, the star-product obtained by deformation
quantization only depends on the equivalence class of the Poisson
bivector under (formal) diffeomorphisms \cite{Kontsevich:1997vb}.
In fact, this extends to the deformation quantization
of \emph{homotopy Poisson manifolds} obtained as a corollary
of the \emph{relative formality} theorem proved by
Cattaneo and Felder \cite{Cattaneo:2003dp, Cattaneo:2005zz}
(see also \cite{Cattaneo:2007er} for a review),
and from a different perspective, by Lyakhovich and Sharapov
\cite{Lyakhovich:2004xd}.

Recall that homotopy Poisson manifolds are $\Z$-graded
manifolds $\mc M$ equipped with a family of polyvectors
$\{\Pi_{r}\}$ of rank $r\geq1$ and of degree $2-r$
(see e.g. \cite{Schatz:2009, Bandiera:2017}),
which are in involution, meaning their Schouten--Nijenhuis
bracket vanish,
\[
    \sum_{i+j=r}[\Pi_i, \Pi_j]_{\text{SN}} = 0\,,
    \qquad 
    r \geq 1\,.
\]
A Poisson $\Z$-graded manifold is an example of homotopy
Poisson manifold, with the family of polyvectors
reducing to a single Poisson bivector. A slightly less trivial
example would be that of a \emph{dg-Poisson manifold},
which is a special case with homotopy Poisson structure
consisting of an homological vector field $Q$,
and a $Q$-invariant Poisson bivector $\Pi$, as the involution
condition reads
\[
    Q^2 = 0\,,
    \qquad 
    {\cal L}_{Q}\Pi = 0\,,
    \qquad 
    [\Pi,\Pi]_{\text{SN}} = 0\,.
\]
This is exactly the type of structure that we considered
on $\mc M \cong E[0,1]$, where $Q$ is the restriction
of $\cQ_{S}$ to $\mc M$ (more specifically the pushforward
of $\cQ_{S}$ by the projection
$T^{\ast}[1,0]\mc M \twoheadrightarrow \mc M$),
and $\Pi$ the super-Poisson bivector $\mc P$.
This means that the parity-shifted tangent bundle can,
in fact, also be considered as a homotopy Poisson manifold,
which in particular is a $\Z$-graded manifold.
In this case, the restriction of the Poisson bracket
having the form \eqref{eq:Poisson_tangent} becomes justified
as it is the most general form allowed by the requirement
that it be of $\Z$-degree $0$.

Homotopy Poisson structures on $\mc M$ are in bijection
with self-commuting functions of degree $2$
on $T^{\ast}[1]\mc M$ \cite{Voronov:2001qf, Mehta:2010}
(in complete analogy with the result of \cite{Roytenberg:2002nu}).
From the AKSZ perspective, this function which contains
all polyvectors of the homotopy Poisson structure,
should be used as the interaction term of the sigma model.
This suggests that, in the dg-Poisson case,
the homological vector field should be part 
of the data to be quantized. More concretely, this would mean
that for differential forms, not only the wedge product
is deformed into a star-product, but also the de Rham differential
is deformed into a new operator, subject to compatibility 
conditions with the star-product (in general forming a possibly
curved $A_\infty$-algebra). Concerning the models discussed
in this paper, this would amount to adding the Hamiltonian
function for $\cQ_{S}$ to the action, but this does not
seem possible, due to the fact that the latter is parity odd.

To summarize, there seems to be several ways of quantizing
differential forms, depending on what type of structure
is used to encoded them, be it a Poisson supermanifold,
with Poisson structure possibly invariant under a parity-odd
homological vector field, or a homotopy Poisson manifolds.
The path integral quantization of the corresponding
sigma models may shed light on these possible differences
and similarities.

\paragraph{An outlook towards higher dimensions.}
As mentioned in the introduction, the Poisson sigma model
is \emph{the} AKSZ sigma model in two dimensions,
while in three dimensions, the generic AKSZ-type model
is the Courant sigma model
\cite{Ikeda:2002wh, Roytenberg:2006qz}. 
As suggested by the name, it is completely characterized
by the data of a Courant algebroid. Although little is known 
about its quantization (with the notable exception
of the works of Hofman and Park \cite{Hofman:2002jz, Hofman:2002rv})
the Courant sigma model has been extensively studied in recent years
at the classical level (see e.g. \cite{Cattaneo:2009zx, Bessho:2015tkk, Severa:2016prq, Mylonas:2012pg, Chatzistavrakidis:2018ztm, Chatzistavrakidis:2023lwo}).
A first step towards its supersymmetrization,
which would couple nontrivially super Chern--Simons
and super-BF theories, was taken in \cite{Salnikov:2016bny}.
It would be interesting to analyze the construction
of supersymmetric Courant sigma models in detail and identify
the higher-dimensional analogon of the differential
Poisson sigma model and its contravariant cousin.  
A related question is whether the fact that the coadjoint 
representation of a Lie algebroid appears as
a supersymmetry-generating vector field in two dimensions 
generalizes somehow to higher dimensions. In this respect,
note that representations up to homotopy for the split case
of Lie 2- and $n$-algebroids were studied in Ref. \cite{Jotz:2020}, 
whereas the concept of basic curvature tensor for connections
on Courant algebroids and its relation to the BV/BRST formulation 
of \emph{bosonic} Courant sigma models was investigated in Refs. 
\cite{Chatzistavrakidis:2023otk,Chatzistavrakidis:2023lwo}.

\section*{Acknowledgments}
We are grateful to Nicolas Boulanger, Noriaki Ikeda, Lara Jonke, Kevin Morand, Jan Rosseel, Dima Roytenberg, Per Sundell and Ping Xu for insightful discussions. We also thank Dima Roytenberg for comments on the manuscript. This work is funded by the Croatian Science Foundation project ``Higher Structures and Symmetries in Gauge and Gravity Theories'' (IP-2024-05-7921) and by the European Union — NextGenerationEU. The work of Ath. Ch. is also supported by the Ulam Programme of the Polish National Agency for Academic Exchange. 
The work of T.B. was supported by the European Union’s
Horizon 2020 research and innovation programme
under the Marie Sk{\l}odowska Curie grant agreement No 101034383,
and the European Research Council grant agreement No 101002551.
T.B. would like to thank the Rudjer Bo\v{s}kovi\'c Institute
for hospitality and support through an MZ2-24 Mobility of Researchers grant.

\appendix 
\section{General conditions}
\label{appa}

In this appendix we spell out the expanded form
of the conditions that appear in the main body of the paper.
In the first three subsections of the appendix,
we work on a general shifted super-vector bundle
$\mc V[1,0] \twoheadrightarrow \mc M$,
over the supermanifold $\mc M=E[0,1]$, with coordinates 
$(x^{\mu},\theta^{a},a^{m},\chi^{I})$ of $\Z \times\Z_2$ degrees 
$(0,0), (0,1), (1,0)$ and $(1,1)$ respectively,
and index ranges as explained in Section \ref{sec:Q-smfd}.
We report the conditions that are encoded in $[\cQ,\cQ]=0$,
$[\cQ_{S},\cQ_{S}]=0$ and $[\cQ,\cQ_{S}]=0$ in that order. 
In the last two subsections, we specialize
to $T^{\ast}[1,0]E[0,1]$, which has a canonical
graded symplectic structure and report the conditions
encoded in the compatibility of $\cQ$ and $\cQ_{S}$ with it. 
Note that all coefficients shown are functions
of $x^{\mu}$ and $\theta^{a}$, specifically quadratic
in the latter, and the derivatives~$\partial_{\mu}$
and $\partial_{a}$ are, respectively,
with respect to $x^{\mu}$ and $\theta^{a}$.

\subsection{Homological vector field for  gauge symmetries}
\label{app:Qgauge}
The general vector field $\cQ$ of $\Z\times \Z_2$ degree $(1,0)$ 
on the graded super-vector bundle $\mc V[1,0]$
has the following components in terms of yet undetermined 
functions of $x$ and $\theta^2$: 
\begin{subequations}
\begin{align}
    \cQ^\mu & = a^m\,\rho_m{}^\mu
    + \chi^I \theta^a\,\rho_{I\,a}{}^\mu\,, \\[4pt]
    \cQ^a & = a^m \theta^b\,\rho_{m\,b}{}^a
    + \chi^I\,\rho_I{}^a\, \\[4pt]
    \cQ^p & = -\tfrac12\,a^m\,a^n\,f_{mn}{}^p
    - a^m \chi^I \theta^a\,f_{m\,I\,a}{}^p
    -\tfrac12\,\chi^I \chi^J\,f_{IJ}{}^p\,, \\[4pt]
    \cQ^I & = -\tfrac12\,a^m a^n \theta^a\,
    f_{mn\,a}{}^I
    - a^m \chi^J\,f_{m\,J}{}^I
    -\tfrac12\,\chi^J \chi^K \theta^a\,f_{JK\,a}{}^I\,,
\end{align}
\end{subequations}
Demanding that this vector field is homological
yields fourteen independent conditions. For completeness,
we report them below, even though they are not so illuminating 
without any further assumptions, such as restricting
to the case of $\mc V = T^{\ast} \mc M$ as we eventually assume
in the text. From $\cQ^2x^{\mu}=0$ we obtain three conditions:
\begin{subequations}
\begin{align}
    & \rho_{[m}{}^{\nu}\partial_{\nu}\rho_{n]}{}^{\mu}
    + \theta^{a}\rho_{[m|a}{}^{b}\partial_{b}\rho_{|n]}{}^{\mu}
    - \tfrac12\,f_{mn}{}^{p}\rho_{p}{}^{\mu}
    - \tfrac12\,\theta^{a}\theta^{b}
        f_{mn[a|}{}^{I} \rho_{I|b]}{}^{\mu}=0\,,\\[4pt]
    & \rho_{I}{}^{a}\partial_{a}\rho_{m}{}^{\mu}
    - \theta^{a}\big(\rho_{m}{}^{\nu}\partial_{\nu}\rho_{Ia}{}^{\mu}
    + \rho_{ma}{}^{b}\rho_{Ib}{}^{\mu}
    - \rho_{Ia}{}^{\nu}\partial_{\nu}\rho_{m}{}^{\mu}
    - f_{mIa}{}^{p}\rho_{p}{}^{\mu} - f_{mI}{}^{J}\rho_{Ja}{}^{\mu}\big) \\[4pt]
    & \hspace{280pt} + \theta^{a}\theta^{b} \rho_{ma}{}^{c}
        \partial_{c}\rho_{Ib}{}^{\mu}=0\,, \nonumber\\[4pt] 
    & \rho_{(I}{}^{c}\rho_{J)c}{}^{\mu}
        - \tfrac12\,f_{IJ}{}^{m}\rho_{m}{}^{\mu}
    - \theta^{a}\rho_{(I}{}^{b}\partial_{b}\rho_{J)a}{}^{\mu}
    + \theta^{a}\theta^{b}
        \big(\rho_{(I|a}{}^{\nu}\partial_{\nu}\rho_{|J)b}{}^{\mu}
        -\tfrac12\,f_{IJa}{}^{K}\rho_{Kb}{}^{\mu}\big) = 0\,,
\end{align}
\end{subequations}
where (anti)symmetrizations are with weight $1$,
and they refer only to the two surrounded indices
in the present set of equations. Next, from $\cQ^2\theta^{a}=0$,
we obtain three additional conditions:
\begin{subequations}
\begin{align}
    & \theta^b\big(\rho_{[m}{}^{\mu}\partial_{\mu}\rho_{n]b}{}^{a}
    + \rho_{[m|b}{}^{c}\rho_{|n]c}{}^{a}
    - \tfrac12\,f_{mn}{}^{p}\rho_{pb}{}^{a}
    - \tfrac12\,f_{mnb}{}^{I}\rho_{I}{}^{a}\big)
    - \theta^{b}\theta^{c}
        \rho_{[m|b}{}^{d}\partial_{d}\rho_{|n]c}{}^{a} = 0\,,\\[4pt] 
    & \rho_{m}{}^{\mu}\partial_{\mu}\rho_{I}{}^{a}
    - \rho_{I}{}^{b}\rho_{mb}{}^{a} - f_{mI}{}^{J}\rho_{J}{}^{a}
    + \theta^{b}\big(\rho_{I}{}^{c}\partial_{c}\rho_{mb}{}^{a}
    + \rho_{mb}{}^{c}\partial_{c}\rho_{I}{}^{a}\big) \\[4pt] 
    & \hspace{200pt} -\,\theta^{b}\theta^{c}
        \big(\rho_{Ib}{}^{\mu}\partial_{\mu}\rho_{mc}{}^{a}
        + f_{mIb}{}^{p}\rho_{pc}{}^{a}\big) = 0\,, \nonumber \\[4pt]
    & \rho_{(J}{}^{b}\partial_{b}\rho_{I)}{}^{a}
    + \theta^{b}\big(\rho_{(I|b}{}^{\mu}\partial_{\mu}\rho_{|J)}{}^{a}
    -\tfrac12 f_{IJ}{}^{m}\rho_{mb}{}^{a}
    -\tfrac12 f_{IJb}{}^{K}\rho_{K}{}^{a}\big) = 0\,.
\end{align}
\end{subequations}
The next set of independent conditions is obtained from $\cQ^2a^{m}=0$ and it comprises four equations as follows:
\begin{subequations}
\begin{align}
    & \rho_{[n}{}^{\mu}\partial_{\mu}f_{pq]}{}^{m}
    + f_{[np}{}^{r}f_{q]r}{}^{m} 
    + \theta^{a} \rho_{[n|a}{}^{b}\partial_{b}f_{|pq]}{}^{m}
    + \theta^{a}\theta^{b} f_{[np|a}{}^{I} f_{|q]Ib}{}^{m} = 0\,, \\[4pt]
    & \rho_{I}{}^{a}\partial_{a}f_{np}{}^{m}
    + \theta^{a}\big(2\rho_{[n}{}^{\mu}\partial_{\mu}f_{p]Ia}{}^{m}
    + 2\rho_{[n|a}{}^{b} f_{|p]Ib}{}^{m}
    + \rho_{Ia}{}^{\mu}\partial_{\mu}f_{np}{}^{m}
    - 2f_{[n|Ia}{}^{q}f_{p]q}{}^{m} \\[4pt]
    & \hspace{80pt} - 2f_{[n|I}{}^{J}f_{|p]Ja}{}^{m}
    - f_{np}{}^{q}f_{qIa}{}^{m} - f_{npa}{}^{J}f_{IJ}{}^{m}\big)
    - 2\theta^{a}\theta^{b}
    \rho_{[n|a}{}^{c}\partial_{c}f_{|p]Ib}{}^{m} = 0\,, \nonumber \\[4pt]
    & 2\rho_{(I|}{}^{a} f_{n|J)a}{}^{m}
    - \rho_{n}{}^{\mu}\partial_{\mu}f_{IJ}{}^{m}
    + 2f_{n(I}{}^{K}f_{J)K}{}^{m} - f_{IJ}{}^{p}f_{np}{}^{m} \\[4pt]
    & \hspace{100pt} \nonumber
    - \theta^{a}\big(\rho_{na}{}^{b}\partial_{b}f_{IJ}{}^{m}
    + 2\rho_{(I|}{}^{b}\partial_{b}f_{n|J)a}{}^{m}\big) \\[4pt]
    & \hspace{120pt} \nonumber + 2\theta^{a}\theta^{b}
    \big(\rho_{(I|a}{}^{\mu}\partial_{\mu}f_{n|J)b}{}^{m}
    + f_{n(I|a}{}^{p}f_{p|J)b}{}^{m}
    - \tfrac12\,f_{IJa}{}^{K}f_{nKb}{}^{m}\big) = 0\,, \\[4pt]
    & \rho_{(I}{}^{a}\partial_{a}f_{JK)}{}^{m}
    + \theta^{a}\big(\rho_{(I|a}{}^{\mu}\partial_{\mu}f_{|JK)}{}^{m}
    - f_{(IJ|}{}^{n}f_{n|K)a}{}^{m}
    - f_{(IJ|a}{}^{L}f_{|K)L}{}^{m}\big) = 0\,,
\end{align}
\end{subequations}
Finally, a last set of four conditions stems from $\cQ^2\chi^{I}=0$,
and reads
\begin{subequations}
\begin{align}
    & \theta^{a}\big(\rho_{[m}{}^{\mu}\partial_{\mu}f_{np]a}{}^{I}
    + \rho_{[m|a}{}^{b}f_{|np]b}{}^{I} + f_{[mn}{}^{q}f_{p]qa}{}^{I}
    + f_{[mn|a}{}^{J}f_{|p]J}{}^{I}\big) - \theta^{a}\theta^{b}
    \rho_{[m|a}{}^{c}\partial_{c}f_{|np]b}{}^{I} = 0\,, \\[4pt]
    & 2\rho_{[m}{}^{\mu}\partial_{\mu}f_{n]J}{}^{I}
    + \rho_{J}{}^{a}f_{mna}{}^{I} - 2f_{[m|J}{}^{K}f_{|n]K}{}^{I}
    - f_{mn}{}^{p}f_{pJ}{}^{I} \\[4pt] & \hspace{100pt} \nonumber
    + \theta^{a}\big(2\rho_{[m|a}{}^{b}\partial_{b}f_{|n]J}{}^{I}
    - \rho_{J}{}^{b}\partial_{b}f_{mna}{}^{I}\big) \\[4pt]
    & \hspace{120pt} \nonumber + \theta^{a}\theta^{b}
    \big(\rho_{Ja}{}^{\mu}\partial_{\mu}f_{mnb}{}^{I}
    - 2f_{[m|Ja}{}^{p}f_{|n]pb}{}^{I}
    - f_{mna}{}^{K}f_{JKb}{}^{I}\big) = 0\,, \\[4pt]
    & 2\rho_{(J|}{}^{a}\partial_{a}f_{m|K)}{}^{I}
    + \theta^{a}\big(2\rho_{(J|a}{}^{\mu}\partial_{\mu}f_{m|K)}{}^{I}
    - \rho_{m}{}^{\mu}\partial_{\mu}f_{JKa}{}^{I}
    - \rho_{ma}{}^{b}f_{JKb}{}^{I}
    + 2f_{m(J|a}{}^{n}f_{n|K)}{}^{I} \\[4pt]
    & \hspace{100pt} + 2f_{m(J}{}^{L}f_{K)La}{}^{I}
    - f_{JK}{}^{n} f_{mna}{}^{I} - f_{JKa}{}^{L}f_{mL}{}^{I}\big)
    + \theta^{a}\theta^{b}\rho_{ma}{}^{c}\partial_{c}f_{JKb}{}^{I}
    = 0\,, \nonumber \\[4pt]
    & \rho_{(J}{}^{a}f_{KL)a}{}^{I} - f_{(JK|}{}^{m}f_{m|L)}{}^{I}
    - \theta^{a}\rho_{(J|}{}^{b}\partial_{b}f_{|KL)a}{}^{I} \\[4pt]
    & \hspace{120pt} + \theta^{a}\theta^{b}
    \big(\rho_{(J|a}{}^{\mu}\partial_{\mu}f_{|KL)b}{}^{I}
    - f_{(JK|a}{}^{M}f_{|L)Mb}{}^I\big) = 0\,. \nonumber  
\end{align}
\end{subequations}

\subsection{Homological vector field for supersymmetry}
\label{app:Q_susy}
We recall from the main text that the general vector field $\cQ_{S}$
of degree $(0,1)$ in the chosen coordinate system has components 
\begin{equation}  
    \cQ_{S}^{\mu} = \theta^{a} t_{a}{}^{\mu}\,,
    \quad 
    \cQ_{S}^{a} = V^{a}\,,
    \quad 
    \cQ_{S}^{m} = a^{n}\theta^{a} U_{na}{}^{m} +\chi^{I}W_{I}{}^{m}\,,
    \quad 
    \cQ_{S}^{I} = a^{m} Y_{m}{}^{I} + \chi^{J}\theta^{a} Z_{Ja}{}^{I}\,, 
\end{equation}  
with all undetermined coefficients being functions of $x$
and $\theta^2$. Requiring that this vector field is homological
gives the following six conditions, each of them organised
in increasing order of $\theta$: 
\begin{subequations}
\begin{align}
    & V^{a} t_{a}{}^{\mu} - \theta^{a}V^{b}\partial_{b}t_{a}{}^{\mu}
    - \theta^{a}\theta^{b}
        t_{b}{}^{\nu}\partial_{\nu}t_{a}{}^{\mu} = 0\,, \\[4pt] 
    & V^{b}\partial_{b}V^{a}
        + \theta^{b} t_{b}{}^{\mu}\partial_{\mu}V^{a} = 0\,,\\[4pt]
    & Y_{n}{}^{I}W_{I}{}^{m} - V^{a}U_{na}{}^{m}
    + \theta^{a} V^{b} \partial_{b} U_{na}{}^{m}
    - \theta^{a}\theta^{b}\big(U_{nb}{}^{l}U_{la}{}^{m}
    -t_{b}{}^{\lambda}\partial_{\lambda}U_{na}{}^{m}\big) = 0\,, \\[4pt]
    & V^{a}\partial_{a} W_{I}{}^{m}
    + \theta^{a}\big(W_{I}{}^{n}U_{na}{}^{m} + Z_{Ia}{}^{J}W_{J}{}^{m}
    + t_{a}{}^{\mu}\partial_{\mu} W_{I}{}^{m}\big) = 0\,, \\[4pt]
    & V^{a}\partial_{a}Y_{m}{}^{I}
        - \theta^{a}\big(U_{ma}{}^{n}Y_{n}{}^{I}
        - t_{a}{}^{\mu}\partial_{\mu}Y_{m}{}^{I}
        + Y_{m}{}^{J}Z_{Ja}{}^{I}\big) = 0\,, \\[4pt] 
    & V^{a} Z_{Ja}{}^{I} + W_{J}{}^{m}Y_{m}{}^{I}
    - \theta^{a} V^{b} \partial_{b} Z_{Ja}{}^{I}
    - \theta^{a}\theta^{b}\big(Z_{Jb}{}^{K} Z_{Ka}{}^{I}
    + t_{b}{}^{\mu}\partial_{\mu} Z_{Ja}{}^{I}\big) = 0\,.
\end{align}
\end{subequations}

\subsection{Commutator of gauge symmetry and supersymmetry}
\label{app:QQ}
Next we examine under what conditions the two homological vector fields
$\cQ$ and $\cQ_{S}$ commute. Due to their grading this means that
we should study the equation
\begin{equation}
    \cQ\cQ_{S} + \cQ_{S}\cQ = 0\,,
\end{equation}
which amounts to
\begin{subequations}
\begin{align}
    0 & = \rho_{A}{}^{\beta}\partial_{\beta} V^{\alpha}
    + (-1)^{|A|}\,V^{\beta}\partial_{\beta}\rho_{A}{}^{\alpha}
    + U_{A}{}^{B}\rho_{B}{}^{\alpha}\,, \\[4pt]
    0 & = (-1)^{|C|(|B|+1)}\,
    \big(\rho_{B}{}^{\alpha}\partial_{\alpha} U_{C}{}^{A}
    -U_{B}{}^{D}f_{CD}{}^{A}\big)
    + (-1)^{|B|}\,
    \big(\rho_{C}{}^{\alpha}\partial_{\alpha} U_{B}{}^{A}
    -U_{C}{}^{D}f_{BD}{}^{A}\big) \\[4pt]
    & \hspace{50pt} \nonumber - f_{BC}{}^{D}U_{D}{}^{A}
    - (-1)^{|B|+|C|}\,V^{\alpha}\partial_{\alpha}f_{BC}{}^{A}\,,
\end{align}
\end{subequations}
in the condensed notation of Section \ref{sec:Q-smfd}.
This gives a set of ten conditions, which we organize
as follows. From the action on $x^{\mu}$, we obtain
\begin{subequations}
\begin{align}
    & V^{a}\partial_{a}\rho_{m}{}^{\mu}
    - \theta^{b}\big(\rho_{mb}{}^{a} t_{a}{}^{\mu}
    + \rho_{m}{}^{\nu}\partial_{\nu} t_b{}^{\mu}
    + U_{mb}{}^{n}\rho_{n}{}^{\mu}
    - t_{b}{}^{\nu}\partial_{\nu}\rho_{m}{}^{\mu}
    + Y_{m}{}^{I}\rho_{Ib}{}^{\mu}\big) \label{QQ1} \\[4pt] 
    & \hspace{320pt} \nonumber - \theta^{a}\theta^{b}
    \rho_{mb}{}^{c}\partial_{c} t_{a}{}^{\mu} = 0\,, \\[4pt]
    & \rho_{I}{}^{a} t_{a}{}^{\mu} + W_{I}{}^{m}\rho_{m}{}^{\mu}
    + V^{a}\rho_{Ia}{}^{\mu}
    - \theta^{a}\big(\rho_{I}{}^{b}\partial_{b} t_{a}{}^{\mu}
    + V^{b}\partial_{b}\rho_{Ia}{}^{\mu}\big) \label{QQ2}\\[4pt]
    & \hspace{185pt} \nonumber -\,\theta^{a}\theta^{b}
    \big(\rho_{Ib}{}^{\nu}\partial_{\nu}t_{a}{}^{\mu}
    + Z_{Ib}{}^{J}\rho_{Ja}{}^{\mu}
    + t_{b}{}^{\nu}\partial_{\nu}\rho_{Ia}{}^{\mu}\big) = 0\,.
\end{align}
\end{subequations}
From the action on $\theta^{a}$, we obtain
two additional conditions:
\begin{subequations}
\begin{align}
    & \rho_{m}{}^{\mu}\partial_{\mu}V^{a}
    - V^{b}\rho_{mb}{}^{a} + Y_{m}{}^{I}\rho_{I}{}^{a}
    + \theta^{b}\big(\rho_{mb}{}^{c}\partial_{c}V^{a}
    + V^{c}\partial_{c}\rho_{mb}{}^{a}\big) \label{QQ3} \\[4pt] 
    & \hspace{200pt} \nonumber -\,\theta^{b}\theta^{c}
    \big(U_{mc}{}^{m}\rho_{mb}{}^{a}
    - t_{c}{}^{\mu}\partial_{\mu}\rho_{mb}{}^{a}\big) = 0\,,\\[4pt] 
    & \rho_{I}{}^{b}\partial_{b}V^{a}
    + V^{b}\partial_{b}\rho_{I}{}^{a}
    + \theta^{b}\big(\rho_{Ib}{}^{\mu}\partial_{\mu}V^{a}
    + Z_{Ib}{}^{J}\rho_{J}{}^{a} + W_{I}{}^{m}\rho_{mb}{}^{a}
    + t_{b}{}^{\mu}\partial_{\mu}\rho_{I}{}^{a}\big) = 0\,. 
    \label{QQ4}
\end{align} 
\end{subequations}
The action on $a^m$ gives three more equations:
\begin{subequations}
\begin{align}
    & V^{a} \partial_{a} f_{np}{}^{m}
    +\theta^{a}\big(f_{np}{}^{q} U_{qa}{}^{m}
    - 2\rho_{[n|a}{}^{b} U_{|p]b}{}^{m}
    - 2\rho_{[n}{}^{\mu} \partial_{\mu} U_{p]a}{}^{m}
    + f_{npa}{}^{I} W_{I}{}^{m} \\[4pt] 
    &\hspace{55pt} \nonumber +\,2 U_{[n|a}{}^{q}f_{|p]q}{}^{m}
    + t_{a}{}^{\mu} \partial_{\mu} f_{np}{}^{m}
    + 2Y_{[n}{}^{I} f_{p]Ia}{}^{m}\big) + 2\theta^{a}\theta^{b}
    \rho_{[n|a}{}^{c} \partial_{c} U_{|p]b}{}^{m}=0\,, \\[4pt] 
    & \rho_{n}{}^{\mu} \partial_{\mu} W_{I}{}^{m}
    - \rho_{I}{}^{a} U_{na}{}^{m} + V^{a} f_{nIa}{}^{m}
    - f_{nI}{}^{J} W_{J}{}^{m} + W_{I}{}^{p} f_{np}{}^{m}
    - Y_{n}{}^{J} f_{IJ}{}^{m} \\[4pt] 
    & \qquad +\,\theta^{a}\big(\rho_{I}{}^{b} \partial_{b} U_{ma}{}^{m}
    + \rho_{na}{}^{b} \partial_{b} W_{I}{}^{m}
    - V^{b} \partial_{b} f_{nIa}{}^{m}\big) \nonumber \\[4pt] 
    & \qquad\qquad \nonumber +\, \theta^{a}\theta^{b}
    \big(t_{a}{}^{\mu} \partial_{\mu} f_{nIb}{}^{m}
    - \rho_{Ia}{}^{\mu} \partial_{\mu} U_{nb}{}^{m}
    - f_{nIa}{}^{p} U_{pb}{}^{m} - U_{na}{}^{p} f_{pIb}{}^{m}
    + Z_{Ia}{}^{J}f_{nJb}{}^{m}\big) = 0\,,\\[4pt]
    & 2\rho_{(I}{}^{a} \partial_{a} W_{J)}{}^{m}
    - V^{a} \partial_{a} f_{IJ}{}^{m}
    - \theta^{a}\big(f_{IJ}{}^{n} U_{na}{}^{m}
    + f_{IJa}{}^{K} W_{K}{}^{m}
    - 2\rho_{(I|a}{}^{\mu} \partial_{\mu} W_{|J)}{}^{m} \\[4pt] 
    & \hspace{155pt} \nonumber +\,2W_{(I|}{}^{n} f_{n|J)a}{}^{m}
    + 2 Z_{(I|a}{}^{K}f_{|J)K}{}^{m}
    + t_{a}{}^{\mu} \partial_{\mu} f_{IJ}{}^{m}\big) = 0\,.
\end{align}
\end{subequations}
Finally, the action on $\chi^I$ gives the final three equations: 
\begin{subequations}
\begin{align}
    & 2\rho_{[m}{}^{\mu} \partial_{\mu} Y_{n]}{}^{I}
    - V^{a} f_{mna}{}^{I} - f_{mn}{}^{p} Y_{p}{}^{I}
    - 2Y_{[m}{}^{J} f_{n]J}{}^{I}
    + \theta^{a}\big(2\rho_{[m|a}{}^{b} \partial_{b} Y_{|n]}{}^{I}
    + V^{b} \partial_{b} f_{mna}{}^{I}\big) \\[4pt] 
    & \hspace{150pt} \nonumber -\, \theta^{a}\theta^{b}
    \big(f_{mna}{}^{J} Z_{Jb}{}^{I} - 2U_{[m|a}{}^{p}f_{|n]pb}{}^{I}
    - t_{a}{}^{\mu} \partial_{\mu} f_{mnb}{}^{I}\big) = 0\,, \\[4pt] 
    & \rho_{J}{}^{a} \partial_{a} Y_{m}{}^{I}
    - V^{a} \partial_{a} f_{mJ}{}^{I}
    -\theta^{a}\big(\rho_{m}{}^{\mu} \partial_{\mu} Z_{Ja}{}^{I}
    + \rho_{ma}{}^{b} Z_{Jb}{}^{I}
    - \rho_{Ja}{}^{\mu} \partial_{\mu} Y_{m}{}^{I}
    + t_{a}{}^{\mu} \partial_{\mu} f_{mJ}{}^{I} \\[4pt]
    & \qquad \nonumber + W_{J}{}^{n}f_{mna}{}^{I}
    + Z_{Ja}{}^{K} f_{mK}{}^{I} - U_{ma}{}^{n}f_{nJ}{}^{I}
    - Y_{m}{}^{K} f_{JKa}{}^{I} - f_{mJa}{}^{n} Y_{n}{}^{I}
    - f_{mJ}{}^{K}Z_{Ka}{}^{I}\big) \\[4pt] 
    & \qquad\qquad \nonumber +\, \theta^{a}\theta^{b}
    \rho_{ma}{}^{c}\partial_{c}Z_{Jb}{}^{I} = 0\,,\\[4pt] 
     & 2\rho_{(J}{}^{a} Z_{K)a}{}^{I} - f_{JK}{}^{m}Y_{m}{}^{I} 
    - 2W_{(J|}{}^{m}f_{m|K)}{}^{I} - V^{a} f_{JKa}{}^{I} \\[4pt] 
    & \quad -\, \theta^{a}\big(2\rho_{(J}{}^{b}\partial_{b}Z_{K)a}{}^{I}
    - V^{b} \partial_{b} f_{JKa}{}^{I}\big) \nonumber\\[4pt] 
    &\qquad \nonumber +\,\theta^{a}\theta^{b}
    \big(2\rho_{(J|a}{}^{\mu} \partial_{\mu} Z_{|K)b}{}^{I}
    - 2Z_{(J|a}{}^{L} f_{|K)Lb}{}^{I} - f_{JKa}{}^{L} Z_{Lb}{}^{I}
    - t_{a}{}^{\mu} \partial_{\mu} f_{JKb}{}^{I}\big) = 0\,.
\end{align}
\end{subequations}

\subsection{Compatibility of gauge symmetry and symplectic form}
\label{app:Q_omega}
We consider now the symplectic case $T^{\ast}[1,0]E[0,1]$, 
recalling that the symplectic form of $\Z$-degree $1$
is taken to be 
\begin{equation}
    \omega = \dd a_{\mu} \wedge \dd x^{\mu}
        +\dd \chi_{a} \wedge \dd \theta^{a}\,,
\end{equation}
in therms of the coordinates $x^{\mu}, \theta^{a}$
and the conjugate momenta $a_{\mu}, \chi_{a}$.
The compatibility of the homological vector field $\cQ$
that generates the gauge symmetries of the supersymmetric 
Poisson sigma model and the symplectic form reads
\begin{equation}
    {\mc L}_{\cQ} \omega = 0\,.
\end{equation}
In detail, this imposes the following conditions
on the coefficients that appear in $\cQ$: 
\begin{subequations}
\begin{eqnarray}
    && \rho^{\mu\nu}=-\rho^{\nu\mu}\,,
    \quad \rho^{a\mu}=\rho^{\mu a}\,,
    \quad \rho^{ab}=\rho^{ba}\,, \\[4pt] 
    && f^{\kappa\lambda}{}_{\mu}
        = \partial_{\mu}\rho^{\kappa\lambda}\,,
    \quad f^{ab}{}_{\mu} = -\partial_{\mu}\rho^{ab}\,,
    \quad f^{\kappa b}{}_{a}
        = \partial_{a}\rho^{\kappa b}\,, \\[4pt] 
    && \theta^{b} f^{\kappa a}{}_{b\mu}
        = -\partial_{\mu}\rho^{\kappa a}\,,
    \quad \theta^{b}f^{\kappa\lambda}{}_{ba}
    = \partial_{a}\rho^{\kappa\lambda}\,,
    \quad \theta^{d}f^{bc}{}_{da} = -\partial_{a}\rho^{bc}\,.
\end{eqnarray}
\end{subequations}
These correspond to the more compact Eq. \eqref{eq:Q_omega}
of the main text.

\subsection{Compatibility of supersymmetry and symplectic form}
\label{app:QS_omega}
Finally, we ask whether the homological vector field $\cQ_{S}$ associated to supersymmetry transformations is compatible
with the symplectic form $\omega$, i.e. 
\begin{equation}
    {\mc L}_{\cQ_{S}}\omega=0\,.
\end{equation}
This gives rise to the following 
independent conditions: 
\begin{align}\label{QS omega compatibility}
    Y^{\mu}{}_{a} = - t_{a}{}^{\mu}
    + \theta^{b} \partial_{a} t_{b}{}^{\mu}\,,
    \qquad
    U^{\mu}{}_{a\nu} = \partial_{\nu} t_{a}{}^{\mu}\,,
    \qquad
    W^{a}{}_{\mu} = \partial_{\mu} V^{a}\,,
    \qquad
    \theta^{c} Z^{a}{}_{cb} = \partial_{b} V^{a}\,.
\end{align}

Evidently this fixes all additional coefficients in terms
of the basic coefficients $t_{a}{}^{\mu}$ and $V^{a}$.

\providecommand{\href}[2]{#2}\begingroup\raggedright\endgroup

\end{document}